\documentclass[preprint]{aastex63}
\usepackage{CJK}
\usepackage{amsmath,amstext, amssymb}
\usepackage[T1]{fontenc} 
\usepackage{graphicx}
\usepackage{color}
\usepackage{tablefootnote}
\usepackage{subfigure}
\usepackage{natbib}
\usepackage{bm}

\received{June 1, 2019}
\revised{January 10, 2019}
\accepted{\today}
\submitjournal{ApJ}

\shorttitle{Elliptical accretion disks}
\shortauthors{Liu et al.}

\begin{document}

\title{Elliptical accretion disk as a model for tidal disruption events}


\correspondingauthor{F.K. Liu}
\email{fkliu@pku.edu.cn}

\author{F.K. Liu}
\affiliation{Department of Astronomy, School of Physics, Peking University, Beijing 100871, China}
\affiliation{Kavli Institute for Astronomy and Astrophysics, Peking University, Beijing 100871, China}

\author{C.Y. Cao}
\affiliation{Department of Astronomy, School of Physics, Peking University, Beijing 100871, China}

\author{M.A. Abramowicz}
\affiliation{Nicolaus Copernicus Astronomical Center, Polish Academy of Sciences, Bartycka 18, PL-00-716 
    Warszawa, Poland}
\affiliation{Physics Department, University of Gothenburg, 412-96 G\"oteborg, Sweden}
\affiliation{Kavli Institute for Astronomy and Astrophysics, Peking University, Beijing 100871, China}
\affiliation{Institute of Physics, Silesian University in Opava, Czech Republic}

\author[0000-0002-8635-4242]{M. Wielgus}
\affiliation{Black Hole Initiative at Harvard University, 20 Garden St., Cambridge, MA 02138, USA}
\affiliation{Center for Astrophysics | Harvard \& Smithsonian, 60 Garden Street, Cambridge, MA 02138, USA}
\affiliation{Kavli Institute for Astronomy and Astrophysics, Peking University, Beijing 100871, China}

\author{R. Cao}
\affiliation{Department of Astronomy, School of Physics, Peking University, Beijing 100871, China}

\author{Z.Q. Zhou}
\affiliation{Department of Astronomy, School of Physics, Peking University, Beijing 100871, China}



\begin{abstract}
Elliptical accretion disk models for tidal disruption events (TDEs) have been recently proposed and 
independently developed by two groups. Although these two models are characterized by a similar geometry, 
their physical properties differ considerably. In this paper, we further investigate the properties of the elliptical 
accretion disk of the nearly uniform distribution of eccentricity within the disk plane. Our results show that the elliptical 
accretion disks have distinctive hydrodynamic structures and spectral energy distributions, associated with TDEs.
The soft X-ray photons generated at pericenter and nearby are trapped in the disk and advected around the 
ellipse because of large electron scattering opacity. They are absorbed and reprocessed into emission lines and 
low-frequency continuum via recombination and bremsstrahlung emission. Because of the rapid increase of 
bound-free and free-free opacities with radius, the low-frequency continuum photons become trapped in 
the disk at large radius and are advected through apocenter and back to the photon-trapping 
radius.  Elliptical accretion disks predict sub-Eddington luminosities and emit mainly at the photon-trapping 
radius of thousands of Schwarzschild radii with a blackbody spectrum of nearly single temperature of typically 
about $3\times 10^4 \, {\rm K}$. Because of the self-regulation, the photon-trapping radius expands and contracts 
following the rise and fall of accretion rate. The radiation temperature is nearly independent of BH mass and 
accretion rate and varies weakly with the stellar mass and the viscosity parameter. Our results are well consistent 
with the observations of optical/UV TDEs. 
\end{abstract}



\section{Introduction} \label{sec:intro}

When a star closely passing by a supermassive black hole (SMBH) is tidally disrupted, the subsequent accretion 
of stellar debris ignites the otherwise quiescent BH \citep{hil75,ree88}. In the classical work of \citet{ree88}, 
the bound stellar debris is expected to be circularized rapidly because of strong relativistic apsidal precession
and to form a circular accretion disk of about twice the size of the orbital pericenter of the star. The hydrodynamic 
simulations of stellar tidal disruptions show that rapid circularization of the debris streams and the formation of 
a compact accretion disk are allowed when the orbital pericenter of the star is of order of the BH Schwarzschild 
radius and the self-intersection of the streams occurs near the orbital pericenter, which is caused by
 the strong relativistic 
apsidal precession \citep{ram09,hay13,gui14,bon16,sad16,ste19,kro20,ryu20}. Because the accretion disk is hot and 
radiates dominantly in soft X-rays, the emission in optical/UV wave bands is the extremely weak Rayleigh-Jeans 
tail, and no strong broad optical emission line is expected \citep{str09}. 

The observations of tidal disruption events (TDEs) and candidates discovered in the optical/UV sky-surveys are 
inconsistent with the
expectations and challenge the classical circular accretion disk model \citep[see][for a recent review of 
observations]{kom15}. Most optical/UV TDEs and candidates are much more luminous in optical/UV wave bands 
than expected and with little or no radiation in soft X-rays \citep[e.g.,][]{gez12,hol14,hol16a,hol16b,
bla19,lel19,van20} and have strong broad optical emission lines of complex and diverse profiles 
\citep{kom08,gez12,wan12,arc14,hol14,hol16b,hol16a,hol19,sho20,van20} and of peculiar chemical 
abundance \citep{gez12,arc14,sho20}. The spectral energy distributions (SEDs) of all optical/UV TDEs are 
blackbodies of nearly single and constant temperature at about $(1-6) \times 10^4 \, {\rm K}$ and the effective 
spherical blackbody radii of optical/UV radiations are as large as a few thousand Schwarzschild radii of 
SMBHs and much larger than the expected tidal disruption radius of main-sequence stars \citep{gez12,hol14,
hol16a,hol16b,hol19,nic19,van20}. To solve the problem, it is proposed that an optically thick envelope of a 
few thousand Schwarzschild radii, enshrouding the compact hot accretion disk, forms and reprocesses the 
soft X-ray photons to low frequency at large radius \citep{str09,lod11,met16,rot16,dai18}. The optically thick 
envelope may be formed by the disk outflows or winds during the super-Eddington accretion at 
the early time, and the photosphere would recede with the decay of the disk outflows and winds following 
the fallback rate. The broad optical emission lines form in the surface layers of the optically thick envelopes 
\citep{rot16}. A top-capped light curve regulated by the Eddington luminosity would be expected with 
the super-Eddington disk-envelope model for TDEs and significantly decoupled from the fallback rate during 
the super-Eddington accretion.

The close follow-up of the bolometric luminosity to the fallback rate of TDEs \citep[e.g.][]{moc19} challenges the 
model of the super-Eddington accretion and strong outflows. The observations of sub-Eddington peak 
bolometric luminosity and the extremely small amount of total accreted matter of TDEs challenge the circular 
accretion disk model \citep{sax18}. The hydrodynamic simulations of tidal disruption of a star with orbital 
pericenter radius much larger than the BH Schwarzschild radius show that the circularization of the bound debris 
streams is quite inefficient and the circularization timescale is much longer than the fallback timescale, because
the relativistic apsidal precession is little and the self-intersection of the streams occurs nearly at the apocenter 
of the most bound stellar debris \citep{shi15}. Inspired by the hydrodynamic simulations, \citet{pir15} 
proposed that the optical and UV radiation of optical/UV TDEs is powered by the shocks at the self-intersection 
of streams, rather than accretion onto the SMBH. Provided that the fallback materials, including the most bound 
stellar debris, have zero initial bound energy and the thermal energy converted from the orbital kinetic energy 
by the shocks can be radiated efficiently with little of the heat being converted back into kinetic energy, 
they showed that the shock model can approximately explain the observations of optical luminosities, the 
low temperature, and the large radiation radius of optical/UV TDEs at the peak brightness. Because the shock 
model neglects the radiation contributions of disk accretion, it has to explain what happens to this radiation 
\citep{pir15}. It is argued that the radiation originating from the accretion disk may be much weaker than 
that originating from the shocks at the apocenter, if the angular momentum transport in the accretion disk
is much faster than the orbital energy dissipation, and the eccentricity of the inner parts of the debris disk rises 
so much that the pericenter radius of the orbits of the inner accretion disk reaches the marginally bound 
orbit, and the matter plunges directly onto the BH without significant decrease of the orbital semimajor 
axis \citep{svi17,cha18}. 

The evolution of eccentricity of accretion disks is complex. The hydrodynamic investigations of eccentric 
accretion disks show that the eccentricity of an elliptical accretion disk may or may not decrease with time 
because the viscosity may not lead to the circularization of a disk \citep{sye92,lyu94,ogi01,ogi14}. 
The investigations of magnetohydrodynamic (MHD) stress and the magnetorotational instability (MRI) in 
eccentric accretion disks show that in some situations the angular momentum transport is more efficient,
but in other cases the orbital energy dissipation is more significant. It is unclear which one is preferred in the 
accretion disk of TDEs, and more investigations are needed \citep{svi17,cha18}. 

Liu and colleagues \citep{liu17,cao18} recently proposed an elliptical accretion disk model for TDEs, whose 
size and eccentricity are mainly determined by the orbital pericenter of a star and the relativistic apsidal 
precession of the most bound stellar debris. The orbital eccentricity of the disk fluid elements is nearly the same 
across different radii. The inner 
edge is determined by the marginal bound orbit and the eccentricity of the inner disk fluids. They suggested that 
the double-peaked broad H$\alpha$ emission line of the TDE candidate PTF09djl, as a reminiscent  of the 
double-peaked broad Balmer emission lines of active galactic nuclei (AGNs), originates in the eccentric accretion 
disk \citep{liu17}. 
They suggested to observationally determine the structure of the accretion disk of TDEs by modeling the 
complex and asymmetric profiles of broad optical emission lines of optical/UV TDEs \citep{liu17,cao18}. 
The disk origination of broad optical emission lines can naturally explain the complexities, asymmetries,
and diversities of the line profiles with the random disk inclinations and pericenter orientations 
\citep{liu17,cao18,hol19,sho20}, the peculiar intensity ratios of broad emission lines of the He and H elements 
with the chromosphere of the optically thick disk as in cataclysmic variables and hot main-sequence stars 
\citep{gas14,gui14,rot16,liu17,cao18}, and the flat Balmer decrement of some optical/UV TDEs 
\citep{sho20}. Modeling the double-peaked line profiles of PTF09djl suggests that the peculiar substructures 
of the line profiles with one peak at the line rest wavelength and the other redshifted to about $3.5\times 
10^4\, {\rm km\; s^{-1}}$ are mainly due to the orbital motion of the emitting matter within the disk plane 
of size a few hundred Schwarzschild radii and of roughly uniform eccentricity of about $0.966$ \citep{liu17}. 
Modeling the optical emission lines of the TDE ASASSN-14li with radically different profiles also reveals
a large accretion disk of size up to 1700 Schwarzschild radii and roughly uniform eccentricity about $0.977$ 
\citep{cao18}. 

Elliptical accretion disk models for the TDEs have been independently suggested by two groups 
-- \citet{pir15} and \citet{liu17}. The similarities of these models follow from the fact that the TDEs originate from 
matter with a lot of energy (of order of the binding energy at the apocenter) and very little angular momentum
(of order of the angular momentum of rotation at the pericenter). Unless there 
is a rapid process of energy dissipation, this will lead to an elliptical disk. At the same time, there is an "inverse energy
crisis" as the energy observed in TDEs is around $10^{51}\, {\rm erg}$ (or at most $10^{52} \, {\rm erg}$) while the 
energy reservoir has $> 10^{53} \, {\rm erg}$. This suggests that the elliptical disk does not circularize quickly, if at all. 
Although the two models are characterized by a similar geometry, their physical properties differ considerably.

Piran and colleagues argued that the energy dissipation at the disk is not the energy source of the observed 
radiation. The energy source of the radiation is the stream-stream interaction that takes place at around the 
apocenter. The dissipation processes that take place in the elliptical disk are mostly unimportant given the 
fact that it loses some small fraction of its angular momentum and then the matter falls ballistically onto the BH without 
energy production \citep{pir15,svi17}. \citet{pir15a} coined the name "Jerusalem bagel" for these disks because of 
their oval shape -- in contrast to the circular thick accretion disks, often referred to as Polish doughnuts \citep{abr78}.
The dynamics and thermal emission of the elliptical accretion disk heated by the self-crossing shocks at about 
apocenter have been investigated recently \citep{zan20}.

In contrast, the elliptical accretion disk model of the Beijing group assumes that the energy dissipation occurs 
mainly in the accretion of the matter into the BH, because of the uniform eccentricity of the fluid orbits. The 
energy dissipation from the shocks at apocenter is unimportant \citep{liu17,cao18,zho20}. To stress similarities and 
differences, we call these disks "Jerusalem bagels from Beijing."

The accretion of matter with large orbital eccentricity onto the central BH would convert a small but 
significant amount of the orbital kinetic energy into heat \citep{liu17,cao18,zho20}. Based on the
relativistic elliptical accretion disk model, we recently calculated the conversion efficiency of matter
into radiation and compared the model expectations of peak luminosity and total radiation energy with 
the observations of a sample of TDEs \citep{zho20}. The results showed that the expectations of both the 
peak luminosity and total radiation energy of TDEs are well consistent with the observations. The masses 
of SMBHs of the TDEs, derived by comparing the model expectations and the observations, are well consistent 
with the estimates of BH masses of the sample TDE sources obtained with the $M_{\rm BH}-\sigma_*$ 
relation of the BH mass and the stellar velocity dispersion of host galaxy \citep{zho20}. 

In this paper, we investigate the dynamic structure and SED of an elliptical accretion disk of uniform eccentricity. 
The relationship of the TDEs discovered in the optical/UV and X-ray transient surveys will be discussed in the next 
work. Following \citet{liu17} and \citet{cao18}, for simplicity we assume that the elliptical accretion disk consists 
of an aligned ellipse of uniform eccentricity. Because the orbital energy dissipation happens mainly at pericenter 
and near regions with radius $r\sim r_{\rm p}$ and $-\pi/2 \la \phi \la \pi/2$, with $\phi$ the azimuthal angle 
starting at pericenter \citep{svi17,cha18}, and the angular momentum transport near apocenter may 
be less efficient \citep{cha18}, we assume for simplicity that the heat generation and angular momentum 
transfer occur only at pericenter and nearby and that both effects can be described by introducing an 
effective viscosity parameter $\alpha$ of a step function with $\alpha = \alpha_{\rm p}$ for $ r  \sim 
r_{\rm p}$ and $-\pi/2 \la \phi \la \pi/2$ and $\alpha = 0$ otherwise.

The paper is organized as follows. In section~\ref{sec:diskmodel} we briefly introduce the elliptical accretion 
disk model for TDEs. Sections~\ref{sec:mass}-\ref{sec:vischeating} discuss, respectively, 
the mass conservation, angular momentum conservations, and heat generation of a vertically integrated 
quasi-stationary elliptical accretion disk. In section~\ref{sec:structellipse}, we investigate the hydrodynamic 
structures around a highly eccentric ellipse. The vertical hydro-equilibrium and the distributions of mass density, 
temperature, and radiation around the ellipse are distinctively different from those in circular annulus of the 
circular accretion disk. In section~\ref{sec:compobs}, we compare the expectations of the elliptical accretion 
disk model and the observations of optical/UV TDEs. It is shown that the expectations are well consistent with the 
observations. A brief discussion and conclusions are given in section~\ref{sec:dis}.

\section{The elliptical accretion disk model for TDEs} \label{sec:diskmodel}

\subsection{The elliptical accretion disk}

In this section, we summarize the properties of the elliptical accretion disk model. A more detailed and 
complete description can be found in \citet{liu17} and \citet{cao18}.

A star of radius $R_*$ and mass $M_*$ is tidally disrupted by an SMBH of mass $M_{\rm 
BH}$, when the orbital pericenter radius of the star, $r_{\rm p*}$, is less than the tidal disruption radius 
\begin{eqnarray}
       r_{\rm t} & = & f_{\rm T} R_* (M_{\rm BH}/M_*)^{1/3} \cr
       & \simeq & 23.545 f_{\rm T} r_* m_*^{-1/3} M_6^{-2/3} r_{\rm S} ,
\label{eq:tidalr}
\end{eqnarray}
where $r_* = R_* / R_\odot$, $m_* = M_* / M_\odot$,  $M_6 =  M_{\rm BH} / 10^6  M_\odot$, and
$r_{\rm S} = 2 G M_{\rm BH}/c^2$. The correction factor $f_{\rm T}$ depends on the internal stellar structure
\citep{phi89,gui13,ryu20,ryu20b} and relativistic effects \citep{iva06,ryu20}. The general relativistic hydrodynamic 
simulations of tidal disruptions of a main sequence star give the correction factor $f_{\rm T} = f_{\rm BH} f_*$ with 
$ f_{\rm BH} \simeq 0.80 + 0.26 M_6^{0.5} $ and $f_* \simeq 1.47$ for a star of mass $m_* \la 0.5 $ and $f_* \simeq 
1/2.34$ for star of mass $m_* \ga 1$ \citep{ryu20}. For star of typical mass $m_* = 0.3$ and BH of typical mass $M_6 
= 1$, $f_{\rm T} \simeq 1.56$. We notice that here we adopt the Latin letters $f_{\rm T}$, $f_*$ and $f_{\rm BH}$ 
to note the correction factors, which are, respectively, denoted by the Greek letters $\Psi$, $\Psi_*$ 
and $\Psi_{\rm BH}$ in \citet{ryu20}. After tidal disruption, about half of the stellar debris becomes bound and 
returns to the orbital pericenter of the star. The fallback rate of the bound stellar debris after peak is approximately 
a power-law, 
\begin{equation}
\dot{M} \simeq \dot{M}_{\rm p}  \left({t+\Delta{t}_{\rm p} \over \Delta{t}_{\rm p}}\right)^{-n} ,
\label{eq:accr}
\end{equation}
where time $t$ starts at peak fallback rate and the power-law index $n$ depends on both the structure 
and age of the star \citep{ree88,eva89,phi89,lod09,gui13,sto13} and the orbital penetration factor $\beta_* 
= r_{\rm t}/r_{\rm p*}$ \citep{gui13,cou19,ryu20}. For full tidal disruptions, $n=5/3$ is a very good approximation 
except for at about the peak time \citep{gui13}, and for partial disruptions $n=9/4$ is more typical 
\citep{gui13,cou19,ryu20}. The peak time $\Delta{t}_{\rm p} $ and peak fallback rate $\dot{M}_{\rm p}$ are, 
respectively, 
\begin{eqnarray}
\Delta{t}_{\rm p} & \simeq & 2\pi \left({a_{\rm mb}^3 \over G M_{\rm BH}}\right)^{1/2} \cr
       &\simeq & 0.1122 f_{\rm T}^3 r_*^{3/2} m_*^{-1} M_6^{1/2} \, {\rm yr}
\end{eqnarray}
with $a_{\rm mb} \simeq {r_{\rm t}^2 / (2 R_*)}$ the orbital semi-major axis of the most tightly bound stellar debris
and 
\begin{eqnarray}
\dot{M}_{\rm p} & \simeq&  {M_* \over 3 \Delta{t}_{\rm p} } \left[{3\over 2} (n-1)\right]  \cr
      & \simeq & 2.972 \left[3(n-1)/2\right] f_{\rm T}^{-3} r_*^{-3/2} m_*^2  M_6^{-1/2} \; M_\odot/{\rm yr} .
\label{eq:pmass}
\end{eqnarray}
When it is needed in this paper, we adopt the mass-radius relation for a main-sequence star to convert the radius
to the mass of a star: $r_* \simeq m_*^{1-\zeta}$ with $\zeta =0.21$ for $0.1 \leq m_* \leq 1$ and $\zeta =0.44$ for 
$1 \leq m_* \leq 150$ \citep{kip12}. 

In the popular circular accretion disk model for TDEs in the literature, the typical radiation
efficiency $\eta=0.1$ is adopted and the peak mass fallback rate given by equation~(\ref{eq:accr}) leads to 
hyper-Eddington luminosities for TDEs with the BH mass $M_{\rm BH} \sim 10^6 M_\odot$ and strong outflows may
be driven by the radiation pressure \citep{str09,lod11,met16,rot16,dai18} or may not (see the theoretical 
arguments given by \citet{abr00}).  The real accretion rate of matter onto the BH is expected to significantly 
deviate from the mass fallback rate given by equation~(\ref{eq:accr}). Because the radiation efficiency of the 
elliptical accretion disk is as small as $\sim 10^{-3}$ 
\citep[][see also equation~(\ref{eq:coneff})]{liu17,cao18,zho20} and the peak luminosity is sub-Eddington for TDEs 
with the BH mass $M_{\rm BH} \ga 10^6 M_\odot$ (see also equation~(\ref{eq:coneff1})), no strong outflow is 
expected for the elliptical accretion disks, and the accretion rate of matter onto the BH would closely follow the 
mass fallback rate given by equation~(\ref{eq:accr}). In this paper, the results are given as functions of the 
accretion rate $\dot{M}$ and the peak fallback rate $\dot{M}_{\rm p}$ given by equation~(\ref{eq:pmass}) is 
adopted mainly for scaling the accretion rate. The results are mainly determined by the accretion rate and 
nearly independent of the power-law index $n$. However, when it is needed, we will assume that the accretion 
rate is the mass fallback rate given by equation~(\ref{eq:accr}) and present the results as functions of time for the 
typical value $n=5/3$.  As an example,  we will discuss the results obtained for both $n=5/3$ and $n=9/4$ in 
Sec.~\ref{sec:obs_BB_rad}.

The semi-major axis of the elliptical orbit of the bound stellar debris after fallback is reduced to form an
accretion disk mainly due to the shocks of the intersection of the newly inflowing and post-pericenter outflowing
fluid streams because of the relativistic apsidal precession \citep{ree88,eva89,koc94,hay13,dai15,shi15,bon16,hay16}.
The semimajor axis $a_{\rm d}$ of the accretion disk determined by the location of the self-intersections 
is approximately 
\begin{eqnarray}
a_{\rm d} & \simeq & {r_{\rm p*} \over 1-e_{\rm mb} + 2^{-1}\sin^2(\Omega_{\rm dS}/2)} \cr
      & \simeq & {2 \beta_*^{-1} r_{\rm t} \over 2\delta + \sin^2(\Omega_{\rm dS}/2)}
      \label{eq:dsize}
\end{eqnarray}
and the eccentricity of accretion disk given by the conservation of angular momentum of the streams is
\begin{eqnarray}
e_{\rm d} & \simeq & \left[1 - {(1-e_{\rm mb}^2) a_{\rm mb} \over a_{\rm d}}\right]^{1/2} \cr
      &\simeq & \left[1 - 2\delta (1+\Delta_*) \right]^{1/2}
     \label{eq:decc}
\end{eqnarray}
with $\Delta_* \equiv \sin^2(\Omega_{\rm dS}/2)/2\delta$ and $\delta = 2{R_* r_{\rm p*} / r_{\rm t}^2} \simeq 
0.02 f_{\rm T}^{-1} \beta_*^{-1} m_*^{1/3} M_6^{-1/3}$ \citep{liu17,cao18}, where $e_{\rm mb} = 1- \delta$ is the 
orbital eccentricity of the most bound stellar debris and $\Omega_{\rm dS}$ is the instantaneous de Sitter 
precession at periapse of the most bound stellar debris \citep{des16},
\begin{eqnarray}
\Omega_{\rm dS} & \simeq & {6\pi G M_{\rm BH} \over c^2 (1-e_{\rm mb}^2) a_{\rm mb}}  \cr
      & \simeq & {3\pi  r_{\rm S} \over (1+e_{\rm mb}) r_{\rm p*}} \cr
      & \simeq & {3\pi \over 2}  {r_{\rm S} \over r_{\rm t}} \beta_* .
      \label{eq:deSitter}
\end{eqnarray}
From equation~(\ref{eq:deSitter}), we have $\Delta_* \simeq {1\over 4} f_{\rm T}^{-1} \beta_*^3 r_*^{-2} 
m_*^{1/3} M_6^{5/3}$ for $r_{\rm p*} \gg r_{\rm S}$.

Following \citet{liu17} and \citet{cao18},  we assume for simplicity that the eccentric accretion disk 
consists of a nested aligned ellipse of semimajor axis $a$ and uniform eccentricity $e$ with $e=e_{\rm d}$ and 
that the fluid elements in the cylindrical coordinates ($r$, $\phi$, $z$) have trajectories 
\begin{equation}
r = {a (1 - e^2) \over 1+ e \cos(\phi )} ,
\label{eq:streamline}
\end{equation}
where $r$ is the radius from the center of the BH and $\phi$ starts from the orientation of pericenter. 

To include the general relativistic effects in our Newtonian treatments, we adopt the generalized Newtonian 
potential in the low-energy limit 
\begin{eqnarray}
\Phi_{\rm G}(r,\dot{r}, \dot{\phi}) &= &  -{G M_{\rm BH} \over r} - \left({2r_{\rm g} \over r - 2 r_{\rm 
          g}}\right) \times \cr
&& \left[\left({r-r_{\rm g} \over r- 2 r_{\rm g}}\right) \dot{r}^2 + {r^2\dot{\phi}^2 \over 2}\right] 
\label{eq:gNP}
\end{eqnarray}
with $r_{\rm g} =r_{\rm S}/ 2 = G M_{\rm BH}/c^2$ the gravitational radius \citep{tej13}, which is a good 
approximation for  
particles with large eccentricity and low bound energy. With the generalized Newtonian potential, the 
trajectories of particles in Schwarzschild space-time and the radial dependences of the specific binding 
energy and angular momentum of the elliptical orbits can be  reproduced exactly. Provided the semimajor 
axis $a$ and eccentricity $e$ of the elliptical orbit, the specific angular momentum $l_{\rm G}$ and binding 
energy $e_{\rm G}$ are, respectively, 
\begin{eqnarray}
l_{\rm G}  & = & {(1-e^2) a  c \sqrt{r_{\rm S}}  \over \sqrt{2 (1-e^2) a   - (3 + e^2) r_{\rm S}}} \cr
    & = & {(1+e) r_{\rm p}  c \sqrt{r_{\rm S}}  \over \sqrt{2 (1+e) r_{\rm p}   - (3 + e^2) r_{\rm S}}} ,
     \label{eq:ang} \\
e_{\rm G} & = & {c^2 \over 2} {[(1-e^2)a - 2r_{\rm S}]r_{\rm S} \over a \left[2 (1-e^2) a -
	(3+e^2)r_{\rm S}\right]} \cr
	& = & {c^2 \over 2} {[(1+e)r_{\rm p} - 2r_{\rm S}]r_{\rm S} \over a \left[2 (1+e) 
	r_{\rm p}- (3+e^2)r_{\rm S}\right]}
\label{eq:eng}
\end{eqnarray}
\citep{liu17,cao18}, where $r_{\rm p}=(1-e) a$ is the pericenter radius of the elliptical orbit. Noticing 
$(3+e^2)/2(1+e) = 1+[(1-e)/2]^2[2/(1+e)]$ and neglecting the terms $[(1-e)/2]^2$ or higher, we have
\begin{eqnarray}
l_{\rm G} &\simeq& \left({1+e\over 2}\right)^{1/2} \left({r_{\rm p}\over r_{\rm S}}\right)^{1/2}
      \left(1 - {r_{\rm S} \over r_{\rm p}}\right)^{-1/2} r_{\rm S} c 
       \label{eq:ang1} \\
e_{\rm G} &\simeq& {c^2 \over 4} \left({r_{\rm S} \over a}\right) \left[1 - \left({2 \over 1+e}\right) 
       \left({r_{\rm S} \over r_{\rm p}}\right)\right] \left[1 - \left({r_{\rm S} \over r_{\rm p}}\right)\right]^{-1} .
\label{eq:eng1}
\end{eqnarray}
With the specific angular momentum and binding energy, the radial and azimuthal velocities at $r$ 
and $\phi$ are, respectively, 
\begin{eqnarray}
v_{\rm r} &= &   c \left( 1 - {r_{\rm S} \over r}\right) \sqrt{2 {e_{\rm G} \over
       c^2} + {r_{\rm S} \over r} - {l_{\rm G}^2 \over r^2 c^2 } \left(1 - 
       {r_{\rm S} \over r}\right)} , \label{eq:vradial0}\\
v_{\phi} &=& r \dot{\phi} =  l_{\rm G} {r - r_{\rm S} \over r^2} = \left(1 - {r _{\rm S}\over r}\right) 
      {l_{\rm G} \over r} 
 \label{eq:vphi0}
\end{eqnarray}
\citep{tej13}. From equations~(\ref{eq:vradial0}), (\ref{eq:vphi0}), (\ref{eq:ang}), and (\ref{eq:eng}), 
we obtain
\begin{eqnarray}
v_{\rm r} &\simeq &   c \left({r_{\rm S} \over r}\right)^{1/2} \left(1 - {r_{\rm S} \over r}\right) 
      \left(1-{r\over 2a}\right)^{1/2} \left[1-\left({1+e \over 2}\right)\left({r_{\rm p}\over r}\right){2a 
      \over 2a - r}\right]^{1/2} ,       
       \label{eq:vradial}\\
v_{\phi} & \simeq & c \left({1+e \over 2}\right)^{1/2} \left({r_{\rm p}\over r_{\rm S}}\right)^{-1/2}  
      \left(1 - {r_{\rm S} \over r_{\rm p}}\right)^{-1/2}  \left(1 - {r_{\rm S} \over r}\right)  \left({r_{\rm 
      p} \over r}\right)  ,
 \label{eq:vphi}
\end{eqnarray}
and the angular velocity 
\begin{equation}
  \Omega = \dot{\phi} = {v_{\phi} \over r} = \left({1+e\over2}\right)^{1/2} \left({r_{\rm p}\over
       r_{\rm S}}\right)^{1/2} \left( 1 - {r_{\rm S} \over r_{\rm p}}\right)^{-1/2}  \left( 1 - {r_{\rm S} \over 
       r}\right)  \left({r_{\rm S} \over r^2}\right) c .
       \label{eq:omega}
\end{equation}
To obtain the Equations~(\ref{eq:vradial})-(\ref{eq:omega}), we neglect the terms $[(1-e)/2]^2$ or higher. 
From equations~(\ref{eq:vradial}) and (\ref{eq:vphi}), we have the fluid velocity
\begin{eqnarray}
v &=&\left(v_{\rm r}^2 + v_\phi^2\right)^{1/2} \cr
   &\simeq& c\left(1-{r_{\rm S} \over r}\right) \left({r_{\rm S}\over 
    r}\right)^{1/2} \left(1-{r\over 2a}\right)^{1/2} \left[1+\left({1+e \over 2}\right)\left({r_{\rm S} \over r}\right)
    \left(1 - {r_{\rm S} \over r_{\rm p}}\right)^{-1}\left({2a \over 2a -r }\right)\right]^{1/2} .
\label{eq:velocity}
\end{eqnarray}
At $r=r_{\rm p}$, we have
\begin{eqnarray}
v_{\rm r,p} & \simeq & 0 \\ 
v_{\rm p} & \simeq & v_{\rm \phi,p}  \simeq  c \left({1+e \over 2}\right)^{1/2} \left({r_{\rm p}\over r_{\rm S}}\right)^{-1/2}  
                            \left(1 - {r_{\rm S} \over r_{\rm p}}\right)^{1/2}                              
                              \label{eq:velocityp}
\end{eqnarray}
and 
\begin{equation}
\Omega_{\rm p} \simeq \left({1+e\over2}\right)^{1/2} \left( 1 - {r_{\rm S} \over r_{\rm p}}\right)^{1/2} 
    \left({r_{\rm S} \over r_{\rm p}}\right)^{3/2} {c \over r_{\rm S}} .
\label{eq:omegap}
\end{equation}

\begin{figure*}
	\plotone{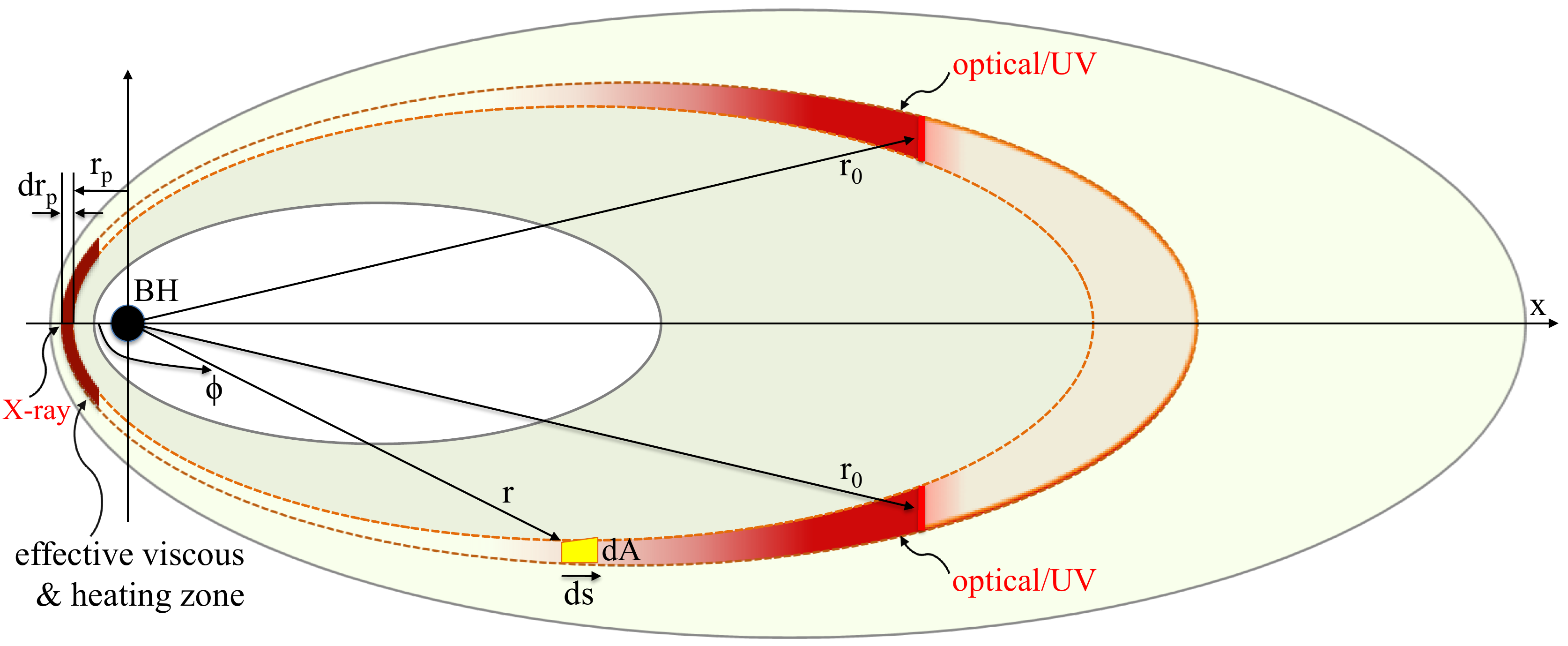}
	\caption{Schematics of an elliptical accretion disk. The effective viscous and heating zone is the 
	region at the pericenter and nearby with $r \sim r_{\rm p}$ and $-\pi/2 \la \phi \la \pi/2$. Soft X-ray 
	photons are produced in the heating zone and are trapped inside the disk because of large 
	electron scattering opacity.  They are absorbed owing to bound-free and free-free opacities 
	and reprocessed into emission lines and low-frequency continuum via recombinations and 
	bremsstrahlung radiation. The low-frequency continuum is emitted mainly at radiation radius 
	$r\la r_0$. The low-frequency photons become trapped at radius $r>r_0$ and advected through 
	the apocenter and back to the radiation radius $r\la r_0$ because of the large bound-free 
	and free-free opacities. No significant radiation is emitted from the disk area with $r > r_0$. 
		\label{fig:disksch}}
\end{figure*}
%

\subsection{Effective viscous torque and heating mechanisms} \label{sec:heating}

When the newly inflowing fluid stream intersects the post-pericenter outflowing matter at apocenter
because of the general relativistic apsidal precession, the self-intersection of the streams forms shocks and would 
convert a fraction of the orbital kinetic energy into heat. In their elliptical accretion disk model, \citet{pir15} proposed 
that the observed luminosity of the optical/UV light of TDEs is powered by the self-crossing shocks at about apocenter 
at the formation of an elliptical accretion disk rather than during the subsequent accretion of matter onto the BH. For 
the elliptical accretion disk model of roughly uniform eccentricity proposed by \citet{liu17}, optical/UV TDEs are 
powered by the accretion of matter onto the BH, and the radiation emitted by the self-crossing shocks is small 
\citep{liu17,cao18,zho20}. The energy of the orbital kinetic dissipated by the self-crossing shocks at formation 
of the elliptical accretion disk is $\Delta{e}_{\rm sh} \simeq e_{\rm G}(a_{\rm d}) - e_{\rm G}(a_{\rm mb}) \la
e_{\rm G}(a_{\rm d}) \sim e_{\rm G}(a_{\rm mb})$ for $a_{\rm d} \sim a_{\rm mb}$, where $e_{\rm G}(a_{\rm 
d})$ and $e_{\rm G}(a_{\rm mb})$ are given by equation~(\ref{eq:eng1}) for semimajor axis $a_{\rm d}$ and 
$a_{\rm mb}$, respectively. Comparing the energy $\Delta{e}_{\rm sh}$ with the total radiation of an elliptical 
accretion disk of uniform eccentricity, $\Delta{e}_{\rm tot} \simeq e_{\rm G}(a_{\rm in})$, we have 
$\Delta{e}_{\rm sh}/\Delta{e}_{\rm tot}\la a_{\rm in}/a_{\rm mb} \la r_{\rm ms}/ r_{\rm p*}$. For  a typical 
tidal disruption of star with orbital pericenter $r_{\rm p*} \sim r_{\rm t} \sim 23 r_{\rm S}$, we have $\Delta{e}_{\rm 
sh}/\Delta{e}_{\rm tot} \la 1/10$. Taking into account that a fraction of the kinetic energy dissipated by the 
self-crossing shocks may be converted back to kinetic energy by adiabatic expansion \citep{jia16}, we could
neglect the radiation of the self-crossing shocks and assume that the optical/UV TDEs are powered by the accretion 
of matter onto the BH. Because no significant intersection shock is expected for the eccentric ellipse with semimajor
$a< a_{\rm d}$, it is reasonable to assume that no strong shock forms at apocenter of the eccentric ellipse with 
$a_{\rm in}\leq a \leq a_{\rm d}$ and that the eccentric ellipse of the elliptical accretion disk is symmetric with 
respect to the major axis.

At the pericenter region, the streams in different orbital planes in the $z$-direction converge to form a "nozzle 
shock" \citep{eva89,koc94,ogi14,shi15},  which is too weak to be important for the stream circularization but 
strong enough to heat the matter to radiate in soft X-rays \citep{gui14,kro16}. In addition to the convergence
nozzle shock, interaction shocks would be introduced at about pericenter by the relativistic apsidal precession 
of the orbits and dissipate some part of the orbital kinetic into heat \citep{svi17,cha18}. The MRI evolves differently 
in an eccentric accretion disk, and the strong magnetic stresses can be efficiently developed \citep{cha18}. 
The strong shear viscous torques in the pericenter region would efficiently dissipate the orbital energy 
\citep{svi17,cha18}. Both the nozzle and interaction shocks and the shear viscous torques work together to 
efficiently dissipate the orbital kinetic energy and transfer angular momentum outward at pericenter and 
nearby \citep{svi17,cha18}.

Because of the complexity and the nonlinearity of the physical processes at pericenter and nearby with 
$r \sim r_{\rm p}$ \citep{svi17,cha18}, we do not discuss the physical processes and the structures of density, 
pressure, temperature, and entropy in that region. Instead, we assume that the physics effects of the transfer of 
the angular momentum and the dissipation of kinetic energy can be approximated with effective shear viscous 
torque at the pericenter and nearby and the complexity and uncertainties can be effectively absorbed by the 
viscosity  parameter $\alpha$ as for those in the standard thin $\alpha$-disk \citep{sha73}. In a geometrically thin 
elliptical accretion disk, the velocity in the $z$-direction is much smaller than the azimuthal velocity at pericenter 
and the nozzle shock is weak. The energy dissipation and the angular momentum at the pericenter and nearby 
are dominated by the magnetic stresses, and the assumption of the effective shear viscous torque would be 
reasonable. We approximate the effective viscosity with a step function:  $\alpha=\alpha_{\rm p}$ for $r\sim 
r_{\rm p}$ and $-\pi/2 \la \phi \la \pi/2$, and $\alpha=0$ for $r\gg r_{\rm p}$. We call the pericenter and 
nearby the "heating zone." The "heating zone" generates the soft X-ray photons, and has a radial size $r\sim r_{\rm 
p}$, and extends azimuthally between $-\pi/2 \la \phi \la \pi/2$, as is schematically shown in Figure~\ref{fig:disksch}.

\section{Mass conservation}\label{sec:mass}

We consider a vertically integrated quasi-stationary elliptical accretion disk. We assume that the radius of 
the apocenter $r_{\rm ap}$ is much larger than the pericenter $r_{\rm p}$, $r_{\rm ap}/r_{\rm p} \gg 1$ or 
specifically $e \ga 0.6$.  The mass element of disk material at radius ($r$,$\phi$) along an arbitrary 
ellipse lying between $a$ and $a+\mathrm{d}a$ for $a_{\rm in} \leq a \leq a_{\rm d}$ is 
\begin{equation}
\mathrm{d}M = 2H\rho \mathrm{d}{A} \mathrm{d}{s}  = \Sigma \mathrm{d}{A} \mathrm{d}{s} ,
\label{eq:delMass0}
\end{equation}
where $\rho$ is the mass density on the midplane of the disk, $H$ is the disk half-thickness in the 
$z$-direction, $\Sigma = 2H\rho$ is the
surface density of the accretion disk, $\mathrm{d}{s}$ is the arc-length element along the fluid streamline, and
$\mathrm{d}A$ is the cross section of the ellipse between $a$ and $a+\mathrm{d}a$ at radius $r$ (see
Fig.~\ref{fig:disksch}). In the cylindrical coordinates ($r$, $\phi$, $z$), the vector arc-length element in the
disk equatorial plane is $\mathrm{d}\vec{s}= (\partial{s}/\partial{r}) \mathrm{d}r \vec{e}_{\rm r} + 
(\partial{s}/\partial{\phi}) \mathrm{d}\phi \vec{e}_{\rm \phi}$. From Equation~(\ref{eq:streamline}), we 
have the arc-length element
\begin{equation}
\mathrm{d}s=\left({1+e\over2}\right)^{-1/2} \left({r\over r_{\rm p}}\right)^{3/2} \left(1-{r\over 2a}
    \right)^{1/2} r_{\rm p} \mathrm{d}\phi ,
\label{eq:arclen_phi}
\end{equation}
or 
\begin{equation}
\mathrm{d}s =\left[1 - \left({1+e\over 2}\right) \left({r_{\rm p} \over r}\right) \left(1-{r\over 
    2a}\right)^{-1} \right]^{-1/2} \mathrm{d} r ,
\label{eq:arclen_simr}
\end{equation}
and the cross section of the stream lines
\begin{eqnarray}
\mathrm{d}A &= & {\mathrm{d}{r} \times r \mathrm{d}{\phi} \over \sqrt{(\mathrm{d}{r})^2 + ( r 
        \mathrm{d}{\phi})^2}} \cr
        & = & \left({1+e \over 2}\right)^{1/2} \left({r \over r_{\rm p}}\right)^{1/2} \left(1 - {r \over 2 
       a}\right)^{-1/2} \mathrm{d}r_{\rm p}  .
\label{eq:crosec}
\end{eqnarray}
From Equations~(\ref{eq:arclen_phi}) and (\ref{eq:crosec}), we have 
\begin{equation}
    \mathrm{d}{A} \mathrm{d}{s} = \left({r\over r_{\rm p}}\right)^2 r_{\rm p} \mathrm{d}{r}_{\rm p} 
    \mathrm{d}{\phi} ,
\label{eq:area_element_phi}
\end{equation}
or from Equations~(\ref{eq:arclen_simr}) and (\ref{eq:crosec}) we have
\begin{eqnarray}
 \mathrm{d}{A} \mathrm{d}{s} & = & \left({1+e \over 2}\right)^{1/2} \left({r \over r_{\rm p}}\right)^{1/2} 
         \left(1 - {r \over 2 a}\right)^{-1/2}  \times \cr
 &&    \left[1 - \left({1+e\over 2}\right) \left({r_{\rm p} \over r}\right)\left(1 - {r \over 2 a}\right)^{-1} \right]^{-1/2} 
           \mathrm{d}r_{\rm p} \mathrm{d}r .
\label{eq:area_element_r}
\end{eqnarray}

Equations~(\ref{eq:delMass0}) and (\ref{eq:area_element_phi}) give 
\begin{equation}
\mathrm{d}M = \Sigma \left({r\over r_{\rm p}}\right)^2 r_{\rm p} \mathrm{d}{r}_{\rm p} 
    \mathrm{d}{\phi} 
\label{eq:delMass}
\end{equation}
and 
\begin{equation}
\Delta{M} = \oint_{\rm a} \Sigma \mathrm{d}{A} \mathrm{d}{s} =
{\mathrm{d}{r}_{\rm p} \over r_{\rm p}} \oint_{\rm a} \Sigma r^2  \mathrm{d}{\phi} ,
\label{eq:annMass}
\end{equation}
where the integration is around the ellipse of semimajor axis $a$. 
The conservation of mass around the ellipse between $a$ and $a+\mathrm{d}a$ reads 
\begin{eqnarray}
{\partial{\Delta{M}} \over \partial t} & = & \oint_{\rm a+\mathrm{d}a} \Sigma v_{\rm a} \mathrm{d}s - 
      \oint_{\rm a} \Sigma v_{\rm a} \mathrm{d}s \cr
      & = & \mathrm{d}a \times {\partial \over \partial a} \oint_{\rm a} \Sigma v_{\rm a} \mathrm{d}s ,
\end{eqnarray}
where $v_{\rm a} = \mathrm{d}a/\mathrm{d}t$ is the radial drift velocity of matter and is a function of position 
($r$, $\phi$). For a quasi-stationary elliptical accretion disk, 
\begin{equation}
{\partial{\Delta{M}} \over \partial t} =  0
\end{equation}
and 
\begin{equation}     
      \mathrm{d}a \times {\partial \over \partial a} \oint_{\rm a} 
      \Sigma v_{\rm a} \mathrm{d}s = 0 .
\label{eq:mass_cons}
\end{equation}
From equation~(\ref{eq:mass_cons}), we have 
\begin{equation}     
 \oint_{\rm a} \Sigma v_{\rm a} \mathrm{d}s = {\rm constant}  .
\end{equation}
We define the mass accretion rate of an elliptical disk as
\begin{equation}     
 \dot{M} \equiv - \oint_{\rm a} \Sigma v_{\rm a} \mathrm{d}s .
\label{eq:accrate}
\end{equation}

\section{Angular momentum conservation} \label{sec:angular}

 In an elliptical accretion disk, the angular momentum is constant along the 
streamline of semimajor axis $a$ and eccentricity $e$.  
The vertically integrated torque per $\mathrm{d}s$ arc length along the ellipse with semimajor axis $a$ is 
\begin{equation}
    \mathrm{d}\vec{\mathcal{G}}  = - 2 H r \vec{\vec{\sigma}} \cdot \mathrm{d} \vec{s} ,
\label{eq:torelement}
\end{equation}
where $\vec{\vec{\sigma}}$ is the shear stress tensor. The total torque exerted on the outer ring by the inner 
ring is given by
\begin{eqnarray}
\mathcal{G}(a) &=& -  \oint_{\rm a}{\mathrm{d} \mathcal{G} \over \mathrm{d}s} \mathrm{d}s \cr %
&\simeq& \int_{\rm -\pi/2}^{\pi/2} 2 H r  \rho \nu \left[{1\over r} {\partial{v_{\rm r}} \over \partial{\phi}} 
         + r {\partial{\Omega} \over \partial{r}}\right]
         \left({\partial{s} \over \partial{\phi}}\right) \mathrm{d}\phi     \cr %
&\simeq& \int_{\rm -\pi/2}^{\pi/2} r  \nu \Sigma \left[{1\over r} {\partial{v_{\rm r}} \over \partial{\phi}} 
         + r {\partial{\Omega} \over \partial{r}}\right]  \left({1+e\over2}\right)^{-1/2} \left({r\over r_{\rm 
         p}}\right)^{3/2} \left(1-{r\over 2a} \right)^{1/2} r_{\rm p} \mathrm{d}\phi \cr %
 &\simeq& \int_{\rm -\pi/2}^{\pi/2} r  \nu \Sigma \left[{1\over r} {\partial{v_{\rm r}} \over \partial{r}} 
         \left({r e \sin{\phi} \over 1+ e \cos{\phi}}\right)  
         + r {\partial{\Omega} \over \partial{r}}\right]  \times \cr
&& \qquad \left({1+e\over2}\right)^{-1/2} \left({r\over r_{\rm p}}\right)^{3/2} \left(1-{r\over 2a} \right)^{1/2} 
         r_{\rm p} \mathrm{d}\phi ,
         \label{eq:tortot}
\end{eqnarray}
where $\nu$ is the effective kinematic viscous coefficient and the integration is along the ellipse with $r$ given 
by equation~(\ref{eq:streamline}). In equation~(\ref{eq:tortot}), we neglected the bulk component of viscosity 
$\sigma_{\rm rr}$, because for the geometrically thin elliptical accretion disk the vertical velocity is small, and 
the compressing "nozzle shocks" are weak and negligible in dissipation of the orbital kinetic energy, and the 
turbulent shear viscosity is dominating (See Section~\ref{sec:heating} for more discussion). We adopt the standard 
$\alpha$-prescription for the kinematic viscous coefficient $\nu = \alpha c_{\rm s} H$, with $c_{\rm s}$ the 
sound speed at $r$ \citep{sha73}, and a step function for the effective viscosity parameter, $\alpha=\alpha_{\rm 
p}$ for $r\sim r_{\rm p}$ and $\alpha=0$ for $r\gg r_{\rm p}$. From equation~(\ref{eq:tortot}), we have
\begin{eqnarray}
\mathcal{G}(a) & \simeq & \nu_{\rm p} \Sigma_{\rm p} r_{\rm p}^2 \left(r_{\rm p} {\partial{\Omega_{\rm 
      p}} \over \partial{r_{\rm p}}}\right) \Delta{\phi} \cr
& \simeq & \nu_{\rm p} \Sigma_{\rm p} r_{\rm p}^2 \left(r_{\rm p} {\partial{\Omega_{\rm p}} \over 
      \partial{r_{\rm p}}}\right) \pi  ,
      \label{eq:tortotp}
\end{eqnarray}
where $\nu_{\rm p}$ and $\Sigma_{\rm p}$  are, respectively, the kinematic viscous coefficient and 
the surface mass density of a disk at pericenter $r\simeq r_{\rm p}=(1-e)a$. In equation~(\ref{eq:tortotp}),
we adopt $\Delta{\phi}\sim \pi$. However, we will show that our results are insensitive to the exact value of
$\Delta{\phi}$. From equations~(\ref{eq:tortotp}) and (\ref{eq:omegap}), we have
\begin{equation}
\mathcal{G}(a) \simeq -{3\over2} \pi D_{\rm p} \nu_{\rm p} \Sigma_{\rm p} r_{\rm p}^2 
     \Omega_{\rm p} ,
      \label{eq:tortotf}
\end{equation}
where $D_{\rm p} = 1-{1\over 3} \left({r_{\rm S} \over r_{\rm p}}\right) \left(1-{r_{\rm S} \over r_{\rm p}}
      \right)^{-1}$. 

The net torque on an ellipse of a gas streamline per unit arc length between $a$ and $a+\mathrm{d}{a}$ (or 
between $r_{\rm p}$ and $r_{\rm p} + \mathrm{d}{r_{\rm p}}$ at pericenter) is 
\begin{eqnarray}
\mathrm{d}\mathcal{G} & = & {\mathrm{d}\mathcal{G} (a+\mathrm{d}a) \over \mathrm{d}s} - 
          {\mathrm{d}\mathcal{G}(a) \over \mathrm{d}s} \cr
&=&   {\partial \over \partial A} \left({\partial \mathcal{G} \over \partial s}\right) \mathrm{d}A .
\label{eq:delG}
\end{eqnarray}
Because the angular momentum per $\mathrm{d}s$ arc length along the streamline is
\begin{equation}
\mathrm{d} L = l_{\rm G} \Sigma \mathrm{d}s \mathrm{d}A  ,
\end{equation}
we have the total angular momentum of the ellipse between $a$ and $a+\mathrm{d}a$ 
\begin{eqnarray}
\Delta{L} & = & \oint_{\rm a} l_{\rm G} \Sigma \mathrm{d}s \mathrm{d}A \cr
      & = & {\mathrm{d}{r}_{\rm p} \over r_{\rm  p}} \oint_{\rm a} l_{\rm G} \Sigma r^2 \mathrm{d}\phi .
\end{eqnarray}
The conservation of angular momentum gives 
\begin{equation}
{\partial({\Delta{L}}) \over \partial{t}} = \oint_{\rm a} l_{\rm G} \Sigma v_{\rm a} \mathrm{d}s - \oint_{\rm a 
    +\mathrm{d}a} l_{\rm G} \Sigma v_{\rm a} \mathrm{d}s + \oint_{\rm a} \left({\partial^2\mathcal{{G}} \over \partial{s} 
    \partial{A}}\right) \mathrm{d}s \mathrm{d}A.
\end{equation}
For quasi-stationary accretion disk ${\partial({\Delta{L}}) /\partial{t}} = 0$, we have 
\begin{equation}
 -\mathrm{d}a \times {\partial \over \partial{a}} {\oint_{\rm a} l_{\rm G} \Sigma v_{\rm a} \mathrm{d}s} +
    \mathrm{d}a \times {\partial \over \partial{a}} \oint_{\rm a} \left({\partial{\mathcal{G}}
    \over \partial{s}}\right) \mathrm{d}s = 0 .
\label{eq:angu_cons_stat}
\end{equation}
Integrating equation~(\ref{eq:angu_cons_stat}), we have 
\begin{eqnarray}
 {\oint_{\rm a} l_{\rm G} \Sigma v_{\rm a} \mathrm{d}s}  &= &  \oint_{\rm a} \left({\partial{\mathcal{G}}
    \over \partial{s}}\right) \mathrm{d}s + C \cr
    & = & \mathcal{G}(a) + C ,
\label{eq:inte_ang}
\end{eqnarray}
where $C$ is the integration constant. Because $l_{\rm G}$ is constant around the ellipse of semimajor axis $a$, 
we have
\begin{eqnarray}
{\oint_{\rm a} l_{\rm G} \Sigma v_{\rm a} \mathrm{d}s} & = & l_{\rm G} {\oint_{\rm a}  \Sigma v_{\rm a} 
          \mathrm{d}s} \cr
  & = & l_{\rm G} (-\dot{M}) .
\label{eq:angu_mdot}
\end{eqnarray}
From equations~(\ref{eq:inte_ang}) and (\ref{eq:angu_mdot}), we have 
\begin{equation}
-l_{\rm G} \dot{M} = \mathcal{G}(a) + C .
\label{eq:agn_mdot1}
\end{equation}
When the disk fluids migrate toward the BH and the orbital pericenter of the fluids reaches the marginally 
stable orbit $r_{\rm ms}$, the matter passing through $r_{\rm ms}$ falls freely onto the BH \citep{abr78}. For 
a parabolic orbit with $e=1$, $r_{\rm ms}= 2 r_{\rm S}$, and for a circular orbit $r_{\rm ms} = 3 r_{\rm S}$. For 
an elliptical orbit with $0 < e <1$, we have $2 r_{\rm S} < r_{\rm ms} < 3 r_{\rm S}$. We adopt the corresponding 
elliptical orbit of fluid as the inner edge of the elliptical accretion disk and take the simple inner boundary 
condition $\mathcal{G}_{\rm in} = \mathcal{G}(a_{\rm in}) = 0$, where $a_{\rm in}= r_{\rm ms}/(1-e)$ is 
the semimajor axis of the inner boundary of an elliptical accretion disk.  From equation~(\ref{eq:ang1}), 
we have the specific orbital angular momentum of the fluid at the inner boundary
\begin{eqnarray}
l_{\rm in} & = & l_{\rm G} (a_{\rm in}) \cr
    &\simeq & \left({1+e\over 2}\right)^{1/2} \left({r_{\rm ms}\over r_{\rm S}}\right)^{1/2}
      \left(1 - {r_{\rm S} \over r_{\rm ms}}\right)^{-1/2} r_{\rm S} c 
\end{eqnarray}
From the inner boundary condition, equation~(\ref{eq:agn_mdot1}) gives
\begin{equation}
\mathcal{G}(a) = -\dot{M} l_{\rm G} \left(1-{l_{\rm in} \over l_{\rm G}}\right) .
\label{eq:tormdot}
\end{equation}
From equations~(\ref{eq:tortotf}) and (\ref{eq:tormdot}), we have 
\begin{equation}
    \nu_{\rm p} \Sigma_{\rm p} = {2 \dot{M} \over 3\pi}  f C_{\rm p}^{-1} D_{\rm p}^{-1}  ,
\label{eq:nusigma}
\end{equation}
where $C_{\rm p} = 1- ({r_{\rm S}/r_{\rm p}})$ and
\begin{equation}
f=1-{l_{\rm in} \over l_{\rm G}} \simeq 1- \left({r_{\rm ms} \over r_{\rm p}}\right)^{1/2} \left[{1-
     \left(r_{\rm S}/r_{\rm ms}\right) \over 1- \left(r_{\rm S}/r_{\rm p}\right)}\right]^{-1/2} .
\end{equation}

\section{Heat generation in the pericenter region}\label{sec:vischeating}

The net torque on an ellipse of a gas streamline per unit arc length
\begin{equation}
\mathrm{d}\mathcal{G}
      = {\partial \over \partial A} \left({\partial \mathcal{G} \over \partial s}\right) \mathrm{d}A .
\end{equation}
(see Equation~(\ref{eq:delG})) is acting on the ring of gas per $\mathrm{d}s$ arc length at radius $r$ in the sense 
of angular velocity $\Omega(a,r)$ and dissipates the orbital kinetic energy at a rate of work
\begin{eqnarray}
\mathrm{d}^2 Q^+ &=& \vec{\Omega}\cdot \mathrm{d} \vec{\mathcal{G}} \cr
       &=& {\partial \over \partial{A}}\left(\vec{\Omega}\cdot {\partial \vec{\mathcal{G}} \over 
       \partial{s}}\mathrm{d}{s}\right) \mathrm{d}{A} - \left({\partial \vec{\mathcal{G}} \over 
       \partial{s}}\mathrm{d}{s}\right) \cdot \left({\partial\vec{\Omega}\over \partial{A}}\right) \mathrm{d}{A} \cr
       & \simeq & \nu \Sigma  r \left({1\over r} {\partial 
      v_{\rm r} \over \partial \phi} + r{\partial
       \Omega\over \partial r}\right) \left({\partial\Omega \over \partial{A}}\right)
       \left({\partial{s} \over \partial\phi}\right)\mathrm{d}\phi
       \mathrm{d}A ,
\label{eq:workrate}
\end{eqnarray}
where $\left[{\partial \left(\vec{\Omega}\cdot {\partial\vec{\mathcal{G}} \over \partial{s}}\mathrm{d}{s}
\right) / \partial{A}}\right] \mathrm{d}{A}$ is the convection term of the rotational energy through the gas by 
the torques, is determined solely by the inner and outer disk edges, and does not contribute to the local 
rate of loss of mechanical energy to the gas. We drop it from the equation. The total viscous dissipation rate 
within the gas between $a$ and $a+\mathrm{d}a$ caused by the effective viscous torques is 
\begin{eqnarray}
\Delta{Q}^+ & = & \oint_{\rm a} \mathrm{d}^2 Q^+ \cr
      & = & \oint_{\rm a} \mathrm{d}A \cdot \left[\nu \Sigma  r \left({1
      \over r} {\partial v_{\rm r} \over \partial \phi} + r{\partial \Omega\over \partial r}\right) \left({\partial
      \Omega \over \partial{A}}\right) \left({\partial{s} \over \partial\phi}\right)\mathrm{d}\phi \right] .
\label{eq:visheating}
\end{eqnarray}
Because  $\nu \sim \nu_{\rm p}$ for $r\sim r_{\rm p}$ and $\nu\sim 0$ for $r\gg r_{\rm p}$, 
equation~(\ref{eq:visheating}) gives 
\begin{eqnarray}
\Delta{Q}^+ & \simeq &  \nu_{\rm p} \Sigma_{\rm p}  r_{\rm p} \left(r_{\rm p}{\partial \Omega_{\rm 
          p} \over \partial{r}_{\rm p}}\right)^2 \Delta{\phi} \mathrm{d}r_{\rm p} \cr
 &\simeq &  {9\over 4} \pi D_{\rm p}^2 \nu_{\rm p} \Sigma_{\rm p}  r_{\rm p} \Omega_{\rm p}^2  
        \mathrm{d}r_{\rm p} ,
\label{eq:heatrate}
\end{eqnarray}
where $r_{\rm p} \left(\partial\Omega_{\rm p} /\partial{r}_{\rm p}\right) \simeq -{3\over2} D_{\rm p} 
\Omega_{\rm p}$ and $\Delta\phi\sim \pi$. 

Equations~(\ref{eq:heatrate}) and (\ref{eq:nusigma}) give 
\begin{eqnarray}
\Delta{Q}^+ &\simeq&  {3\over 2} \dot{M} f D_{\rm p} C_{\rm p}^{-1}  r_{\rm p}  \Omega_{\rm p}^2  
          \mathrm{d}r_{\rm p} \cr
          &\simeq& {3\over 2} f D_{\rm p} \left[{(1+e)\over2} {r_{\rm S} c^2 \over r_{\rm p}^2}\right] \dot{M}  
        \mathrm{d}r_{\rm p} ,
\label{eq:heatrate1}
\end{eqnarray}
which is independent of $\Delta{\phi}$ and depends on the radial distribution of $\Omega$ and angular 
momentum $l_{\rm G}$ at $r \sim r_{\rm p}$, and on the assumption of uniform eccentricity. With
the Eddington luminosity 
\begin{equation}
L_{\rm Edd} =  {4\pi G M_{\rm BH} c \over \kappa_{\rm es}} = 1.44\times 10^{44} M_6 \, {\rm erg\, s^{-1}} ,
\end{equation}
we define the critical accretion rate  
\begin{equation}
\dot{M}_{\rm Edd} = {L_{\rm Edd} \over 0.1 c^2} = {20\pi r_{\rm S} c \over \kappa_{\rm es}} \simeq 2.54
      \times 10^{-2} M_6 \, {\rm M_\odot\; yr^{-1}} ,
\label{eq:eddaccrate}
\end{equation}
where $\kappa_{\rm es} = 0.2 (1+X) \, {\rm cm^2 \;  g^{-1}} = 0.348 \, {\rm cm^2 \;  g^{-1}}$ is the 
electron scattering opacity of fully 
ionized plasma of solar abundance. We adopt the solar abundance of the mass fractions of hydrogen, helium, 
and metallicity, respectively, $X=0.740$, $Y=0.2466$, and $Z=0.0134$ \citep{asp09}. The viscous dissipation heating
rate of energy at $r \sim r_{\rm p}$ is 
\begin{equation}
\Delta{Q}^+ \simeq \left[{30\pi \over \kappa_{\rm es}} f\right] D_{\rm p} \left({1+e\over2}\right) \left({r_{\rm S} 
    \over r_{\rm p}}\right)^2  \left({\dot{M} \over \dot{M}_{\rm Edd}}\right)   c^3 \mathrm{d}r_{\rm p} .
\label{eq:energyrate}
\end{equation}
It is well known that although the Eddington luminosity depends only on the mass of an SMBH \citep{edd18}, the
Eddington accretion rate is a function of both the mass of an SMBH and the radiation efficiency $\eta$. The radiation
efficiency of a circular accretion disk in AGNs and X-ray BH binaries is typically $\eta \simeq 0.1$, and the 
corresponding critical accretion rate $\dot{M}_{\rm Edd}$ is also called the Eddington accretion rate in the 
literature. However, the radiation efficiency of TDEs inferred from TDE observations is about $\eta \simeq 
2.7\times 10^{-3}$ and much smaller than the typical value in AGNs and galactic BH X-ray binaries 
\citep{pir15,liu17,cao18,moc19,zho20}. The corresponding Eddington accretion rate of TDEs is about 37 times 
higher than the critical accretion rate of equation~(\ref{eq:eddaccrate}). Although the accretion rate of 
equation~(\ref{eq:eddaccrate}) may not have much physical meaning in an elliptical accretion disk, we use it to 
scale the accretion rate in this paper.

\section{Structure of the elliptical disk}\label{sec:structellipse}

\subsection{Vertical hydrostatic equilibrium and laminar flows}

We discuss the hydrostatic equilibrium of an eccentric accretion disk in the $z$-direction. 
The Euler equation for the flows around the ellipse in the $z$-direction reads
\begin{equation}
\left(\vec{v}\cdot\nabla\right)v_{\rm z} = -{1\over\rho}{\partial{p}\over \partial{z}} -{\partial{\Phi}_{\rm G} 
    \over \partial{z}} ,
\label{eq:eulerz}
\end{equation}
where $v_{\rm z}$ is the vertical velocity and $p$ is the total pressure of gas. The surface boundary condition is no
mass flux to cross the disk surface, which gives
\begin{equation}
{v_{\rm H} \over v_{\rm r}} \simeq {\partial{H} \over \partial{r}} \simeq {H \over r} ,
\label{eq:boundcond}
\end{equation}
where $v_{\rm H}$ is the velocity in the $z$-direction at the disk surface $z=H$. The vertical integration of 
equation~(\ref{eq:eulerz}) gives 
\begin{equation}
 {v_{\rm H}^2  \over H} \simeq {1\over \rho}{p \over H} - {G M_{\rm BH} H \over r^3} \left[1+ \left({r_{\rm S} 
         \over r}\right)\left(2-{r\over 2a}\right)\right] ,
     \label{eq:vertrp}
\end{equation}
where $p$ and $\rho$ are, respectively, the pressure and density of the disk center. To obtain 
equation~(\ref{eq:vertrp}), we have neglected the terms of $(r_{\rm S}/r)^2$ or higher order in the brackets on
the right-hand side related to the 
gravity in $z$-direction. Together with the surface boundary conditions, equation~(\ref{eq:vertrp}) gives
\begin{equation}
{H \over r} \simeq {c_{\rm s} \over r \Omega_{\rm K}} f_{\rm H}^{-1/2} , 
\label{eq:scaleheight_r}
\end{equation}
where $c_{\rm s}^2 = p/\rho$ is the isothermal sound speed, $\Omega_{\rm K} = (G M_{\rm BH}/ r^3)^{1/2}$
is the Keplerian angular velocity, and 
\begin{equation}
f_{\rm H} = 1+ \left({v_{\rm r} \over r \Omega_{\rm K}}\right)^2 + \left({r_{\rm S} \over r}\right)
         \left(2-{r\over 2a}\right)
\end{equation}
and $f_{\rm H}\sim 1$. Equation~(\ref{eq:scaleheight_r}) suggests that any deviation from hydrostatic 
equilibrium in the $z$-direction would be smoothed out on the timescale 
\begin{equation}
t_{\rm z} = H/c_{\rm s} \simeq {r \over r \Omega_{\rm K}} f_{\rm H}^{-1/2} .
\end{equation} 
Because of the radial movement of particles, the vertical gravity varies on the timescale 
\begin{equation}
t_{\rm dyn} \simeq {r \over v_{\rm r}} .
\end{equation}
To respond to the variations of vertical gravity and establish hydrostatic equilibrium, $t_{\rm z} \ll t_{\rm dyn}$ 
is required. Because
\begin{eqnarray}
{t_{\rm z} \over t_{\rm dyn}} &\simeq& {v_{\rm r} \over r\Omega_{\rm K}} f_{\rm H}^{-1/2} \cr
       &\simeq & {v_{\rm r} \over r\Omega_{\rm K}} \left[1+ \left({v_{\rm r} \over r \Omega_{\rm K}}\right)^2 
       + \left({r_{\rm S} \over r}\right) \left(2-{r\over 2a}\right)\right]^{-1/2}
\end{eqnarray}
and 
\begin{equation}
{v_{\rm r} \over r\Omega_{\rm K}} \simeq \sqrt{2} \left(1-{r_{\rm S}\over r}\right) \left[1 - {r\over 2a} - 
        \left({1+e \over 2}\right) \left({r_{\rm p} \over r}\right)\right]^{1/2} ,
\end{equation}
we have $t_{\rm z} \sim t_{\rm dyn}$. The vertical hydrostatic equilibrium cannot be well established 
in an elliptical accretion disk because of the variations of the gravitational potential in the $z$-direction around
the ellipse. The flow is laminar in an eccentric accretion disk\footnote{The calculations show 
that the assumption of vertical hydrostatic equilibrium would change little the results of the 
work.}. The conclusion is consistent with the results of the detailed hydrodynamic simulations \citep{ogi14}.

The laminar flows move around the ellipse with a nearly constant opening angle 
\begin{equation}
\tan{\theta} \simeq {H\over r} \simeq {H_{\rm p} \over r_{\rm p}} ,
\label{eq:scalheightlam}
\end{equation}
where $H_{\rm p}$ is the disk scale height at $r\sim r_{\rm p}$. The convergence of the orbital velocity field 
near the pericenter strongly compresses the plasma in an elliptical accretion disk of eccentricity $e \ga 0.5$,
and the vertical gravity becomes unimportant at the shocks \citep{ogi14}. From equation~(\ref{eq:vertrp}), 
we have $v_{\rm Hp} \simeq c_{\rm sp}$ with $c_{\rm sp}$ the isothermal sound speed at $r_{\rm p}$. We 
assume that the laminar flows move around the ellipse with the same velocity $v$ within the opening angle 
$\theta$. We have 
\begin{equation}
{H\over r} \simeq {H_{\rm p} \over r_{\rm p}} 
 \simeq {v_{\rm Hp} \over v_{\rm p}}  \simeq {c_{\rm sp} \over v_{\rm p}} .
\label{eq:scalheightrp}
\end{equation}

\subsection{Variation of mass density around the ellipse}

The conservation of the mass around the ellipse of $a=constant$ between semimajor axis $a$ and $a+
\mathrm{d}a$ (or between $r_{\rm p}$ and $r_{\rm p} + \mathrm{d}r_{\rm p}$ at $r_{\rm p}$) gives
\begin{equation}
\Sigma_{\rm p} v_{\rm p} dr_{\rm p} = \Sigma v \mathrm{d}A 
\end{equation}
with $\Sigma_{\rm p}$ the surface density of mass at $r_{\rm p}$, and we have
\begin{eqnarray}
\Sigma &= & \Sigma_{\rm p} \left({v_{\rm p} \over v}\right) \left({\mathrm{d}r_{\rm p}\over 
                \mathrm{d}A}\right)\cr
                &\simeq& \Sigma_{\rm p} C_{\rm p}^{1/2} \left(1-{r_{\rm S} \over r}\right)^{-1} \left[1+ \left({1+e\over
                 2}\right)\left({r_{\rm S} \over r}\right)C_{\rm p}^{-1} \left(1-{r \over 2a}\right)^{-1}\right]^{-1/2} \cr
                 &\simeq& {\nu_{\rm p} \Sigma_{\rm p} \over \nu_{\rm p}} C_{\rm p}^{1/2} \left(1-{r_{\rm S} 
                 \over r}\right)^{-1} \left[1+ \left({1+e\over 2}\right)\left({r_{\rm S} \over r}\right)C_{\rm p}^{-1} 
                 \left(1-{r \over 2a}\right)^{-1}\right]^{-1/2} .
\label{eq:masssigannu}                 
\end{eqnarray}
From equations~(\ref{eq:masssigannu}) and (\ref{eq:nusigma}), we have
\begin{equation}
\Sigma \simeq \left({2\dot{M} \over 3\pi}\right) f C_{\rm p}^{-1/2} D_{\rm p}^{-1} \left(1-{r_{\rm S} 
                 \over r}\right)^{-1} \left[1+ \left({1+e\over 2}\right)\left({r_{\rm S} \over r}\right)C_{\rm p}^{-1} 
                 \left(1-{r \over 2a}\right)^{-1}\right]^{-1/2} \nu_{\rm p}^{-1}
\label{eq:sigmamdot}
\end{equation}
and
\begin{eqnarray}
\rho &= & {\Sigma \over 2H} \cr
          &\simeq& \left({\dot{M} \over 3\pi} f\right) C_{\rm p}^{-1/2} D_{\rm p}^{-1} \left(1-{r_{\rm S} 
                 \over r}\right)^{-1} \left[1+ \left({1+e\over 2}\right)\left({r_{\rm S} \over r}\right)C_{\rm p}^{-1} 
                 \left(1-{r \over 2a}\right)^{-1}\right]^{-1/2} {1\over \nu_{\rm p} H} .
\label{eq:rhodot}
\end{eqnarray}
Equations~(\ref{eq:sigmamdot}) and (\ref{eq:rhodot}) show that both the surface density $\Sigma$ and 
mass density $\rho$ depend on $\Delta{\phi}$, but the uncertainties due to $\Delta{\phi}$ can be absorbed
into the effective viscosity parameter $\alpha_{\rm p}$ through $\nu_{\rm p}= \alpha_{\rm p} c_{\rm sp} H_{\rm 
p}$. From equation~(\ref{eq:rhodot}), we have 
\begin{eqnarray}
\rho  &\simeq& \left({\dot{M} \over 3\pi\alpha_{\rm p}} f\right) C_{\rm p}^{-1/2} D_{\rm p}^{-1} \left(1-{r_{\rm S} 
                 \over r}\right)^{-1} 
                 \left[1+ \left({1+e\over 2}\right)\left({r_{\rm S} \over r}\right)C_{\rm p}^{-1} 
                 \left(1-{r \over 2a}\right)^{-1}\right]^{-1/2} {v_{\rm p}^2 \over c_{\rm sp}^3 r_{\rm p} r} ,
\label{eq:rhodotr}
\end{eqnarray}
where we have used $\nu_{\rm p} = \alpha_{\rm p} c_{\rm sp} H_{\rm p}$, $H_{\rm p} = \left({c_{\rm sp}/v_{\rm 
p}}\right) r_{\rm p}$, and $H = \left({c_{\rm sp}/v_{\rm p}}\right) r$. 

For the polytropic process $p\propto \rho^\gamma$, with $\gamma$ the polytropic index, we have the isothermal 
sound speed $c_{\rm s}^2 = p/\rho \propto \rho^{\gamma-1}$ and 
\begin{equation}
{c_{\rm sp} \over c_{\rm s}} = \left({\rho_{\rm p} \over\rho}\right)^{(\gamma-1)/2} .
\end{equation}
From equation~(\ref{eq:rhodotr}), we have 
\begin{eqnarray}
\rho  &\simeq& \left({\dot{M} \over 3\pi\alpha_{\rm p}}f \right) C_{\rm p}^{-1/2} D_{\rm p}^{-1} \left(1-{r_{\rm S} 
                 \over r}\right)^{-1} \times \cr
                 && \left[1+ \left({1+e\over 2}\right)\left({r_{\rm S} \over r}\right)C_{\rm p}^{-1} 
                 \left(1-{r \over 2a}\right)^{-1}\right]^{-1/2} {v_{\rm p}^2 \over c_{\rm s}^3 r_{\rm p} r} \left({\rho 
                 \over \rho_{\rm p}}\right)^{3(\gamma-1)/2} .
\label{eq:rhorho}
\end{eqnarray}
Because 
\begin{eqnarray}
{\rho \over \rho_{\rm p}} &=& \left({\Sigma\over 2H}\right) \left({\Sigma_{\rm p} \over 2H_{\rm p}}\right)^{-1} 
           \simeq \left({\Sigma\over \Sigma_{\rm p}}\right) \left({r_{\rm p} \over r}\right) \cr
           &\simeq& C_{\rm p}^{1/2} \left(1-{r_{\rm S} 
                 \over r}\right)^{-1} \left[1+ \left({1+e\over 2}\right)\left({r_{\rm S} \over r}\right)C_{\rm p}^{-1} 
                 \left(1-{r \over 2a}\right)^{-1}\right]^{-1/2} \left({r_{\rm p} \over r}\right) ,
\label{eq:rhorhor}
\end{eqnarray}
equation~(\ref{eq:rhorho}) gives 
\begin{eqnarray}
\rho  &\simeq& \left({\dot{M} \over 3\pi\alpha_{\rm p}}f \right) C_{\rm p}^{(3\gamma-5)/4} D_{\rm p}^{-1} 
                \left(1-{r_{\rm S} \over r}\right)^{-(3\gamma-1)/2} \times \cr
                 && \left[1+ \left({1+e\over 2}\right)\left({r_{\rm S} \over r}\right)C_{\rm p}^{-1} 
                 \left(1-{r \over 2a}\right)^{-1}\right]^{-(3\gamma-1)/4} {v_{\rm p}^2 \over c_{\rm s}^3 r_{\rm p}^2} 
                 \left({r_{\rm p} \over r}\right)^{(3\gamma-1)/2} .
\label{eq:rhorcs}
\end{eqnarray}

\subsection{Opacities and photon trapping}\label{sec:opacity}

The electron scattering opacity becomes dominated for temperature $T\ga 10^4\, {\rm K}$ to the soft X-ray 
photons \citep{fra02}. Because of the large optical depth in the vertical direction due to the electron 
scattering, the vertical diffusion timescale of soft X-ray photons is much larger than the orbital period of the 
ellipse (see the discussion in Section~\ref{sec:optdepth}). When the soft X-ray photons generated at pericenter 
and nearby are advected with the fluids around the ellipse, they are well trapped and only a small 
fraction of photons could escape from the thin layer of the photosphere of the disk surface in the region $r\sim 
r_{\rm p}$. The soft X-ray photons can be absorbed owing to bound-free (photoionization) and free-free absorptions
and reprocessed into emission lines and low-frequency continuum mainly as a result of recombination and free-free
emission. The optical/UV continuum and emission lines of optical/UV TDEs are powered primarily by the soft X-ray 
photons trapped inside the disk. Collisional excitations would make some contributions to the line emission. 

Because the electron scattering increases the diffusive path of photons and increases the effective bound-free
and free-free absorptions, the effective Rosseland mean opacity is  
\begin{equation}
\kappa_{\rm eff} \simeq \sqrt{\kappa_{\rm es} \kappa_{\rm R}}  ,
\label{eq:effopacity}
\end{equation}
and the Kramers opacity $\kappa_{\rm R}$ is 
\begin{equation}
\kappa_{\rm R} = \kappa_0 \rho T^{-7/2} , 
\label{eq:Kramerfrfr}
\end{equation}
where $\kappa_0$ is a constant depending on the chemical abundance of gas with $\kappa_0 \simeq 3.9\times 
10^{22} (1+X) (1-Z) \, {\rm cm^5\; K^{7/2} \; g^{-2}} \simeq 6.7 \times 10^{22} \, {\rm cm^5\; K^{7/2} \; 
g^{-2}}$ for free-free opacity and $\kappa_{\rm 0} \simeq 4.3 \times 10^{25}  Z (1+X) \, {\rm cm^5\; K^{7/2}
\; g^{-2}} \simeq 1.0 \times 10^{24} \, {\rm cm^5\; K^{7/2} \; g^{-2}}$ for bound-free opacity of gas with the
solar chemical abundances. The bound-free opacity is strongly dominated over the free-free opacity. 
With the effective Rosseland mean opacity, we have the effective optical depth in the vertical direction 
\begin{equation}
\tau_{\rm eff} \simeq \kappa_{\rm eff} \rho H \simeq \left(\kappa_{\rm es}  \kappa_0\right)^{1/2} 
      \rho^{3/2} T^{-7/4} H 
      \label{eq:effoptdepth}
\end{equation}
and the vertical diffusion timescale due to the effective Rosseland mean opacity 
\begin{eqnarray}
t_{\rm diff} \simeq {H \over c} \tau_{\rm eff} \propto \rho^{3/2} T^{-7/4} H^2 \propto \rho^{-(7\gamma-13)/4} r^2 .
\end{eqnarray}
From equation~(\ref{eq:rhodotr}), we have the vertical diffusion timescale 
\begin{eqnarray}
t_{\rm diff} \propto r^{(7\gamma-5)/4} .
\end{eqnarray}
Because the local dynamic timescale is
\begin{equation}
t_{\rm dyn} = {r \over v} \propto r^{3/2} \left(1 - {r \over 2a}\right)^{-1/2} ,
\end{equation}
we have the vertical diffusion time relative to the local dynamic timescale
\begin{equation}
{t_{\rm diff} \over t_{\rm dyn}} \propto r^{(7\gamma-11)/4} \left(1 - {r \over 2a}\right)^{1/2}  .
\end{equation}
Defining the photon-trapping radius $r_{\rm 0}$, at which the vertical diffusion timescale $t_{\rm diff} = 
(H/c)\tau$ because of the effective Rosseland mean opacity equals to the dynamic timescale $t_{\rm 
dyn} = r/v$, or
\begin{equation}
\left({H_{\rm 0} \over c}\right) \tau_{\rm 0} = {r_{\rm 0} \over v_{\rm 0}} ,
\label{eq:tdiffdyneq}
\end{equation}
where $\tau_{\rm 0}$ and $v_{\rm 0}$ are, respectively, the effective optical depth and velocity at $r_{\rm 
0}$, we have 
\begin{equation}
{t_{\rm diff} \over t_{\rm dyn}} \simeq \left({r \over r_{\rm 0}}\right)^{(7\gamma-11)/4} \left[{ 1 - {r \over 2a} 
\over 1 - {r_0 \over 2a}}\right]^{1/2} ,
\end{equation}
which for the typical polytropic index $\gamma=5/3$ gives
\begin{equation}
{t_{\rm diff} \over t_{\rm dyn}} \simeq \left({r \over r_{\rm 0}}\right)^{1/6} \left[{ 1 - {r \over 2a} 
\over 1 - {r_0 \over 2a}}\right]^{1/2} .
 \label{eq:tdifftdyn}
\end{equation}
Equation~(\ref{eq:tdifftdyn}) shows that the vertical diffusion timescales of low-frequency photons are 
smaller than the local dynamic timescale for $r < r_{\rm 0}$. Because for $r > r_{\rm 0}$ the vertical diffusion
timescales due to the effective Rosseland mean opacity are larger than the local dynamic timescale, the 
low-frequency photons are trapped inside the disk and move outward with the fluid. When the trapped photons go
around through apocenter and return to $r \leq r_{\rm 0}$, they would be radiatively transported to the disk 
surface and emitted away. In our elliptical accretion disk model,  most of the dissipation occurs in the 
disk (and not at the shocks). In addition, we do not discuss the elliptical accretion disk model for super-Eddington 
accretion in this work, because the advection cooling of heat across the eccentric ellipse is neglected 
in equation~(\ref{eq:heatcoolingbal}) for energy balance. Therefore, the assumption of $\gamma=5/3$ or $\sim 2$
 is reasonable. We leave the discussion of the elliptical accretion disk model of the polytropic index $\gamma=4/3$ 
 for super-Eddington luminosity for a future work.

We assume that the energy transfer in the $z$-direction is mainly due to the radiation and that the energy transports 
due to turbulence and thermal conductivity are small. The flux of radiant energy in the $z$-direction is
\begin{eqnarray}
F_{\rm rad} & = & - {16 \sigma_{\rm SB} T^3 \over 3 \kappa_{\rm eff} \rho} {\partial{T} \over \partial{z}} \cr
       &= & - {16 \sigma_{\rm SB} T^4 \over 3} \left({1\over T}{ \partial{T} \over \partial{\tau}_{\rm eff}} \right) ,
\label{eq:radfluxz}
\end{eqnarray}
where $\sigma_{\rm SB}$ is the Stefan-Boltzmann constant. Because the strong vertically compressing 
shock near pericenter uniformly heats the plasma, it is expected that the temperature of the shocked gas 
is homogeneous and the gradient of temperature in the $z$-direction in the region $r\sim r_{\rm p}$ is small,
$(\delta{T}/T) \sim 0$. At $r\sim r_{\rm p}$, the disk temperature is $T\ga 10^6\, {\rm K}$, the \ion{He}{2} is 
also photonionized and the absorption due to photoionization is negligible. After the fluids move away from 
the heating region and expand adiabatically, the photons escape from the thin layer of photosphere of the
disk surface. The typical emitted energy at radius $r$ is 
\begin{eqnarray}
\Delta{E^{-}} & \simeq& \sigma_{\rm SB} T^4 t_{\rm dyn} \mathrm{d}{A} \mathrm{d}{s} \cr
      &\propto&  T^4 r^{3/2} \left(1 - {r \over 2a}\right)^{-1/2} r^{3/2}  \left(1 - {r \over 2a}\right)^{-1/2} 
      \mathrm{d}r_{\rm p} \cr
     &\propto& r^{-(4\gamma-7)} \left(1-{r\over 2a}\right)^{-1}  \mathrm{d}r_{\rm p} .
\end{eqnarray}
For the polytropic index $\gamma=5/3$, $\Delta{E^{-}} \propto r^{1/3} \left(1-{r\over 2a}\right)^{-1} 
\mathrm{d}r_{\rm p}$. The radiation emits mainly at large radii.

The emission decreases the temperature of the disk surface, and a vertical gradient of temperature propagates 
toward the disk center in response to the radiative cooling. For a sufficiently large vertical gradient of temperature at 
$r\gg r_{\rm p}$, we define the radiation timescale 
\begin{equation}
t_{\rm rad} = {H a_{\rm rad} T^4 \over \sigma_{\rm SB} T_{\rm s}^4} ,
\end{equation} 
where $T_{\rm s}$ is the surface temperature of the disk and $a_{\rm rad}=4\sigma_{\rm SB}/c$ is the radiation 
constant. At the critical radius $r_{\rm rad}$, the radiation timescale $t_{\rm rad}$ equals the dynamic (advection) 
timescale $t_{\rm dyn,rad}$,
\begin{equation}
 {H_{\rm rad} a_{\rm rad} T_{\rm rad}^4 \over \sigma_{\rm SB} T_{\rm bb}^4} = t_{\rm dyn,rad} .
    \label{eq:defr0}
\end{equation}
In equation~(\ref{eq:defr0}), $T_{\rm bb}$ and $T_{\rm rad}$ are, respectively, the surface and center blackbody
temperature of the disk at $r_{\rm rad}$, $H_{\rm rad}$ is the disk scale height at $r_{\rm rad}$, and $t_{\rm dyn,rad} 
= r_{\rm rad}/ v_{\rm rad}$ (with $v_{\rm rad}$ the velocity at $r_{\rm rad}$) is the dynamical timescale. For $r \la 
r_{\rm rad}$, we have $t_{\rm rad} \la t_{\rm dyn}$. Whether the surface density of radiation contents can be 
efficiently radiated depends on the vertical gradient of temperature, which is established by the surface cooling 
$\Delta{E}^-$. From equation~(\ref{eq:radfluxz}), we have 
\begin{eqnarray}
F_{\rm rad} & = & \sigma_{\rm SB} T_{\rm bb}^4 \cr
    &\simeq& -{16 \sigma_{\rm SB}T^3  \over 3 \kappa_{\rm eff,rad} \rho_{\rm rad}} \left({\partial{T} \over 
         \partial{z}}\right)_{\rm rad} \cr
    &\simeq& {4 \sigma_{\rm SB} T_{\rm rad}^4 \over 3 \tau_{\rm rad}} ,
    \label{eq:temsurfcent}
\end{eqnarray}
where $\tau_{\rm rad} \simeq \kappa_{\rm eff,rad} \rho_{\rm rad} H_{\rm rad}$ is the vertical effective optical depth 
at $r=r_{\rm rad}$ and $ \rho_{\rm rad}$ and $\kappa_{\rm eff,rad}$ are, respectively, the density and effective 
Rosseland mean opacity at $r_{\rm rad}$. To obtain equation~(\ref{eq:temsurfcent}), we have assumed 
$T_{\rm rad}^4 \gg T_{\rm bb}^4$ at $r_{\rm rad}$, although we may have $T_{\rm rad} \ga T_{\rm bb}$. It is 
reasonable that the local dynamic time at $r\gg r_{\rm p}$ is long for the vertical gradient of temperature to 
be established self-consistently in response to the surface cooling emission $\Delta{E}^-$. From 
equations~(\ref{eq:defr0}) and (\ref{eq:temsurfcent}), we have 
\begin{equation}
3 \left({H_{\rm rad} \over c}\right) \tau_{\rm rad}  = \left({r_{\rm rad} \over v_{\rm rad}}\right) .
\label{eq:radradius}
\end{equation}
Equations~(\ref{eq:radradius}) and (\ref{eq:tdiffdyneq}) show that the radiation radius $r_{\rm rad}$ is slightly
larger than the photon-trapping radius $r_{\rm 0}$.  Because the low-frequency photons also become trapped 
owing to the bound-free and free-free absorptions at $r > r_{\rm 0}$, the photon-trapping radius $r_{\rm 0}$ is the 
typical radiation radius.

At the typical radiation radius $r_{\rm 0}$,  we have the velocity
\begin{equation}
v_{\rm 0} \simeq c   A_{\rm 0}^{1/2} C_{\rm 0} \left({r_{\rm S}\over r_{\rm 0}}\right)^{1/2} \left(1-{r_{\rm 0}
        \over 2a}\right)^{1/2} 
\end{equation}
with $C_{\rm 0}=1-{r_{\rm S}\over r_{\rm 0}}$ and $A_{\rm 0} = 1+ C_{\rm p}^{-1} \left({1+e\over 2}\right)
\left({r_{\rm S} \over r_{\rm 0}}\right) \left(1-{r_{\rm 0} \over 2a}\right)^{-1}$, and the disk half-thickness
\begin{eqnarray}
H_{\rm 0} &=&  \left({c_{\rm sp} \over v_{\rm p}}\right) r_{\rm 0} \simeq \left({c_{\rm s0} \over 
            v_{\rm p}}\right) \left({\rho_{\rm p} \over \rho_{\rm 0}}\right)^{(\gamma-1)/2} r_{\rm 0} \cr
            &=& \left({c_{\rm s0} \over  v_{\rm p}}\right)  C_{\rm p}^{-(\gamma-1)/4} C_{\rm 0}^{(\gamma
            -1)/2} A_{\rm 0}^{(\gamma-1)/4} \left({r_{\rm 0} \over r_{\rm p}}\right)^{(\gamma-1)/2} r_{\rm 0} ,
\label{eq:scalheight0}
\end{eqnarray}
where $p_{\rm 0}$ and $\rho_{\rm 0}$ are, respectively, the total pressure and mass density on the midplane 
of the disk at $r_{\rm 0}$ and $c_{\rm s0}^2 = p_{\rm 0}/\rho_{\rm 0}$ is the isothermal sound speed.
From equation~(\ref{eq:effoptdepth}), we have the effective ("true") optical depth at $r_{\rm 0}$ 
\begin{equation}
\tau_{\rm 0} = \left(\kappa_{\rm es} \kappa_0\right)^{1/2} \rho_{\rm 0}^{3/2} T_{\rm 0}^{-7/4} H_{\rm 0} 
\end{equation}
and from equation~(\ref{eq:rhorcs}) the mass density at $r_{\rm 0}$
\begin{eqnarray}
\rho_{\rm 0} &\simeq& \left({\dot{M} \over 3\pi\alpha_{\rm p}} f\right) C_{\rm p}^{(3\gamma-5)/4} 
          D_{\rm p}^{-1} C_{\rm 0}^{-(3\gamma-1)/2} A_{\rm 0}^{-(3\gamma-1)/4} {v_{\rm p}^2 \over c_{\rm 
          s0}^3 r_{\rm p}^2} \left({r_{\rm p} \over r_{\rm 0}}\right)^{(3\gamma-1)/2} .
\label{eq:rhotempr0}
\end{eqnarray}
Finally, equation~(\ref{eq:tdiffdyneq}) gives the first relation of the temperature $T_{\rm 0}$ and the radiation
radius $r_{\rm 0}$
\begin{eqnarray}
T_{\rm 0}^{-3} \left({r_{\rm 0}\over r_{\rm p}}\right)^{-(5\gamma-1)/4} \left\{\left({\beta_{\rm g}^{-1} 
         k_{\rm B} \over \mu m_{\rm H}}\right)^{-5/4} r_{\rm S}^{-1/2} c^{5/2} \kappa_{\rm es}^{-1} 
         \kappa_{\rm 0}^{1/2}\right\} \left({r_{\rm p}\over r_{\rm S}}\right)^{-3}\left({1+e \over 2}\right)^{1/2} 
         \times  \cr         
         C_{\rm p}^{(5\gamma-7)/8} C_{\rm 0}^{-(5\gamma-3)/4} D_{\rm p}^{-3/2} A_{\rm 0}^{-(5\gamma-3)/8} 
         \left({20  f\over 3 \alpha_{\rm p}}\right)^{3/2} \left(1 - {r_0 \over 2a}\right)^{1/2}
          \left({\dot{M} \over \dot{M}_{\rm Edd}}\right)^{3/2} = 1 ,
 \label{eq:RadRadTR}
\end{eqnarray}
where $\beta_{\rm g} = p_{\rm g}/p$ is the ratio of gas pressure to total pressure and $\beta_{\rm g} \sim 1$ in 
the present paper, $k_{\rm B}$ is the Boltzmann constant, $m_{\rm H}$ is the mass of hydrogen, and $\mu$ is the 
mean molecular weight, with $\mu= 0.60$ for fully ionized gas of solar chemical abundance. The equation
of state for a mixture of perfect gas and radiation is adopted, 
\begin{equation}
p = p_{\rm g} + p_{\rm r} = {\rho k_{\rm B} T \over \mu m_{\rm H}} + {a_{\rm rad} \over 3} T^4 ,
\label{eq:eqstate}
\end{equation}
where  $p_{\rm r}$ is the radiation pressure. The isothermal sound speed is  
\begin{equation}
c_{\rm s} = \left({p \over \rho}\right)^{1/2} = \beta_{\rm g}^{-1/2}\left({p_{\rm g} \over  \rho}\right)^{1/2}
     = \left({\beta_{\rm g}^{-1} k_{\rm B} \over \mu m_{\rm H}}\right)^{1/2} T^{1/2} 
\end{equation}
and
\begin{equation}
c_{\rm s0} = \left({\beta_{\rm g}^{-1} k_{\rm B} \over \mu m_{\rm H}}\right)^{1/2} T_{\rm 0}^{1/2} .
\end{equation}

\subsection{Energy balance}

In Section~\ref{sec:opacity}, we showed that the emission of radiation along the ellipse is dominated at large 
radius and mainly at the typical radiation radius $r_{\rm 0}$ with radiation flux 
\begin{equation}
F_{\rm rad,0} = \sigma_{\rm SB} T_{\rm bb}^4  \simeq {4 \sigma_{\rm SB} \over 3 \tau_{\rm 0}} T_{\rm 0}^4 .
\label{eq:radtrans}
\end{equation}
The total cooling rate of the disk around the ellipse is 
\begin{equation}
\Delta{Q}^{-} \simeq 2\times 2\times F_{\rm rad,0} \times \mathrm{d}A \Delta{s}  ,
\label{eq:Qcooling}
\end{equation}
where $\mathrm{d}A = \left({1+e \over 2}\right)^{1/2} \left({r_{\rm 0} \over r_{\rm p}}\right)^{1/2} \left(1-
{r_0 \over 2a}\right)^{-1/2} d r_{\rm p}$ and $\Delta{s} \simeq \left[1+ {1\over2} \left({1+e\over 2}\right) 
\left({r_{\rm p} \over r_{\rm 0}}\right) \left(1- {r_0 \over 2a}\right)^{-1}\right] r_{\rm 0}$. The first "2" on the 
right-hand side of equation~(\ref{eq:Qcooling}) is due to the two sides of the disk surface, and the second "2" 
is because of the symmetry of the ellipse with respect to the major axis. If we assume that the 
energy generating rate $\Delta{Q}^{+}$ is balanced by the radiation cooling rate $\Delta{Q}^{-}$ 
\begin{equation}
\Delta{Q}^{+} = \Delta{Q}^{-} ,
\label{eq:heatcoolingbal}
\end{equation}
we have  
\begin{eqnarray}
\left({r_{\rm 0}\over r_{\rm p}}\right)^{3/2} \left(1- {r_0 \over 2a}\right)^{-1/2} \left({\sigma_{\rm SB} 
                  T_{\rm 0}^4 \over\tau_{\rm 0}}\right) & \simeq &  \left({45\pi \over 8 \kappa_{\rm es}}
                  f\right) \left({1+e \over 2}\right)^{1/2} \left({r_{\rm p} \over r_{\rm S}}\right)^{-3} \times \cr
              && D_{\rm p} B_{\rm 0}^{-1} r_{\rm S}^{-1} c^3 \left({\dot{M}\over \dot{M}_{\rm Edd}}\right) 
 \label{eq:engbal}
\end{eqnarray}
with $B_{\rm 0} = 1 +{1\over2} \left({1+e\over 2}\right)\left({r_{\rm p} \over r_{\rm 0}}\right)\left(1- {r_0 \over 
2a}\right)^{-1} $.  In equation~(\ref{eq:heatcoolingbal}), the energy generating rate 
is locally balanced by the radiation cooling rate, and the cooling due to the advection of heat across 
the eccentric ellipse is neglected. Because the advection cooling may be important in an elliptical accretion 
disk of super-Eddington luminosity in TDEs, e.g., by a BH of mass $M_{\rm BH} \la 10^5 M_\odot$ (see
equation~(\ref{eq:coneff1})), our results cannot be applied to such accretion systems, and an 
elliptical slim disk model with the advective cooling across eccentric ellipse is needed. 
Equations~(\ref{eq:tdiffdyneq}) and (\ref{eq:scalheight0}) give the optical depth at $r_{\rm 0}$,
\begin{eqnarray}
\tau_{\rm 0} &\simeq& \left({c \over v_{\rm 0}}\right) \left({r_{\rm 0}\over H_{\rm 0}}\right) \cr
           &\simeq& C_{\rm p}^{(\gamma+1)/4} C_{\rm 0}^{-(\gamma+1)/2} A_{\rm 0}^{-(\gamma+1)/4}
           \left({1+e\over2}\right)^{1/2}  \left(1 - {r_0 \over 2a}\right)^{-1/2} \times \cr
           && \left\{\left({\beta_{\rm g}^{-1} 
           k_{\rm B} \over \mu m_{\rm H}}\right)^{-1/2} c\right\} T_{\rm 0}^{-1/2} \left({r_{\rm 0} \over r_{\rm 
           p}}\right)^{-(\gamma-2)/2} .
 \label{eq:optdepthr0}
\end{eqnarray}
From Equations~(\ref{eq:engbal}) and (\ref{eq:optdepthr0}), we obtain the second relation of the temperature
$T_{\rm 0}$ and the radiation radius $r_{\rm 0}$,
\begin{eqnarray}
{r_{\rm 0}\over r_{\rm p}}&\simeq & T_{\rm 0}^{-9/(\gamma+1)} \left({45\pi \over 8 }f
              \right)^{2/(\gamma+1)} \left({1+e\over 2}\right)^{2/(\gamma+1)}  \times \cr
        &&A_{\rm 0}^{-1/2} B_{\rm 0}^{-2/(\gamma+1)} C_{\rm 0}^{-1} C_{\rm p}^{1/2} D_{\rm 
        p}^{2/(\gamma+1)} \times \cr
       && \left\{\left({\beta_{\rm g}^{-1} k_{\rm B} \over \mu m_{\rm H}}\right)^{-1/(\gamma+1)} \sigma_{\rm 
             SB}^{-2/(\gamma+1)} \kappa_{\rm es}^{-2/(\gamma+1)} r_{\rm S}^{-2/(\gamma+1)} c^{8/(\gamma
             +1)} \right\} \times \cr
       && \left({r_{\rm p} \over r_{\rm S}}\right)^{-6/(\gamma+ 1)}  
              \left({\dot{M}\over \dot{M}_{\rm Edd}}\right)^{2/(\gamma+1)} . 
\label{eq:engbalTR}
\end{eqnarray}

\subsection{Radiation radius and physics of the disk at $r_0$}

From equations~(\ref{eq:RadRadTR}) and (\ref{eq:engbalTR}), we obtain the temperature of the disk center at 
$r_{\rm0}$ as a function of pericenter $r_{\rm p}$ and accretion rate $\dot{M}$,
\begin{eqnarray}
T_{\rm 0} &\simeq & A_{\rm 0}^{-(\gamma+1)/(33\gamma-21)} B_{\rm 0}^{-2(5\gamma-1)/(33\gamma-21)}
              C_{\rm p}^{3(\gamma+1)/(33\gamma-21)} C_{\rm 0}^{-2(\gamma+1)/(33\gamma-21)}
              D_{\rm p}^{4(4\gamma+1)/(33\gamma-21)} \times \cr
 &&      \left({45\pi \over 8 }f \right)^{2(5\gamma-1)/(33\gamma-
            21)} \left({20 \over 3 \alpha_{\rm p}}f\right)^{-6(\gamma+1)/(33\gamma-21)} \left({1+e\over 
            2}\right)^{4(2\gamma-1)/(33\gamma-21)} \times \cr
 &&    \left\{\left({\beta_{\rm g}^{-1} k_{\rm B} \over \mu m_{\rm H}}\right)^{6/(33\gamma-21)}
              \sigma_{\rm SB}^{-2(5\gamma-1)/(33\gamma-21)} \kappa_{\rm es}^{-6(\gamma-
              1)/(33\gamma-21)}   r_{\rm S}^{-4(2\gamma-1)/(33\gamma-21)} 
              c^{6(5\gamma-3)/(33\gamma-21)} \right. \times \cr
 &&   \left. \kappa_{\rm 0}^{-2(\gamma+1)/(33\gamma-21)}\right\} \beta_*^{18(\gamma-1)/(33\gamma-21)} 
              \left({r_{\rm t} \over r_{\rm S}}\right)^{-18(\gamma-1)/(33\gamma-21)}
              \left({r_{\rm p} \over r_{\rm p*}}\right)^{-18(\gamma-1)/(33\gamma-21)} \times \cr
 &&  \left(1-{r_{\rm 0}\over 2a}\right)^{-2(\gamma+1)/(33\gamma-21)} \left({\dot{M}_{\rm p}\over 
           \dot{M}_{\rm Edd}} \right)^{4(\gamma-2)/(33\gamma-21)} \left({\dot{M}\over \dot{M}_{\rm 
           p}}\right)^{4(\gamma -2)/(33\gamma-21)} ,
\label{eq:radtempcent}
\end{eqnarray}
and from equations~(\ref{eq:engbalTR}) and (\ref{eq:radtempcent}), we have the radiation radius 
\begin{eqnarray}
{r_{\rm 0} \over r_{\rm p}} &\simeq & A_{\rm 0}^{-(11\gamma-13)/2(11\gamma-7)}
              B_{\rm 0}^{8/(11\gamma-7)} C_{\rm p}^{(11\gamma-25)/2(11\gamma-7)} 
              C_{\rm 0}^{-(11\gamma-13)/(11\gamma-7)}  D_{\rm p}^{-26/(11\gamma-7)} \times \cr
&&      \left({45\pi \over 8 }f \right)^{-8/(11\gamma-
            7)} \left({20 \over 3 \alpha_{\rm p}}f\right)^{18/(11\gamma-7)} \left({1+e\over 
            2}\right)^{-2/(11\gamma-7)} \times \cr
 &&     \left\{\left({\beta_{\rm g}^{-1} k_{\rm B} \over \mu m_{\rm H}}\right)^{-11/(11\gamma-7)}  
              \sigma_{\rm SB}^{8/(11\gamma-7)} \kappa_{\rm es}^{-4/(11\gamma-7)} 
              r_{\rm S}^{2/(11\gamma-7)} 
              c^{-2/(11\gamma-7)} \kappa_{\rm 0}^{6/(11\gamma-7)}\right\}  \times \cr
 &&  \beta_*^{12/(11\gamma-7)} 
              \left({r_{\rm t} \over r_{\rm S}}\right)^{-12/(11\gamma-7)}
              \left({r_{\rm p} \over r_{\rm p*}}\right)^{-12/(11\gamma-7)} \left(1-{r_0\over 
              2a}\right)^{6/(11\gamma-7)} \times \cr
	&&\left({\dot{M}_{\rm p}\over \dot{M}_{\rm Edd}}\right)^{10/(11\gamma-7)}   
	\left({\dot{M}\over \dot{M}_{\rm p}}\right)^{10/(11\gamma-7)} ,
\label{eq:radradiuscent}
\end{eqnarray}
where we have used
\begin{equation}
{r_{\rm p} \over r_{\rm S}} = \beta_*^{-1} \left({r_{\rm p} \over r_{\rm p*}}\right) \left({r_{\rm t} \over r_{\rm S}}\right) 
\end{equation}
for $r_{\rm ms} \leq r_{\rm p} \leq r_{\rm p*}$ and 
\begin{equation}
{\dot{M} \over \dot{M}_{\rm Edd}} = \left({\dot{M} \over \dot{M}_{\rm p}}\right) \left({\dot{M}_{\rm p} 
           \over \dot{M}_{\rm Edd}}\right) .
\end{equation}
Equations~(\ref{eq:radtempcent}) and (\ref{eq:radradiuscent}) become, respectively, 
\begin{eqnarray}
T_{\rm 0} &\simeq & A_{\rm 0}^{-(\gamma+1)/(33\gamma-21)} B_{\rm 0}^{-2(5\gamma-1) /(33\gamma-21)} 
              C_{\rm p}^{3(\gamma+1)/(33\gamma-21)} C_{\rm 0}^{-2(\gamma+1)/(33\gamma-21)}
               D_{\rm p}^{4(4\gamma+1)/(33\gamma-21)}  \times \cr
	&& f^{4(\gamma -2)/(33\gamma-21)} \left({45\pi \over  8 } 
	    \right)^{2(5\gamma-1)/(33\gamma-21)} \left({20 \over 3}\right)^{-6(\gamma+1)/(33\gamma-21)} 
	    \alpha_{\rm p}^{6(\gamma+1)/(33\gamma-21)} \times \cr
  &&\left({1+e\over 2}\right)^{4(2\gamma-1)/(33\gamma-21)} \left\{\left({\beta_{\rm g}^{-1} k_{\rm 
	    B} \over \mu m_{\rm H}}\right)^{6/(33\gamma-21)}  \sigma_{\rm 
              SB}^{-2(5\gamma-1)/(33\gamma-21)} \kappa_{\rm es}^{-6(\gamma-1)/(33\gamma-21)} 
               \right. \times \cr
       && \left. r_{\rm 11}^{-4(2\gamma-1)/(33\gamma-21)} c^{6(5\gamma-3)/(33\gamma-21)} \kappa_{\rm 
             0}^{-2(\gamma+1)/(33\gamma- 21)}\right\} 23.545^{-18(\gamma-1)/(33\gamma-21)}  \times \cr
        &&  f_{\rm T}^{-6(5\gamma-7)/(33\gamma-21)}  m_*^{2(7\gamma-11)/(33\gamma-21)} r_*^{-6(4\gamma-5)/(33\gamma-21)} \beta_*^{18(\gamma-1)/(33\gamma-21)} 117^{4(\gamma-
              2)/(33\gamma-21)} \times \cr
        && M_6^{-2(\gamma-2)/(33\gamma-21)} \left[{3 (n-1) \over 2 }\right]^{4(\gamma-2)/(33\gamma-21)}
       \left(1-{r_0 \over 2a}\right)^{-2(\gamma+1) /(33\gamma-21)} \times \cr
       &&              \left({r_{\rm p} \over r_{\rm p*}}\right)^{-18(\gamma-1)/(33\gamma-21)} \left({\dot{M}\over 
              \dot{M}_{\rm p}}\right)^{4(\gamma-2)/(33\gamma-21)} 
\label{eq:radtempcent1}
\end{eqnarray}
and 
\begin{eqnarray}
{r_{\rm 0} \over r_{\rm p*}}  &\simeq & A_{\rm 0}^{-(11\gamma-13)/2(11\gamma-7)}
              B_{\rm 0}^{8/(11\gamma-7)} C_{\rm p}^{(11\gamma-25)/2(11\gamma-7)} 
             C_{\rm 0}^{-(11\gamma-13)/(11\gamma-7)}  D_{\rm p}^{-26/(11\gamma-7)} \times \cr              
  &&   f^{10/(11\gamma-7)} \alpha_{\rm p}^{-18/(11\gamma-7)} \left({45\pi \over 
            8 } \right)^{-8/(11\gamma- 7)} \left({20 \over 3}\right)^{18/(11\gamma-7)} \left({1+e\over 
            2}\right)^{-2/(11\gamma-7)} \times \cr
   &&  \left\{\left({\beta_{\rm g}^{-1} k_{\rm B} \over \mu m_{\rm H}}\right)^{-11/(11\gamma-7)}  
              \sigma_{\rm SB}^{8/(11\gamma-7)} \kappa_{\rm es}^{-4/(11\gamma-7)} r_{\rm 
              11}^{2/(11\gamma-7)} 
              c^{-2/(11\gamma-7)} \kappa_{\rm 0}^{6/(11\gamma-7)}\right\} \times \cr
    && 23.545^{-12/(11\gamma-7)} f_{\rm T}^{-42/(11\gamma-7)} m_*^{24/(11\gamma-7)} 
              r_*^{-27/(11\gamma-7)} 
              \beta_*^{12/(11\gamma-7)} 117^{10/(11\gamma-7)}  \times \cr
    && M_6^{-5/(11\gamma-7)} \left[{3 (n-1) \over 2 }\right]^{10/(11\gamma-7)}
           \left(1-{r_0 \over 2a}\right)^{6/(11\gamma-7)} \times \cr
     &&              \left({r_{\rm p} \over r_{\rm  p*}}\right)^{(11\gamma-19)/(11\gamma-7)} \left({\dot{M}\over 
              \dot{M}_{\rm p}}\right)^{10/(11\gamma-7)}  ,
\label{eq:radradiuscent1}
\end{eqnarray}
where $r_{\rm 11} = r_{\rm S}/M_6 =2.954\times 10^{11} \, {\rm cm}$ and $\dot{M}_{\rm p}\simeq 117 
\left[3(n-1)/2\right] f_{\rm T}^{-3} r_*^{-3/2} m_*^2 M_6^{-3/2} \dot{M}_{\rm Edd}$.  Note that we use $r_{\rm 0}/
r_{\rm p*}$ on the left-hand side of equation~(\ref{eq:radradiuscent1}). 

For typical polytropic index $\gamma=5/3$, equations~(\ref{eq:radtempcent1}) and 
(\ref{eq:radradiuscent1}) give, respectively,
\begin{eqnarray}
T_{\rm 0} &\simeq & A_{\rm 0}^{-4/51} B_{\rm 0}^{-22/51}   C_{\rm p}^{4/17} C_{\rm 0}^{-8/51}
              D_{\rm p}^{46/51} f^{-2/51}  \left({45\pi \over 8} \right)^{22/51} \left({20 \over 
              3}\right)^{-8/17} \alpha_{\rm p}^{8/17} \times \cr
	    && 	
	    \left({1+e\over 2}\right)^{14/51} \left\{\left({\beta_{\rm g}^{-1} k_{\rm B} \over \mu m_{\rm H}}
	    \right)^{3/17} \sigma_{\rm SB}^{-22/51}  \kappa_{\rm es}^{-2/17} r_{\rm 11}^{-14/51} c^{16/17} 
	    \kappa_{\rm 0}^{-8/51}\right\} \cr
	    && 23.545^{-6/17} 117^{-2/51}  f_{\rm T}^{-4/17} m_*^{2/51}  r_*^{-5/17} \beta_*^{6/17} M_6^{1/51} 
	     \times \cr
	    && \left[{3(n-1) \over 2}\right]^{-2/51} \left(1-{r_0 \over 2a}\right)^{-8/51} 
               \left({r_{\rm p} \over r_{\rm p*}}\right)^{-6/17} \left({\dot{M}\over \dot{M}_{\rm p}}\right)^{-2/51} \cr
 &\simeq &2.42\times10^5 \;({\rm K})\; A_{\rm 0}^{-4/51} B_{\rm 0}^{-22/51} C_{\rm 0}^{-8/51}
            C_{\rm p}^{4/17}  D_{\rm p}^{46/51} \times \cr 
              &&f^{-2/51} f_{\rm T}^{-4/17} \alpha_{\rm -1}^{8/17}\left({1+e\over 2}\right)^{14/51} \beta_{\rm g}^{-3/17}
	       m_*^{2/51}  r_*^{-5/17} \beta_*^{6/17}  M_6^{1/51}  \times \cr
              &&  \left[{3(n-1) \over 2}\right]^{-2/51} \left({r_{\rm p} \over r_{\rm p*}}\right)^{-6/17} \left(1-{r_{\rm 0}\over 
              2a}\right)^{-8/51} \left({\dot{M}\over \dot{M}_{\rm p}}\right)^{-2/51}
\label{eq:radtempcent2}
\end{eqnarray}
and 
\begin{eqnarray}
{r_{\rm 0} \over r_{\rm p*}}  &\simeq & A_{\rm 0}^{-4/17}  B_{\rm 0}^{12/17} C_{\rm 0}^{-8/17}
              C_{\rm p}^{-5/17} D_{\rm p}^{-39/17} \times \cr   
&&  f^{15/17} \alpha_{\rm p}^{-27/17} \left({45\pi \over  8 } \right)^{-12/17} \left({20 \over 
           3}\right)^{27/17} \left({1+e\over 2}\right)^{-3/17} \times \cr
              &&\left\{\left({\beta_{\rm g}^{-1} k_{\rm B} \over \mu m_{\rm H}}\right)^{-33/34} 
               \sigma_{\rm SB}^{12/17} \kappa_{\rm es}^{-6/17} r_{\rm 11}^{3/17}
              c^{-3/17}\kappa_{\rm 0}^{9/17}\right\} \times \cr
              &&23.545^{-18/17}  117^{15/17} f_{\rm T}^{-63/17} m_*^{36/17} r_*^{-81/34} 
              \beta_*^{18/17}  \times \cr
              &&M_6^{-15/34} \left[{3(n-1) \over 2}\right]^{15/17} \left(1-{r_{\rm 0}\over 
              2a}\right)^{9/17}  \left({r_{\rm p} \over r_{\rm 
              p*}}\right)^{-1/17} \left({\dot{M}\over \dot{M}_{\rm 
              p}}\right)^{15/17} \cr
   &\simeq& 3.43 \times 10^4 \; A_{\rm 0}^{-4/17} B_{\rm 0}^{12/17} 
               C_{\rm 0}^{-8/17} C_{\rm p}^{-5/17}  D_{\rm p}^{-39/17} \times \cr  
            &&  f^{15/17} f_{\rm T}^{-63/17} \alpha_{\rm -1}^{-27/17} \left({1+e\over 
            2}\right)^{-3/17} \beta_{\rm g}^{33/34} m_*^{36/17} r_*^{-81/34} 
              \beta_*^{18/17}\times \cr
              && M_6^{-15/34} \left[{3(n-1) \over 2}\right]^{15/17}  \left(1-{r_{\rm 0}\over 
              2a}\right)^{9/17} \left({r_{\rm p} \over r_{\rm p*}}\right)^{-1/17} \left({\dot{M}\over \dot{M}_{\rm 
              p}}\right)^{15/17} ,
\label{eq:radradiuscent2}
\end{eqnarray}
where $\alpha_{\rm -1}=\alpha_{\rm p}/0.1$. Equation~(\ref{eq:radtempcent2}) shows that the temperature
of the disk center at the radiation radius $r_{\rm 0}$ is practically independent of both the accretion rate with 
power-law index $-0.039$ and BH mass with power-law index $0.020$, while the radiation radius $r_{\rm 
0}$ given by equation~(\ref{eq:radradiuscent2}) significantly depends on both of them. 
Equations~(\ref{eq:radtempcent2}) and (\ref{eq:radradiuscent2}) show that both the temperature $T_{\rm 0}$ 
and radiation radius $r_{\rm 0}$ depend on the effective viscosity parameter $\alpha_{\rm p}$. The disk-dominated 
late-time UV observations of TDEs show that the disk viscosity parameter is probably in the range $-1.1 \la 
\log{\alpha} \la -0.2$  with average $\langle \log{\alpha} \rangle \simeq -0.46$ \citep[see Table 3 of][]{van19}. 
We notice that they adopted a circular accretion disk of radius $2 r_{\rm p}$ for TDEs which is  different from the 
disk model in this work. Because the viscous torque in an elliptical accretion disk is expected to operate efficiently 
only in the vicinity of the pericenter at $r\sim r_{\rm p}$ and the effective viscous and heating region would be 
expected to be about from $-\pi/2 \la \phi \la \pi/2$ and $r\sim r_{\rm p}$, the size of the effective viscous regions
is not much different from the circular accretion disk. Taking into account the very large uncertainties of the 
measurements of viscosity parameters of TDEs and the simplifications adopted in this work, we do not take into 
account the differences of two viscosity parameters and adopt the range of viscosity parameters $0.01 
\la \alpha_{\rm p} \la 1$ with the typical value $\alpha_{\rm p} = 0.2$. 

From equations~(\ref{eq:rhotempr0}), (\ref{eq:radtempcent}), and (\ref{eq:radradiuscent}), we have the 
mass density at $r_{\rm 0}$
\begin{eqnarray}
\rho_{\rm 0} &\simeq& \left({45\pi \over 8 }f \right)^{(7\gamma-3)/(11\gamma-
	7)} \left({20 \over 3 \alpha_{\rm p}}f\right)^{-(13\gamma-5)/(11\gamma-7)} \left({1+e\over 
	2}\right)^{2(5\gamma-3)/(11\gamma-7)} \times \cr
&&  \left\{A_{\rm 0}^{-2(2\gamma-1)/(11\gamma-7)} B_{\rm 0}^{-(7\gamma-3)/(11\gamma-7)} 
    C_{\rm p}^{6(2\gamma-1)/(11\gamma-7)} C_{\rm 0}^{-4(2\gamma-1)/(11\gamma-7)} 
    D_{\rm p}^{4(5\gamma-2)/(11\gamma-7)}\right\} \times \cr
&&  \left\{\left({\beta_{\rm g}^{-1} k_{\rm B} \over \mu m_{\rm H}}\right)^{2/(11\gamma-7)}
    \sigma_{\rm SB}^{-(7\gamma-3)/(11\gamma-7)} \kappa_{\rm es}^{-2(\gamma-1)/(11\gamma-7)}
    r_{\rm S}^{-2(5\gamma-3)/(11\gamma-7)} c^{(21\gamma-13)/(11\gamma-7)} \right. \times \cr
&&  \left. \kappa_{\rm 0}^{-4(2\gamma-1)/(11\gamma-7)}\right\} \left({r_{\rm p} \over r_{\rm
	S}}\right)^{-6(\gamma-1)/(11\gamma-7)} \left(1-{r_0 \over 2a}\right)^{-4(2\gamma-
	1)/(11\gamma-7)}  \left({\dot{M}\over \dot{M}_{\rm Edd}}\right)^{-2(3\gamma-1)/(11\gamma-7)} .
\label{eq:rhor0}
\end{eqnarray}
For $\gamma=5/3$, equation~(\ref{eq:rhor0}) gives 
\begin{eqnarray}
\rho_{\rm 0} &\simeq& \left({45\pi \over 8 }\right)^{13/17} \left({20 \over 3}\right)^{-25/17} 
    \left({1+e\over 2}\right)^{16/17} f^{-12/17} \alpha_{\rm p}^{25/17} \times \cr
&&   \left\{A_{\rm 0}^{-7/17} B_{\rm 0}^{-13/17} C_{\rm p}^{21/17} C_{\rm 0}^{-14/17} D_{\rm 
      p}^{38/17}\right\} \times \cr
&&  \left\{\left({\beta_{\rm g}^{-1} k_{\rm B} \over \mu m_{\rm H}}\right)^{3/17} 
    \sigma_{\rm SB}^{-13/17} \kappa_{\rm es}^{-2/17} r_{\rm S}^{-16/17} c^{33/17}
     \kappa_{\rm 0}^{-14/17}\right\}\cr
&&  \left({r_{\rm p} \over r_{\rm S}}\right)^{-6/17} \left(1-{r_0 \over 2a}\right)^{-14/17}  
        \left({\dot{M}\over \dot{M}_{\rm Edd}}\right)^{-12/17} \cr
&\simeq& 6.95 \times 10^{-10} \, ({\rm g\; cm^{-3}})\,  \alpha_{\rm -1}^{25/17} f^{-12/17}
    \left({1+e\over 2}\right)^{16/17} \times \cr
&& \left\{A_{\rm 0}^{-7/17} B_{\rm 0}^{-13/17} C_{\rm 0}^{-14/17} C_{\rm p}^{21/17} D_{\rm 
      p}^{38/17}\right\} \times \cr
&&  \beta_{\rm g}^{-3/17} \beta_*^{6/17} f_{\rm T}^{30/17} r_*^{12/17} m_*^{-22/17} M_6^{6/17}
     \times \cr
&& \left[{3(n-1) \over 2}\right]^{-12/17}  \left(1-{r_{\rm 0}\over 2a}\right)^{-14/17}
    \left({r_{\rm p} \over r_{\rm p*}}\right)^{-6/17}  \left({\dot{M}\over \dot{M}_{\rm 
    p}}\right)^{-12/17} .
\label{eq:rhor053}
\end{eqnarray}
Equation~(\ref{eq:rhor053}) shows that the mass density at the radiation radius $r_{\rm 0}$ decreases with
accretion rate $\rho_{\rm 0} \propto \dot{M}^{-12/17}$ and thus increases with time $\rho_0 \propto 
t^{12n/17} \propto t^{60/51}$ for $n=5/3$, mainly because of the receding of the radiation radius $r_{\rm 
0}$ with the decay of accretion rate. The mass density at $r_{\rm 0}$ increases with the mass of SMBHs but 
decreases with stellar mass, $\rho_0 \propto r_*^{12/17} m_*^{-22/17} \propto m_*^{-(10+12\zeta)/17} 
\propto m_*^{-0.736}$ for $\zeta=0.21$. 

From equation~(\ref{eq:rhodotr}), we have the mass density around the ellipse 
\begin{eqnarray}
\rho  &\simeq&  A_0^{1/2} C_0  \left(1-{r_{\rm S} \over r}\right)^{-1} 
                 \left[1+ \left({1+e\over 2}\right)\left({r_{\rm S} \over r}\right)C_{\rm p}^{-1} 
                 \left(1-{r \over 2a}\right)^{-1}\right]^{-1/2} \left({r_0 \over r}\right) \rho_0 .
\label{eq:rhor}
\end{eqnarray}
Equation~(\ref{eq:rhor}), together with equations~(\ref{eq:radradiuscent}) and (\ref{eq:rhor0}), gives 
\begin{eqnarray}
\rho &\simeq & \left(1-{r_{\rm S} \over r}\right)^{-1} \left[1+ \left({1+e\over 
           2}\right)\left({r_{\rm S} \over r}\right) C_{\rm p}^{-1} \left(1-{r \over 2a}\right)^{-1}\right]^{-1/2} 
           \left({r_{\rm p} \over r}\right)  \times \cr
       && \left({45\pi \over 8 }f \right)^{(7\gamma-11)/(11\gamma-7)} 
            \left({20 \over 3 \alpha_{\rm p}}f\right)^{-(13\gamma-23)/(11\gamma-7)}  \left({1+e\over 2}\right)^{2
            (5\gamma-4)/(11\gamma-7)} \times \cr
&& \left\{A_{\rm 0}^{-(4\gamma-5)/(11\gamma-7)} B_{\rm 0}^{-(7\gamma-11)/(11\gamma-7)} 
          C_{\rm 0}^{-2(4\gamma-5)/(11\gamma-7)} C_{\rm p}^{(35\gamma-37)/2(11\gamma-7)} \right. \times \cr
&&  \left. D_{\rm p}^{(20\gamma-34)/(11\gamma-7)}\right\}  \left\{\left({\beta_{\rm g}^{-1} k_{\rm B} \over 
      \mu m_{\rm H}}\right)^{-9/(11\gamma-7)} \sigma_{\rm SB}^{-(7\gamma-11)/(11\gamma-7)} \kappa_{\rm 
      es}^{-2(\gamma+1)/(11\gamma-7)}  \right.     \times \cr
&&  \left. r_{\rm S}^{-2(5\gamma-4)/(11\gamma-7)}  c^{3(7\gamma -5)/(11\gamma-7)} \kappa_{\rm 
       0}^{-2(4\gamma-5)/(11\gamma-7)}\right\} \left({r_{\rm p} \over r_{\rm S}}\right)^{-6(\gamma
        +1)/(11\gamma-7)} \times \cr
 && \left(1-{r_0 \over 2a}\right)^{-2(4\gamma-5)/(11\gamma-7)} \left({\dot{M}\over 
	\dot{M}_{\rm Edd}}\right)^{-6(\gamma-2)/(11\gamma-7)} . 
\label{eq:rhorr}
\end{eqnarray}
For $\gamma=5/3$, we have
\begin{eqnarray}
\rho &\simeq & \left(1-{r_{\rm S} \over r}\right)^{-1} \left[1+ \left({1+e\over 
           2}\right)\left({r_{\rm S} \over r}\right) C_{\rm p}^{-1} \left(1-{r \over 2a}\right)^{-1}\right]^{-1/2} 
           \left({r_{\rm p} \over r}\right)  \times \cr
      && \left({45\pi \over 8 }f \right)^{1/17} \left({20 \over 3 \alpha_{\rm p}}f\right)^{2/17}
             \left({1+e\over 2}\right)^{13/17} \times \cr
&& \left\{A_{\rm 0}^{-5/34}  B_{\rm 0}^{-1/17}  C_{\rm 0}^{-5/17}  C_{\rm p}^{16/17} 
         D_{\rm p}^{-1/17}\right\} \times \cr
   &&  \left\{\left({\beta_{\rm g}^{-1} k_{\rm B} \over \mu m_{\rm H}}\right)^{-27/34}
      \sigma_{\rm SB}^{-1/17} \kappa_{\rm es}^{-8/17} 
       r_{\rm S}^{-13/17} c^{30/17} \right.     \times \cr
&&  \left. \kappa_{\rm 0}^{-5/17} \right\} 
        \left({r_{\rm p} \over r_{\rm S}}\right)^{-24/17} \left(1-{r_0 \over 2a}\right)^{-5/17} 
         \left({\dot{M}\over \dot{M}_{\rm Edd}}\right)^{3/17} \cr
&\simeq& 2.38 \times 10^{-5} \, ({\rm g\; cm^{-3}})\,  f^{3/17} \alpha_{\rm -1}^{-2/17} 
       \left({1+e\over 2}\right)^{13/17} \times \cr
&   &   \left(1-{r_{\rm S} \over r}\right)^{-1}  \left[1+ \left({1+e\over 2}\right)\left({r_{\rm 
         S} \over r}\right) C_{\rm p}^{-1} \left(1-{r \over 2a}\right)^{-1}\right]^{-1/2}  \times \cr
&    &  \left\{A_{\rm 0}^{-5/34}  B_{\rm 0}^{-1/17}  C_{\rm 0}^{-5/17}  C_{\rm p}^{16/17} 
         D_{\rm p}^{-1/17}\right\} \times \cr
&  &    \beta_{\rm g}^{27/34} \beta_*^{24/17} f_{\rm T}^{-33/17} r_*^{-57/34} m_*^{14/17} M_6^{-3/34}\times \cr
&  &  \left[{3(n-1) \over 2}\right]^{3/17}  \left(1-{r_{\rm 0}\over 2a}\right)^{-5/17} 
       \left({r_{\rm p} \over r_{\rm p*}}\right)^{-24/17} 
         \left({\dot{M}\over \dot{M}_{\rm p}}\right)^{3/17}  \left({r_{\rm p} \over r}\right)  .
\label{eq:rhorr53}
\end{eqnarray}
Equation~(\ref{eq:rhorr53}) shows that the mass density at a given radius depends weakly on 
the accretion rate and BH mass and decreases with the mass of a star, $\rho \propto r_*^{-57/34} m_*^{14/17}
\propto m_*^{-(29-57\zeta)/34} \propto m_*^{-0.501}$.

From equations~(\ref{eq:scalheight0}) and (\ref{eq:engbalTR}), we have the disk half-thickness at $r_{\rm 0}$,
\begin{eqnarray}
H_{\rm 0} &=&  \left({45\pi \over 8}f\right) \left({1+e \over 2}\right)^{1/2} 
      \left\{A_{\rm 0}^{-1/2} B_{\rm 0}^{-1} C_{\rm 0}^{-1} 
      D_{\rm p}\right\} \left\{\sigma_{\rm SB}^{-1} \kappa_{\rm es}^{-1} c^3\right\}  \times \cr
      && \left({r_{\rm p}\over r_{\rm S}}\right)^{-3/2} 
      \left({\dot{M}\over \dot{M}_{\rm Edd}}\right) T_0^{-4} ,
\label{eq:scalheight01}
\end{eqnarray}
which suggests a fourth power of the temperature of the disk center. Because the temperature is nearly independent 
of the accretion rate, the disk half-thickness at $r_{\rm 0}$ approximately linearly increases with the accretion
rate. Equations~(\ref{eq:scalheight01}) and (\ref{eq:radtempcent}) give 
\begin{eqnarray}
H_{\rm 0} &\simeq & \left({45\pi \over 8 }f \right)^{-(7\gamma+13)/(33\gamma-
	21)} \left({20 \over 3 \alpha_{\rm p}}f\right)^{24(\gamma+1)/(33\gamma-21)} \left({1+e\over 
	2}\right)^{-(31\gamma-11)/2(33\gamma-21)} \times \cr
&& \left\{ A_{\rm 0}^{-(25\gamma-29)/2(33\gamma-21)} B_{\rm 0}^{(7\gamma+13) /(33\gamma-21)} 
    C_{\rm 0}^{-(25\gamma-29)/(33\gamma-21)} C_{\rm p}^{-12(\gamma+1)/(33\gamma-21)}  \right.
     \times \cr
&&\left. D_{\rm p}^{-(31\gamma+37)/(33\gamma-21)}\right\} \left\{\left({\beta_{\rm g}^{-1} k_{\rm B} \over \mu m_{\rm H}}\right)^{-24/(33\gamma-21)}
   \sigma_{\rm SB}^{(7\gamma+13)/(33\gamma-21)} \kappa_{\rm es}^{-3(3\gamma+1)/(33\gamma-21)}    
     \right. \times \cr
&& \left.  r_{\rm S}^{16(2\gamma-1)/(33\gamma-21)} c^{-3(7\gamma-3)/(33\gamma-21)} \kappa_{\rm 
      0}^{8(\gamma+1)/(33\gamma-21)}\right\} \left({r_{\rm p} \over r_{\rm S}}\right)^{9(5\gamma-
      9)/2(33\gamma-21)} \times \cr
&& \left(1-{r_0 \over 2a}\right)^{8(\gamma+1) /(33\gamma-21)}  \left({\dot{M}\over \dot{M}_{\rm 
      Edd}}\right)^{(17\gamma+11)/(33\gamma-21)} .
\label{eq:scalheight02}
\end{eqnarray}
For $\gamma=5/3$, the disk half-thickness at $r_{\rm 0}$ becomes
\begin{eqnarray}
H_{\rm 0} &\simeq & \left({45\pi \over 8 }f \right)^{-37/51} \left({20 \over 3 \alpha_{\rm
	p}}f\right)^{32/17} \left({1+e\over 2}\right)^{-61/102} \times \cr
&& \left\{ A_{\rm 0}^{-19/102} B_{\rm 0}^{37/51} C_{\rm 0}^{-19/51} C_{\rm p}^{-16/17}
    D_{\rm p}^{-133/51} \right\} \times \cr
&& \left\{\left({\beta_{\rm g}^{-1} k_{\rm B} \over \mu m_{\rm H}}\right)^{-12/17} \sigma_{\rm SB}^{37/51}
   \kappa_{\rm es}^{-9/17} r_{\rm S}^{56/51} c^{-13/17} \kappa_{\rm 0}^{32/51} \right\} \times \cr
&&  \left({r_{\rm p} \over r_{\rm S}}\right)^{-3/34} \left(1-{r_{\rm 0}\over 2a}\right)^{32/51} 
    \left({\dot{M}\over \dot{M}_{\rm Edd}}\right)^{59/51} \cr
&\simeq & 7.23 \times 10^{15} \, ({\rm cm})\, f^{59/51} \alpha_{\rm -1}^{-32/17} \left({1+e\over 
	2}\right)^{-61/102} \times \cr
&&   \left\{A_{\rm 0}^{-19/102} B_{\rm 0}^{37/51} C_{\rm 0}^{-19/51} C_{\rm 
	p}^{-16/17} D_{\rm p}^{-133/51} \right\} \times \cr
&& \beta_{\rm g}^{12/17} \beta_*^{3/34} f_{\rm T}^{-121/34} r_*^{-31/17} m_*^{239/102} M_6^{-59/102} 
        \left({r_{\rm p} \over r_{\rm p*}}\right)^{-3/34} \times \cr
&&  \left[{3(n-1) \over 2}\right]^{59/51} \left(1-{r_{\rm 0}\over 2a}\right)^{32/51} \left({\dot{M}\over 
         \dot{M}_{\rm p}}\right)^{59/51} .
\label{eq:scalheight53}
\end{eqnarray}

From equations~(\ref{eq:rhor0}) and (\ref{eq:scalheight02}), we obtain the surface density at the radiation 
radius $r_0$,
\begin{eqnarray}
\Sigma_0 &\simeq & 2 \rho_0 H_0 \cr
                    &\simeq & 2 \left({45\pi \over 8 }f \right)^{2(7\gamma-11)/3(11\gamma- 7)} 
                    \left({20 \over 3 \alpha_{\rm p}}f\right)^{-(5\gamma-13)/(11\gamma-7)} \left({1+e\over 
	2}\right)^{(29\gamma-25)/6(11\gamma-7)} \times \cr
	&& \left\{A_{\rm 0}^{-(49\gamma-41)/6(11\gamma-7)} B_{\rm 0}^{-2(7\gamma-11)/3(11\gamma-7)} 
    C_{\rm 0}^{-(49\gamma-41)/3(11\gamma-7)} C_{\rm p}^{2(4\gamma-5)/(11\gamma-7)}  \right. \times \cr
&&  \left. D_{\rm p}^{(29\gamma-61)/3(11\gamma-7)}\right\}  \left\{\left({\beta_{\rm g}^{-1} k_{\rm B} \over 
    \mu m_{\rm H}}\right)^{-6/(11\gamma-7)}
    \sigma_{\rm SB}^{-2(7\gamma-11)/3 (11\gamma-7)} \kappa_{\rm es}^{-(5\gamma-1)/(11\gamma-7)}
     \right. \times \cr
&&  \left. r_{\rm S}^{2(\gamma+1)/3(11\gamma-7)} c^{2(7\gamma-5)/(11\gamma-7)} \kappa_{\rm 
        0}^{-4(4\gamma-5)/3(11\gamma-7)}\right\} \left({r_{\rm p} \over r_{\rm S}}\right)^{3(\gamma
        -5)/2(11\gamma-7)} \times \cr
 && \left(1-{r_0 \over 2a}\right)^{-4(4\gamma-5)/3(11\gamma-7)}  \left({\dot{M}\over \dot{M}_{\rm 
        Edd}}\right)^{-(\gamma-17)/3(11\gamma-7)} . 
\label{eq:sigmar0}
\end{eqnarray}
For $\gamma=5/3$, we have
\begin{eqnarray}
\Sigma_0 &\simeq & 2 \left({45\pi \over 8 }f \right)^{2/51} 
                    \left({20 \over 3 \alpha_{\rm p}}f\right)^{7/17} \left({1+e\over 
	2}\right)^{35/102} \times \cr
	&& \left\{A_{\rm 0}^{-61/102} B_{\rm 0}^{-2/51} C_{\rm 0}^{-61/51} C_{\rm p}^{5/17} 
    D_{\rm p}^{-19/51}     \right\} \times \cr
&&  \left\{\left({\beta_{\rm g}^{-1} k_{\rm B} \over \mu m_{\rm H}}\right)^{-9/17} 
    \sigma_{\rm SB}^{-2/51} \kappa_{\rm es}^{-11/17} r_{\rm S}^{8/51} c^{20/17}  \right. \times \cr
&&  \left. \kappa_{\rm 0}^{-10/51}\right\} \left({r_{\rm p} \over r_{\rm
	S}}\right)^{-15/34} \left(1-{r_0 \over 2a}\right)^{-10/51} \left({\dot{M}\over 
	\dot{M}_{\rm Edd}}\right)^{23/51} \cr
&\simeq& 1.00 \times 10^7 \, ({\rm g\; cm^{-2}}) \, f^{23/51} \alpha_{-1}^{-7/17} \left({1+e\over 
	2}\right)^{35/102} \times \cr
&& \left\{A_{\rm 0}^{-61/102} B_{\rm 0}^{-2/51} C_{\rm 0}^{-61/51} C_{\rm p}^{5/17} D_{\rm p}^{-19/51} 
        \right\} \times \cr
        && \beta_{\rm g}^{9/17} \beta_*^{15/34}  f_{\rm T}^{-61/34} r_*^{-19/17} m_*^{107/102} M_6^{-23/102}
         \left({r_{\rm p} \over r_{\rm p*}}\right)^{-15/34} \times \cr
&& \left[{3(n-1) \over 2}\right]^{23/51} \left(1-{r_{\rm 0}\over 2a}\right)^{-10/51} \left({\dot{M}\over 
       \dot{M}_{\rm p}}\right)^{23/51} .
\label{eq:sigmar053}
\end{eqnarray}

From equation~(\ref{eq:masssigannu}), we have the surface density around the ellipse
\begin{eqnarray}
{\Sigma \over \Sigma_0} 
                 &\simeq&  C_0 A_0^{1/2} \left(1-{r_{\rm S} 
                 \over r}\right)^{-1} \left[1+ \left({1+e\over 2}\right)\left({r_{\rm S} \over r}\right)C_{\rm p}^{-1} 
                 \left(1-{r \over 2a}\right)^{-1}\right]^{-1/2} ,
\label{eq:sigmarr0}                 
\end{eqnarray}
which is nearly independent of radius $r$.

\subsection{Geometrically thin  and optically thick disk}
\label{sec:optdepth}

From equation~(\ref{eq:scalheight0}), we have the half-opening angle of the disk
\begin{eqnarray}
{H_{\rm 0} \over r_{\rm 0}} &=& \left({c_{\rm s0}\over v_{\rm p}}\right) A_0^{(\gamma-1)/4} C_0^{(\gamma-1)/2} 
      C_{\rm p}^{-(\gamma-1)/4} \left({r_{\rm 0}\over r_{\rm p}}\right)^{(\gamma-1)/2} \cr
      &=& \left({1+e\over2}\right)^{-1/2} \left\{A_0^{(\gamma-1)/4} C_0^{(\gamma-1)/2} 
      C_{\rm p}^{-(\gamma+1)/4}\right\} \left\{\left({\beta_{\rm g}^{-1} k_{\rm B} \over \mu m_{\rm 
      H}}\right)^{1/2} c^{-1}\right\} \times \cr
  && \left({r_{\rm p} \over r_{\rm S}}\right)^{1/2} T_0^{1/2} 
      \left({r_{\rm 0}\over r_{\rm p}}\right)^{(\gamma-1)/2} .
\label{eq:openingangle0}
\end{eqnarray}
Equation~(\ref{eq:openingangle0}), together with equations~(\ref{eq:RadRadTR}) and (\ref{eq:radradiuscent}),
gives
\begin{eqnarray}
{H_{\rm 0} \over r_{\rm 0}} &=& \left({45\pi \over  8 }f \right)^{-(7\gamma-11)/3(11\gamma-
    7)} \left({20 \over 3 \alpha_{\rm p}}f\right)^{2(4\gamma-5)/(11\gamma-7)} \left({1+e\over 
	2}\right)^{-(31\gamma-23)/6(11\gamma-7)} \times \cr
&&  \left\{A_{\rm 0}^{(4\gamma-5)/3(11\gamma-7)} B_{\rm 0}^{(7\gamma-11)/3(11\gamma-7)} 
        C_{\rm 0}^{2(4\gamma- 5)/3(11\gamma-7)} C_{\rm p}^{-(19\gamma-17)/2(11\gamma-7)} \right. 
         \times \cr
&&  \left. D_{\rm p}^{-(31\gamma-41)/3(11\gamma-7)}\right\} \left\{\left({\beta_{\rm g}^{-1} k_{\rm B} 
      \over \mu m_{\rm H}}\right)^{3/(11\gamma-7)} \sigma_{\rm SB}^{(7\gamma-11)/3(11\gamma-7)} 
      \kappa_{\rm es}^{-3(\gamma-1)/(11\gamma-7)}  \right. \times \cr
&& \left. r_{\rm S}^{-(\gamma+1)/3(11\gamma-7)} c^{-(7\gamma-5)/(11\gamma-7)} \kappa_{\rm 
      0}^{2(4\gamma-5)/3(11\gamma-7)}\right\} \left({r_{\rm p} \over r_{\rm S}}\right)^{-(7\gamma- 
      11)/2(11\gamma-7)}  \times \cr
&& \left(1-{r_0 \over 2a}\right)^{2(4\gamma -5)/3(11\gamma-7)}  \left({\dot{M}\over \dot{M}_{\rm 
      Edd}}\right)^{(17\gamma-19)/3(11\gamma-7)} .
\label{eq:openingangle01}
\end{eqnarray}
For $\gamma=5/3$, the disk opening angle becomes
\begin{eqnarray}
{H_{\rm 0} \over r_{\rm 0}} &=& \left({45\pi \over  8 }f \right)^{-1/51} \left({20 \over 3 
	\alpha_{\rm p}}f\right)^{5/17} \left({1+e\over 2}\right)^{-43/102} \times \cr
&&  \left\{A_{\rm 0}^{5/102} B_{\rm 0}^{1/51} C_{\rm 0}^{5/51} C_{\rm p}^{-11/17} D_{\rm p}^{-16/51}
         \right\} \times \cr
&&  \left\{\left({\beta_{\rm g}^{-1} k_{\rm B} \over \mu m_{\rm H}}\right)^{9/34} \sigma_{\rm SB}^{1/51} 
	\kappa_{\rm es}^{-3/17} r_{\rm S}^{-4/51} c^{-10/17} \kappa_{\rm 0}^{5/51} \right\} \times \cr
&&	\left({r_{\rm p} \over r_{\rm S}}\right)^{-1/34} \left(1-{r_0\over 2a}\right)^{5/51} 
        \left({\dot{M}\over \dot{M}_{\rm Edd}}\right)^{14/51} \cr
&\simeq& 3.03 \times 10^{-2} \, f^{14/51} \alpha_{\rm -1}^{-5/17} \left({1+e\over 2}\right)^{-43/102} \times \cr
&&  \left\{A_{\rm 0}^{5/102} B_{\rm 0}^{1/51} C_{\rm 0}^{5/51} C_{\rm p}^{-11/17} D_{\rm p}^{-16/51}
      \right\} \times \cr
&&  \beta_{\rm g}^{-9/34} \beta_*^{1/34} f_{\rm T}^{-29/34} M_6^{-8/17} r_*^{-15/34} m_*^{19/34}  \times \cr
&& \left[{3(n-1) \over 2}\right]^{14/51} \left({r_{\rm p} \over r_{\rm p*}}\right)^{-1/34} \left(1-{r_{\rm 0}\over 
     2a}\right)^{5/51} \left({\dot{M}\over \dot{M}_{\rm p}}\right)^{14/51} .
\label{eq:openingangle53}
\end{eqnarray}
Because the scale height of an elliptical accretion disk at radius $r$ around the ellipse is $H\simeq (H_0/r_0) r$, 
equation~(\ref{eq:openingangle53}) shows that the elliptical accretion disk is geometrically thin.

From equations~(\ref{eq:optdepthr0}), (\ref{eq:radtempcent}), and (\ref{eq:radradiuscent}), we have the 
vertical optical depth at $r_{\rm 0}$,
\begin{eqnarray}
\tau_{\rm 0} &\simeq& \left({c \over v_{\rm 0}}\right) \left({r_{\rm 0}\over H_{\rm 0}}\right) \cr
  &\simeq& A_{\rm 0}^{-(115\gamma-101)/4(33\gamma-21)} B_{\rm 0}^{-(7\gamma-23) /(33\gamma-21)}
         C_{\rm 0}^{-(115\gamma-101)/2(33\gamma-21)}  \times \cr
  &&  C_{\rm p}^{(49\gamma-59)/4(11\gamma-7)}  D_{\rm p}^{(31\gamma-80)/(33\gamma-21)}
         \left({45\pi \over 8}\right)^{(7\gamma-23)/(33\gamma-21)} \times \cr
&&\left({20\over 3 \alpha_{\rm p}}\right)^{-(8\gamma-19)/(11\gamma-7)} f^{-17(\gamma-2)/(33\gamma-21)} 
  \left({1+e\over 2}\right)^{(31\gamma-29)/2(33\gamma-21)} \times \cr
&& \left\{\left({\beta_{\rm g}^{-1} k_{\rm B} \over \mu m_{\rm H}}\right)^{-17/2(11\gamma-7)} \sigma_{\rm 
	SB}^{-(7\gamma-23)/(33\gamma-21)} \kappa_{\rm es}^{3(3\gamma-5)/(33\gamma-21)} 
    r_{\rm S}^{(\gamma+4)/(33\gamma-21)} \right. \times \cr
&&  \left. c^{3(7\gamma-6)/(33\gamma-21)} \kappa_{\rm 0}^{-(8\gamma-19)/(33\gamma -21)}\right\} 
    \left({r_{\rm p} \over r_{\rm S}}\right)^{3(3\gamma-5)/(11\gamma-7)} \times \cr
&&  \left(1 - {r_0 \over 2a}\right)^{-(49\gamma-59) /2(33\gamma-21)} \left({\dot{M}\over 
     \dot{M}_{\rm Edd}}\right)^{-17(\gamma -2)/(33\gamma-21)} .
\end{eqnarray}
For $\gamma=5/3$, we have
\begin{eqnarray}
\tau_{\rm 0} &\simeq& \left({45\pi \over 8}\right)^{-1/3} \left({20 \over 3 \alpha_{\rm 
		p}}\right)^{1/2} f^{1/6} \left({1+e\over 2}\right)^{1/3} \times \cr
&& A_{\rm 0}^{-2/3} B_{\rm 0}^{1/3} C_{\rm 0}^{-4/3} C_{\rm p}^{1/2} D_{\rm p}^{-85/102} \times \cr
&& \left\{\left({\beta_{\rm g}^{-1} k_{\rm B} \over \mu m_{\rm H}}\right)^{-3/4} \sigma_{\rm 
	SB}^{1/3}  r_{\rm S}^{1/6} c^{1/2} \kappa_{\rm 0}^{1/6} \right\}  \times \cr
&&  \left(1-{r_0 \over 2a}\right)^{-1/3} \left({\dot{M}\over \dot{M}_{\rm Edd}}\right)^{1/6} \cr
& \simeq & 2.96 \times 10^4\; \times \alpha_{-1}^{-1/2} f^{1/6} \left({1+e\over 2}\right)^{1/3} \times \cr
&& \left\{ A_{\rm 0}^{-2/3} B_{\rm 0}^{1/3} C_{\rm 0}^{-4/3} C_{\rm p}^{1/2} D_{\rm p}^{-85/102}\right\} \times \cr
&& \beta_{\rm g}^{3/4} f_{\rm T}^{-1/2} r_*^{-1/4} m_*^{1/3} M_6^{-1/12}  \times \cr
    && \left[{3(n-1) \over 2}\right]^{1/6}  \left(1-{r_0 \over 2a}\right)^{-1/3} \left({\dot{M}\over \dot{M}_{\rm 
        p}}\right)^{1/6} ,
\end{eqnarray}
which is independent of the electron scattering opacity,  orbital penetration factor of a star, and disk pericenter 
radius $r_{\rm p}$ and depends rarely on BH mass. The optical depth $\tau_0$ depends only weakly on the mass 
of a star and accretion rate with $\tau_0 \propto m_*^{(1+3\zeta)/12} \dot{M}^{1/6} \propto m_*^{0.136}
\dot{M}^{0.167}$ for $\zeta=0.21$. The elliptical accretion disk at the radiation radius remains optically thick until 
the event essentially fades away.

The vertical optical depth at radiation radius $r_0$ due to electron scattering is 
\begin{eqnarray}
\tau_{\rm es,0} &\simeq& \kappa_{\rm es} \Sigma_0/2 \cr
          & \simeq& 1.75 \times 10^6 \, f^{23/51} \alpha_{-1}^{-7/17} \left({1+e\over 
	2}\right)^{35/102} \times \cr
&& \left\{A_{\rm 0}^{-61/102} B_{\rm 0}^{-2/51} C_{\rm 0}^{-61/51} C_{\rm p}^{5/17} 
        D_{\rm p}^{-19/51} 
        \right\} \times \cr
        && \beta_{\rm g}^{9/17} \beta_*^{15/34}  f_{\rm T}^{-61/34} r_*^{-19/17} m_*^{107/102} M_6^{-23/102} 
        \left({r_{\rm p} \over r_{\rm p*}}\right)^{-15/34}  \times \cr
&& \left[{3(n-1) \over 2}\right]^{23/51}  \left(1-{r_0 \over 2a}\right)^{-10/51} \left({\dot{M}\over 
      \dot{M}_{\rm p}}\right)^{23/51} 
\label{eq:tau053}
\end{eqnarray}
and $\tau_{\rm es} \simeq \kappa_{\rm es} \Sigma /2 \simeq \tau_{\rm es,0}$ at radius $r$, which is  
independent of radius around the ellipse. The Rosseland mean opacity, $\tau_{\rm R} \simeq \tau_0^2/
\tau_{\rm es,0} \simeq 502$, is about three orders of magnitude smaller than the optical depth due to 
the electron scattering opacity.

When the vertical diffusion timescale due to the electron scattering is longer than the radial dynamic 
timescale, the soft X-ray photons would be trapped in fluids and advected around the ellipse without escape. 
The ratio of the vertical diffusion time $t_{\rm diff,es}$ to the radial dynamic timescale $t_{\rm dyn}$ is 
\begin{eqnarray}
{t_{\rm diff,es} \over t_{\rm dyn}} & \simeq & \left({H \over c} \tau_{\rm es}\right) \left({r \over v}\right)^{-1} \cr
             &\simeq & \left({H_0 \over r_0}\right) \left({\kappa_{\rm es} \Sigma_0 \over 2}\right) {v \over c} \cr
             &\simeq &  1.09 \times 10^4 \,  \alpha_{\rm -1}^{-12/17} f^{37/51} \left({1+e\over 2}\right)^{-4/51} 
             \times \cr
&&  \beta_{\rm g}^{9/34} \beta_*^{33/34} f_{\rm T}^{-107/34} M_6^{-37/102} r_*^{-35/17} m_*^{181/102}  
        \left({r_{\rm p} \over r_{\rm p*}}\right)^{-33/34}  \times \cr
&& \left[{3(n-1) \over 2}\right]^{37/51}  \left(1-{r_{\rm 0}\over 2a}\right)^{-5/51} \left({\dot{M}\over 
        \dot{M}_{\rm p}}\right)^{37/51} \left(1-{r\over 2a}\right)^{1/2} \left({r_{\rm p}\over  r}\right)^{1/2} 
         \label{eq:esdiffusetime}
\end{eqnarray}
for $\gamma=5/3$, where we have neglected all the quantities of order unity. At apocenter of the ellipse 
($r=(1+e) a$), we have
\begin{eqnarray}
{t_{\rm diff,es} \over t_{\rm dyn}} & \simeq &  109 \, \alpha_{\rm -1}^{-12/17}f^{37/51}  f_{\rm T}^{-141/34}
        \left({1+e\over 2}\right)^{-161/102} \times \cr
&&  \beta_{\rm g}^{9/34} \beta_*^{-1/34}    r_*^{-35/17} m_*^{215/102} M_6^{-71/102} 
         (1+\Delta_*)   \times \cr
         && \left[{3(n-1) \over 2}\right]^{37/51} \left({r_{\rm p} \over r_{\rm p*}}\right)^{-33/34} \left(1-{r_{\rm 
         0}\over 2a}\right)^{-5/51} 
         \left({\dot{M}\over \dot{M}_{\rm p}}\right)^{37/51}  ,
 \label{eq:esdiffuseap}
\end{eqnarray}
where we have used $e= [1- 2\delta (1+\Delta_*)]^{1/2}$ and $\delta \simeq 0.02 f_{\rm T}^{-1} \beta_*^{-1} 
m_*^{1/3} M_6^{-1/3}$. 

Equation~(\ref{eq:esdiffuseap}) shows that the vertical diffusion timescale due to electron scattering is much 
longer than the radial dynamic timescale even at apocenter of the ellipse. Equation~(\ref{eq:esdiffuseap}) 
shows that when the accretion rate decreases to less than the critical accretion rate
\begin{eqnarray}
  \left({\dot{M}_{\rm x} \over \dot{M}_{\rm p}}\right)  &\simeq& 1.97 \times 10^{-2} \, 
             \alpha_{\rm -1}^{36/37} f^{-1} \left({f_{\rm T}\over 1.56}\right)^{423/74} \left({1+e\over 2}\right)^{161/74} 
             \times \cr
&&  \beta_{\rm g}^{-27/74} \beta_*^{3/74} r_*^{105/37} m_*^{-215/74} M_6^{71/74} 
           \times \cr
 && \left[{3(n-1) \over 2}\right]^{-1} (1+\Delta_*)^{-51/37}  \left({r_{\rm p} \over r_{\rm p*}}\right)^{99/74} ,
   \label{eq:escritrate}
\end{eqnarray}
the vertical diffusion timescale is smaller than the dynamic timescale. To obtain equation~(\ref{eq:escritrate}), 
we have assumed $r_0 \ll 2 a$ for $\dot{M} \la \dot{M}_{\rm x}$. Equation~(\ref{eq:escritrate}) suggests that 
the elliptical accretion disk with large viscosity parameter $\alpha_{\rm p}$ of TDEs with large BH mass but small 
stellar mass may have a rapid change of radiation 
characteristics before the accretion mode changes from a thin disk to advection-dominated accretion flow.
Because of the energy conservation and the invariance  of the radiation efficiency, $\Delta{Q}^{+} \simeq 
\Delta{Q}^{-} \simeq \Delta{Q}_{\rm optical}^{-} + \Delta{Q}_{\rm X-ray}^{-}$, the rapid brightening of optical/UV 
TDEs in X-rays would be associated with a decrease of optical/UV luminosity, but the total (bolometric) luminosity 
may smoothly follow the accretion rate. Although the real size of the X-ray emission region is large, the effective
blackbody spherical radius of the X-ray luminosity may be small. If both the emission regions of optical/UV and 
X-ray luminosities are spherical, the effective spherical radius $R_{\rm X}$ of the X-ray emission region is $R_{\rm X} 
\simeq R_{\rm bb} (L_{\rm X}/L_{\rm opt})^{1/2} (T_{\rm bb}/T_{\rm X})^2 \sim 3.6\times 10^{-3} R_{\rm bb}$ for 
typical blackbody temperatures $T_{\rm bb} \sim 3\times 10^4 \, {\rm K}$ for optical/UV emission and $T_{\rm 
X}\sim 5\times 10^5 \, {\rm K}$ for X-ray radiation and $L_{\rm X}\sim L_{\rm opt}$. The effective spherical radius 
of the X-ray emission region would be about a few hundred times smaller than that of the optical/UV radiation 
region. Because the model predicts a TDE to be luminous in both optical/UV wave bands and soft X-rays  
at late time, it may be the interpretation of the observational distinction between UV/optical and 
X-ray-dominated TDE candidates. Or, it  may be the explanation of the rapid late-time X-ray brightening of the TDEs 
ASASSN-15oi \citep{gez17b,hol18}, AT2019azh \citep{liu20,van20}, OGLE16aaa \citep{kaj20}, and ASASSN-19dj 
\citep{hin20a}. We will discuss this issue further in a future work.

\subsection{Blackbody temperature and effective blackbody radii of TDEs}
\label{sec:BBTempRadi}

We now derive the surface temperature of the emission regions and the associated effective blackbody radius, 
both of which are measurable. From equations~(\ref{eq:heatcoolingbal}), (\ref{eq:Qcooling}), and 
(\ref{eq:radtrans}), we have
\begin{eqnarray}
T_{\rm bb}^4 &=&  \left({30\pi \over 4}f\right) \left({1+e \over 2}\right)^{1/2} B_{\rm 0}^{-1} D_{\rm p} 
                        \left\{ \kappa_{\rm es}^{-1} \sigma_{\rm SB}^{-1} r_{\rm S}^{-1} c^3\right\} \times\cr
       && \left({r_{\rm p} \over r_{\rm S}}\right)^{-3} \left({r_{\rm 0} \over r_{\rm p}}\right)^{-3/2} 
       \left(1- {r_0 \over 2a}\right)^{1/2} \left({\dot{M}\over \dot{M}_{\rm Edd}}\right) .
\label{eq:tempbb1}
\end{eqnarray}
Equations~(\ref{eq:tempbb1}) and (\ref{eq:radradiuscent}) give 
\begin{eqnarray}
T_{\rm bb} &=&\left({4\over3}\right)^{1/4} \left({45\pi \over 8}\right)^{(11\gamma+
      5)/4(11\gamma-7)} \left({20 \over 3 \alpha_{\rm p}}\right)^{-27/4(11\gamma-7)} \left({1+e 
      	\over 2}\right)^{(11\gamma-1)/8(11\gamma-7)} \times \cr
 && f^{11(\gamma-2)/4(11\gamma-7)}  A_{\rm 0}^{3(11\gamma-13)/16(11\gamma-7)} B_{\rm 
       0}^{-(11\gamma+5)/4(11\gamma-7)} C_{\rm 0}^{3(11\gamma-13)/8(11\gamma-7)}  	\times \cr
 && C_{\rm p}^{-3(11\gamma-25)/16(11\gamma-7)} D_{\rm p}^{(11\gamma+32)/4(11\gamma-7)}  
        \left\{\left({\beta_{\rm g}^{-1} k_{\rm B} \over \mu m_{\rm H}}\right)^{33/8(11\gamma-7)} 
        \sigma_{\rm SB}^{-(11\gamma+5)/4(11\gamma-7)}  \right. \times \cr
 && \left. \kappa_{\rm es}^{-(11\gamma-13)/4(11\gamma -7)} r_{\rm S}^{-(11\gamma -4)/4(11\gamma-7)} 
        c^{3(11\gamma-6)/4(11\gamma-7)} \kappa_{\rm 0}^{-9/4(11\gamma-7)}\right\}  \times \cr
 &&    \beta_*^{3(11\gamma-13)/4(11\gamma-7)} \left({r_{\rm t} \over r_{\rm S}}\right)^{-3(11\gamma
        -13)/4(11\gamma-7)} \left({r_{\rm p} \over r_{\rm p*}}\right)^{-3(11\gamma-13)/4(11\gamma-7)} 
             \times \cr
 & &  \left(1-{r_0 \over 2a}\right)^{(11\gamma-25)/8(11\gamma-7)} \left({\dot{M}_{\rm p}\over  
       \dot{M}_{\rm Edd}}\right)^{11(\gamma -2)/4(11\gamma-7)} \left({\dot{M}\over \dot{M}_{\rm p}} 
       \right)^{11(\gamma -2)/4(11\gamma-7)} .
\label{eq:tempbb2}
\end{eqnarray}
For $\gamma=5/3$, equation~(\ref{eq:tempbb2}) becomes 
\begin{eqnarray}
T_{\rm bb} &=&\left({4\over3}\right)^{1/4} \left({45\pi \over 8}\right)^{35/68}
    \left({20 \over 3 \alpha_{\rm p}}\right)^{-81/136} \left({1+e \over 2}\right)^{13/68} f^{-11/136} \times \cr
	&& A_{\rm 0}^{3/34} B_{\rm 0}^{-35/68} C_{\rm 0}^{3/17}  C_{\rm p}^{15/136} D_{\rm p}^{151/136}
     \left\{\left({\beta_{\rm g}^{-1} k_{\rm B} \over \mu m_{\rm H}}\right)^{99/272} \sigma_{\rm 
	SB}^{-35/68} \right. \times \cr
      && \left. \kappa_{\rm es}^{-2/17} r_{\rm S}^{-43/136} c^{111/136} \kappa_{\rm 0}^{-27/136}\right\} 
         \beta_*^{6/17} \left({r_{\rm t} \over r_{\rm S}}\right)^{-6/17} \left({r_{\rm p} \over r_{\rm p*}}\right)^{-6/17} 
         \times \cr
&& \left(1-{r_0 \over 2a}\right)^{-5/68} \left({\dot{M}_{\rm p}\over \dot{M}_{\rm 
        Edd}}\right)^{-11/136} \left({\dot{M}\over \dot{M}_{\rm p}}\right)^{-11/136} \cr
 &\simeq& 1.98 \times 10^4 \, {\rm K}\, \alpha_{\rm -1}^{81/136} \left({1+e \over 
     2}\right)^{13/68} \times \cr
&& f^{-11/136} A_{\rm 0}^{3/34} B_{\rm 0}^{-35/68} C_{\rm 0}^{3/17}  C_{\rm 
         p}^{15/136} D_{\rm p}^{151/136}      \times \cr
&& \beta_{\rm g}^{-99/272} \beta_*^{6/17} f_{\rm T}^{-15/136} M_6^{11/272} r_*^{-63/272} m_*^{-3/68} \times \cr
 &&  \left[{3(n-1) \over 2}\right]^{-11/136}  \left({r_{\rm p} \over r_{\rm p*}}\right)^{-6/17} \left(1-{r_{\rm 0}\over 
     2a}\right)^{-5/68} 
 \left({\dot{M}\over \dot{M}_{\rm p}}\right)^{-11/136} ,
\label{eq:tempbb3}
\end{eqnarray}
where the radiation radius $r_0$ is given with equation~(\ref{eq:radradiuscent2}).

\begin{figure} 
	\centering 
	\subfigure{
		\begin{minipage}[b]{1\textwidth} 
			\includegraphics[width=1\textwidth]{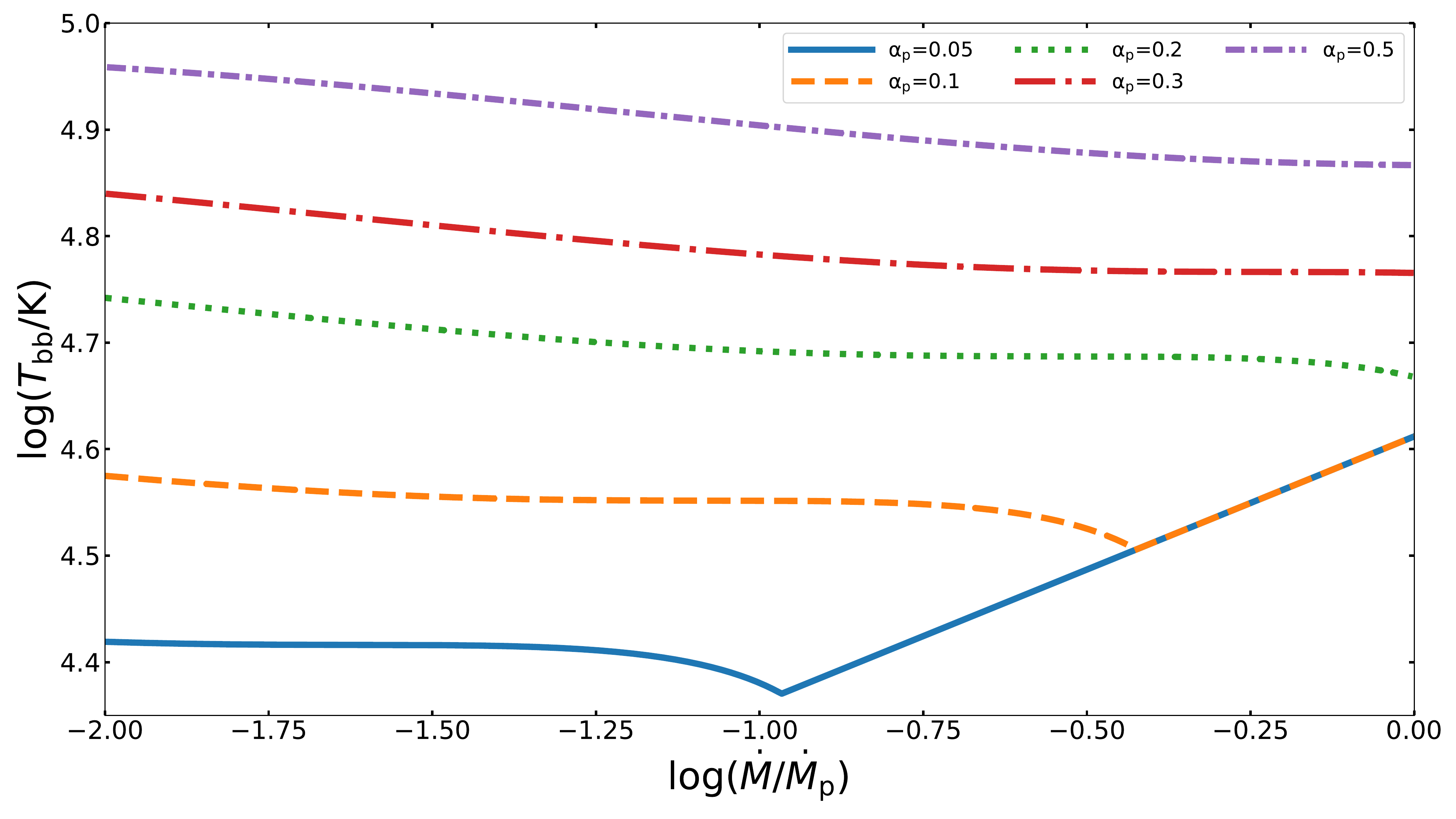} 
		\end{minipage} 
	}
	\subfigure{
		\begin{minipage}[b]{1\textwidth} 
			\includegraphics[width=1\textwidth]{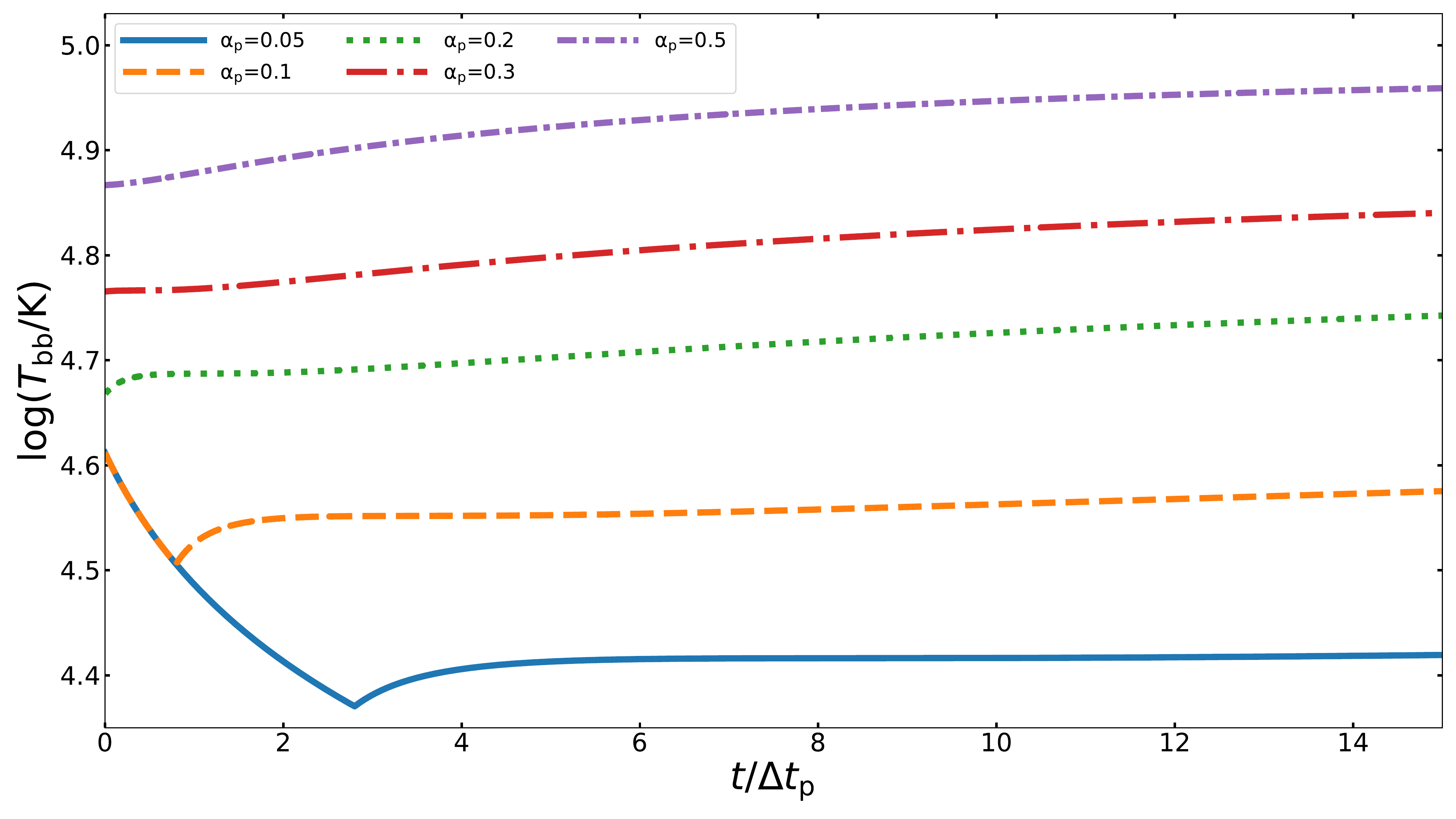} 	
		\end{minipage}	
	} 
\caption{Radiation blackbody temperature as a function of accretion rate (top panel) and
         time (bottom panel) for different viscosity parameters $\alpha_{\rm p}$. The blackbody temperature is 
         calculated for full disruption with $\beta_*=1.0$, $M_{\rm BH} = 10^6 M_\odot$, and $m_* = 0.3$. The other 
         parameters are $r_{\rm ms}=2 r_{\rm S}$, $\beta_{\rm g}=1$, $r_{\rm p} = r_{\rm p*}$, $r_* = m_*^{1-\zeta}$ 
         with $\zeta=0.21$, $f_{\rm T} = 1.56$, and $n=5/3$. The time starts at the peak of the accretion rate. The 
         radiation blackbody temperature is nearly constant except for the low viscosity parameter at time of about 
         the peak. 
		\label{fig:tbbt}
	}
\end{figure}
\begin{figure}
	\centering 
			\includegraphics[width=1\textwidth]{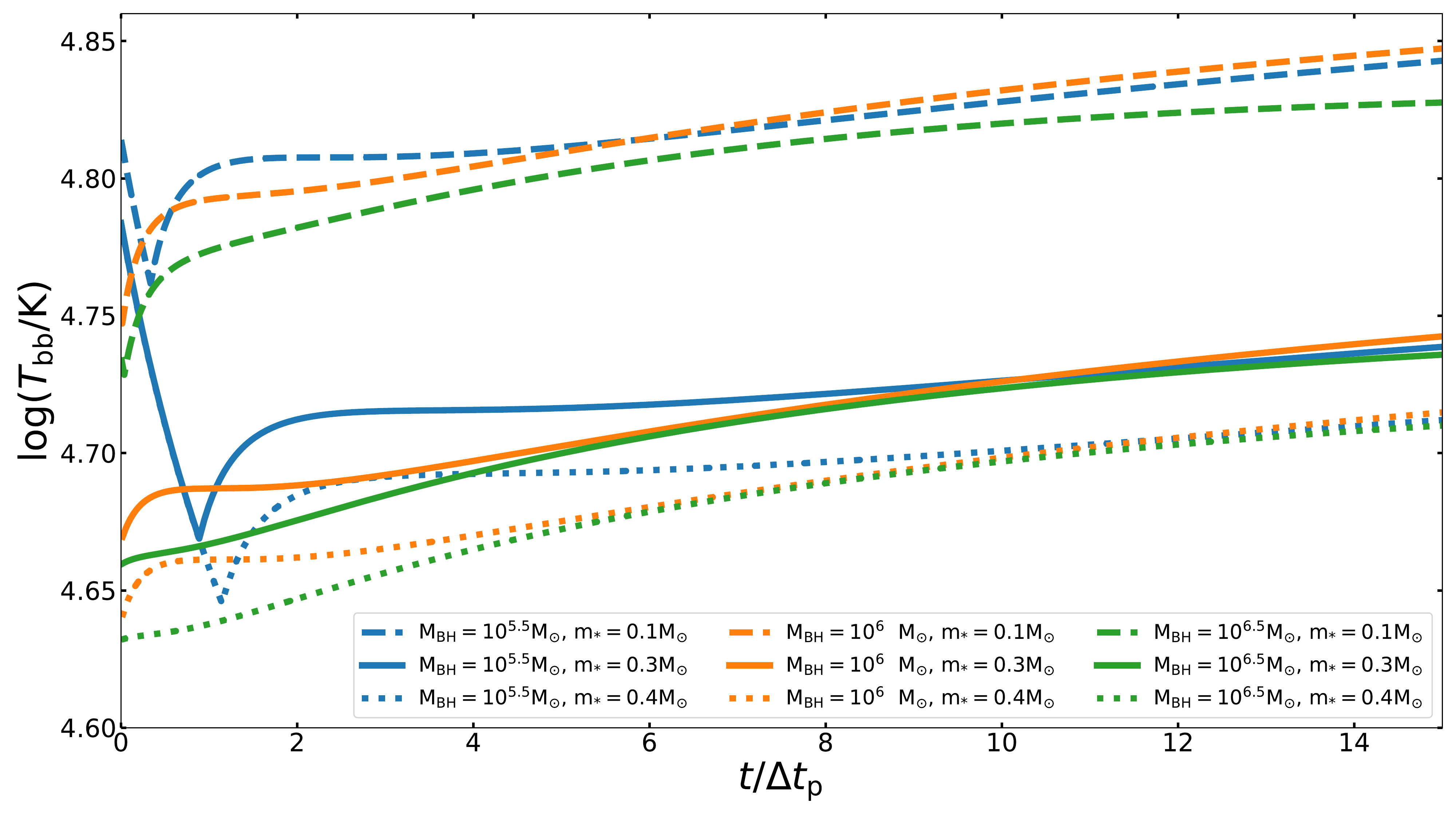} 
	\caption{Radiation blackbody temperature as a function of time for BH mass $M_{\rm 
	BH} =10^{5.5} M_\odot$, $10^6 M_\odot$, and $10^{6.5} M_\odot$, and mass of the star $m_* = 0.1$, $0.3$, 
	and $0.4$. The blackbody temperature for the viscosity parameter $\alpha_{\rm p}=0.2$ is calculated with 
	$\beta_*=1.0$, $\beta_{\rm g}=1$, $r_{\rm ms}=2 r_{\rm S}$, $r_{\rm p} = r_{\rm p*}$, $r_* = m_*^{1-\zeta}$ 
	with $\zeta=0.21$, $f_{\rm T} =1.47(0.80+0.26 M_6^{0.5})$, and $n=5/3$. Time starts from the peak of the accretion 
	rate. The radiation blackbody temperature is nearly constant with time except at about the time of the peak. For 
	low BH mass $M_{\rm BH} = 10^{5.5} M_\odot$, the radiation blackbody temperature decreases at about the 
	time of the peak to the minimum, followed by rapid increase to a nearly constant value. 
	The radiation blackbody temperature decreases with the mass of the star.
		\label{fig:tbbrp}}
\end{figure}
%

Equation~(\ref{eq:tempbb3}) shows that the elliptical accretion disk radiates with a typical effective 
blackbody temperature $T_{\rm bb}\simeq 3 \times 10^4\, {\rm K} \, (\alpha_{\rm p}/0.2)^{81/136}$.
Figure~\ref{fig:tbbt} gives the blackbody temperature as a function of accretion rate (top panel) and 
time (bottom panel) for $n=5/3$ and different viscosity parameter $\alpha_{\rm p}=0.05$, 0.1, 0.2, 0.3, and 0.5.
Equation~(\ref{eq:tempbb3}) and Figure~\ref{fig:tbbt} show that the blackbody temperature depends on
on the viscosity parameter $\alpha_{\rm p}$ and is nearly independent of the accretion rate, $T_{\rm bb} 
\propto \alpha_{-1}^{81/136} \dot{M}^{-0.081}$ except at about the time of the peak accretion rate, when the 
blackbody temperature evolves rapidly (see discussion in Sec.~\ref{sec:critaccrate}). For the reasonable range 
of the viscosity parameter \citep{van19}, the blackbody temperature of optical/UV TDEs is typically $T_{\rm bb}
\simeq 3\times 10^4 \, {\rm K}$ for $\alpha_{\rm p}\simeq 0.2$ and would be in the range of $T_{\rm bb} 
\simeq 1\times 10^4\, {\rm K}$ for $\alpha_{\rm p}=0.05$ and $8\times 10^4\, {\rm K}$ for $\alpha_{\rm p}=0.5$ 
and $m_* \simeq 0.3$. For the typical fallback rate $\dot{M} \propto t^{-5/3}$, we have $T_{\rm bb} \propto 
t^{0.135}$. If the radiation cooling in soft X-ray and EUV in the region of the $r_{\rm p} \la r \ll r_{\rm 0}$ is significant,
the polytropic index is larger than 
the adiabatic index $\gamma_{\rm ad} =5/3$. For example, if the polytropic index is $\gamma=2$, 
typical for the gas giant planets, equation~(\ref{eq:tempbb2}) shows that the radiation temperature 
would be completely independent of the accretion rate, $T_{\rm bb} \propto \dot{M}^0$.  Therefore, 
a prediction of the elliptical accretion disk model is that optical/UV TDEs with comparable X-ray radiation 
should have rather steady or even decreasing blackbody temperature with decay of the accretion rate. 
However, when the radiation cooling in soft X-rays at $r \ll r_{\rm 0}$ is significant, we cannot simply use 
equation~(\ref{eq:Qcooling}) to estimate the cooling rate $\Delta{Q}^{-}$ and instead have to integrate 
the emission of the disk surface from $r_{\rm p} \la r \la r_{\rm 0}$. 

Figure~\ref{fig:tbbrp} gives the blackbody temperature for the typical viscosity parameter $\alpha_{\rm p} = 
0.2$ as a function of time for  BH mass $M_{\rm BH} = 10^{5.5} M_\odot$, $10^6 M_\odot$, and 
$10^{6.5} M_\odot$ and  mass of the star $m_* = 0.1$, 0.3, and 0.4. Figure~\ref{fig:tbbrp} and 
equation~(\ref{eq:tempbb3}) show that the typical effective blackbody temperature decreases weakly with
the mass of the star, $T_{\rm bb}\propto  r_*^{-63/272} m_*^{-3/68} \propto m_*^{-(75-63\zeta)/272} \propto 
m_*^{-0.23}$ for $\zeta=0.21$, and is practically independent of the BH mass $T_{\rm bb}\propto M_6^{0.04}$
except around the time of the peak. At about the time of the peak, the blackbody temperature for low BH mass with 
$M_{\rm BH} \simeq 10^{5.5} M_\odot$ decreases rapidly first to a minimum and is followed by the swift increase 
to  a constant value. Our results suggest that the variations of observed blackbody temperature of optical/UV 
TDEs are mainly due to the differences of the viscosity parameters and partly to the variations of the 
orbital penetration factor among TDEs.

Equation~(\ref{eq:tempbb3}) shows that the radiation temperature $T_{\rm bb}$  is a weak function of 
pericenter radius $r_{\rm p}$ and has a much smaller dependence of radius than the typical power law 
$r^{-3/4}$ ($r_{\rm p}=r$ in the circular disk) either in the 
standard thin disk \citep{sha73} or $r^{-1/2}$ in the slim accretion disk \citep{abr88,str09}. It changes by 
up to 70\% from $r_{\rm p*} =23.545 r_{\rm S}$ to $3r_{\rm S}$ and gives  an SED of emission very close to 
a single-temperature blackbody, significantly different from the SEDs of either the standard thin or slim 
accretion disk.

The typical radiation radius $r_{\rm 0}$ given with equation~(\ref{eq:radradiuscent2}) is not directly 
measured in the literature. Model-independent effective blackbody radius is observationally obtained
by assuming that the observed bolometric luminosity $L_{\rm bol}$ is emitted by a spherical envelope 
with blackbody of single temperature $T_{\rm bb}$, 
\begin{equation}
4\pi R_{\rm bb}^2 \sigma_{\rm SB} T_{\rm bb}^4 = L_{\rm bol} .
\label{eq:effredius}
\end{equation}
In the elliptical accretion disk, the total radiation energy can be calculated with $L_{\rm bol} = \eta \dot{M} 
c^2$, where $\eta$ is the radiation efficiency. Letting $r_{\rm p} = r_{\rm ms}$, we can calculate the radiation
efficiency with equation~(\ref{eq:eng1})  \citep[see also][]{liu17,cao18,zho20},
\begin{eqnarray}
\eta &=& {e_{\rm G} \over c^2} \cr
       &\simeq& \delta {(1+\Delta_*) \over 8} \left({1+e \over 2}\right)^{-1} \left({2 r_{\rm S} \over r_{\rm ms}}\right) 
       \left[1 - {\delta(1+\Delta_*)  \over (1+e)}
       \left({r_{\rm S} \over r_{\rm ms} - r_{\rm S}}\right) \right] .
\label{eq:coneff}
\end{eqnarray}
Here we have neglected the $[(1-e)/2]^2$ terms and higher. Provided the radiation efficiency, we have the total 
luminosity
\begin{eqnarray}
L_{\rm bol} &=& \eta \dot{M} c^2 \cr
      &\simeq& \delta{(1+\Delta_*) \over 8} \left({1+e \over 2}\right)^{-1} \left({2 r_{\rm S} \over r_{\rm 
      	ms}}\right) \left[1 - {\delta(1+\Delta_*) \over (1+e)} \left({r_{\rm S} \over r_{\rm ms} - r_{\rm 
      	S}}\right) \right] \left({\dot{M} \over \dot{M}_{\rm p}}\right) \dot{M}_{\rm p}  c^2 \cr
&\simeq& 7.11 \times 10^{43} \, ({\rm erg\; s^{-1}}) \, \beta_*^{-1} \left({f_{\rm T}\over 1.56}\right)^{-4} 
        r_*^{-3/2} m_*^{7/3} M_6^{-5/6} (1+\Delta_*)  \times \cr
&& \left[{3(n-1) \over 2}\right] \left({1+e \over 2}\right)^{-1} \left({2 r_{\rm S} \over r_{\rm ms}}\right) \left[1 - 
      {\delta(1+\Delta_*) \over (1+e)} \left({r_{\rm S} \over r_{\rm ms} - r_{\rm S}}\right) \right] \left({\dot{M} \over 
    \dot{M}_{\rm p}}\right) \cr
&\simeq& 0.494 \beta_*^{-1}  \left({f_{\rm T}\over 1.56}\right)^{-4} r_*^{-3/2} m_*^{7/3} M_6^{-11/6} 
        (1+\Delta_*) \left({1+e \over 2}\right)^{-1} \times \cr
&& \left[{3(n-1) \over 2}\right] \left({2 r_{\rm S} \over r_{\rm ms}}\right) \left[1 - {\delta(1+\Delta_*) \over (1+e)} 
      \left({r_{\rm S} \over r_{\rm ms} - r_{\rm S}}\right) \right] \left({\dot{M} \over 
\dot{M}_{\rm p}}\right) L_{\rm Edd} .
\label{eq:coneff1}
\end{eqnarray}
TDEs with stellar mass $m_* \la 0.5 $ or BH mass $M_{\rm BH} \ga 10^6 M_\odot$ have sub-Eddington 
luminosities even at peak luminosity, whereas for TDEs with BHs of mass $M_{\rm BH}\la 10^5 M_\odot$, 
the expected peak luminosity is highly super-Eddington, and the light curve would have an extended plateau 
top-capped by the Eddington luminosity.

From equations~(\ref{eq:effredius}) and (\ref{eq:coneff1}), we have the effective blackbody radius   
\begin{eqnarray}
R_{\rm bb} &=& \left({5 \over 4}\right)^{1/2} \left({1+e \over 2}\right)^{-1/2} \left({\delta \over 2}\right)^{1/2}
         \left({2 r_{\rm S} \over r_{\rm ms}}\right)^{1/2} 
        \left[1 - \delta {(1+\Delta_*)  \over (1+e)} \left({r_{\rm S} \over r_{\rm ms} - r_{\rm S}}\right) \right]^{1/2} 
        \times \cr
 &&   \left(1+\Delta_*\right)^{1/2}  \left\{\sigma_{\rm SB}^{-1/2} \kappa_{\rm es}^{-1/2} r_{\rm S}^{1/2} 
        c^{3/2}\right\}  \left({\dot{M}\over \dot{M}_{\rm Edd}}\right)^{1/2}  T_{\rm bb}^{-2} ,
        \label{eq:envradius0}
\end{eqnarray}
which, together with equation~(\ref{eq:tempbb2}), gives 
\begin{eqnarray}
R_{\rm bb} &=&\left({320\over3}\right)^{-1/2} \left({45\pi \over 
	8}\right)^{-(11\gamma+5)/2(11\gamma-7)} \left({20 \over 3 \alpha_{\rm p}}\right)^{27/2 
	(11\gamma-7)}  f^{-11(\gamma-2)/2(11\gamma-7)} \times \cr
 && A_{\rm 0}^{-3(11\gamma-13)/8(11\gamma-7)}  B_{\rm 0}^{(11\gamma+5)/2(11\gamma-7)} 
      C_{\rm 0}^{-3(11\gamma-13)/4(11\gamma-7)}  C_{\rm p}^{3(11\gamma-25)/8(11\gamma-7)}       
	\times \cr
  && D_{\rm p}^{-(11\gamma+32)/2(11\gamma-7)} \left({2r_{\rm S} \over r_{\rm ms}}\right)^{1/2}  
        \left({1+e\over 2}\right)^{-3(11\gamma- 5)/4(11\gamma-7)} \left[1-\delta {(1+\Delta_*) \over 
        (1+e)} \left({r_{\rm S} \over r_{\rm ms} - r_{\rm S}}\right) \right]^{1/2} \times \cr
&&  \left(1+\Delta_*\right)^{1/2}  \left\{\left({\beta_{\rm g}^{-1} k_{\rm B} \over \mu m_{\rm 
        H}}\right)^{-33/4(11\gamma-7)} \sigma_{\rm SB}^{6/(11\gamma-7)} \kappa_{\rm 
        es}^{-3/(11\gamma-7)} r_{\rm S}^{11(2\gamma-1)/2(11\gamma-7)} c^{-3/2(11\gamma-7)} 
        \right.  \times \cr
&& \left. \kappa_{\rm 0}^{9/2(11\gamma-7)}\right\}  \left({r_{\rm p} \over r_{\rm p*}}\right)^{3(11
        \gamma-13)/2(11\gamma-7)}  \left({r_{\rm t} \over r_{\rm S}}\right)^{3(11\gamma-13)/2(11
        \gamma-7)} \beta_*^{-(22\gamma-23)/(11\gamma-7)} f_{\rm T}^{-1/2} \times \cr
&&  m_*^{1/6} M_6^{-1/6} \left(1-{r_0 \over 2a}\right)^{-(11\gamma-25)/4(11\gamma-7)} 
       \left({\dot{M}_{\rm p} \over \dot{M}_{\rm Edd}}\right)^{15/2(11\gamma-7)} 
        \left({\dot{M}\over \dot{M}_{\rm p}}\right)^{15/2(11\gamma-7)} .
\label{eq:envradius1}
\end{eqnarray}
For the polytropic index $\gamma=5/3$, the effective blackbody radius is
\begin{eqnarray}
R_{\rm bb} &=&\left({320\over3}\right)^{-1/2} \left({45\pi \over 8}\right)^{-35/34} \left({20 \over 
       3 \alpha_{\rm p}}\right)^{81/68} f^{11/68} \times \cr
&& A_{\rm 0}^{-3/17} B_{\rm 0}^{35/34} C_{\rm 0}^{-6/17} C_{\rm p}^{-15/68}  D_{\rm p}^{-151/68} 	\times \cr
&& \left({2r_{\rm S} \over r_{\rm ms}}\right)^{1/2} \left({1+e\over
	 2}\right)^{-15/17} \left[1-\delta {(1+\Delta_*) \over (1+e)} \left({r_{\rm 
		S} \over r_{\rm ms} - r_{\rm S}}\right)\right]^{1/2} \times \cr
&&	 \left(1+\Delta_*\right)^{1/2} \left\{\left({\beta_{\rm g}^{-1} k_{\rm B} \over \mu m_{\rm H}}\right)^{-99/136}  
     \sigma_{\rm SB}^{9/17} \kappa_{\rm es}^{-9/34} r_{\rm S}^{77/68} c^{-9/68} \kappa_{\rm 0}^{27/68}\right\} 
    \left({r_{\rm p} \over r_{\rm p*}}\right)^{12/17}  \times \cr
&&   \left({r_{\rm t} \over r_{\rm S}}\right)^{12/17} \beta_*^{-41/34} f_{\rm T}^{-1/2} m_*^{1/6} M_6^{-1/6} 
     \left(1-{r_{\rm 0}\over 2a}\right)^{5/34} \left({\dot{M}_{\rm p}\over \dot{M}_{\rm Edd}}\right)^{45/68} 
      \left({\dot{M}\over \dot{M}_{\rm p}}\right)^{45/68} \cr
&\simeq& 1.87 \times 10^{15}\, ({\rm cm})\, \alpha_{\rm -1}^{-81/68} f^{11/68} 
       A_{\rm 0}^{-3/17} B_{\rm 0}^{35/34} C_{\rm 0}^{-6/17} C_{\rm p}^{-15/68}  D_{\rm p}^{-151/68} 
		 \times \cr
&&   \left({2 r_{\rm S} \over r_{\rm ms}}\right)^{1/2} \left({1+e\over 2}\right)^{-15/17} \left[1-\delta {(1+
         \Delta_*) \over (1+e)} \left({r_{\rm S} \over r_{\rm ms} - r_{\rm S}}\right)\right]^{1/2} \left(1+
         \Delta_*\right)^{1/2} \times \cr
&& \beta_{\rm g}^{99/136} \beta_*^{-41/34} f_{\rm T}^{-121/68}   r_*^{-39/136} m_*^{64/51}  M_6^{-203/408} 
          \times \cr
&&      \left[{3(n-1) \over 2}\right]^{45/68} \left(1-{r_{\rm 0}\over 2a}\right)^{5/34}
        \left({r_{\rm p} \over r_{\rm p*}}\right)^{12/17} \left({\dot{M}\over \dot{M}_{\rm p}}\right)^{45/68} ,
\label{eq:envradius53}
\end{eqnarray}
where the radiation radius $r_0$ is given with equation~(\ref{eq:radradiuscent2}). Equation~(\ref{eq:envradius53}) 
suggests that the effective blackbody radius of optical/UV TDEs increases with accretion rate and decreases with 
time.

Equation~(\ref{eq:envradius53}) shows that the effective blackbody radius significantly depends on the 
accretion rate, the BH mass, the mass and orbital penetration factor of the star, and the viscosity parameter. 
The effective blackbody radius of TDEs depends nearly linearly on the mass of the star, $R_{\rm bb}\propto 
r_*^{-39/136} m_*^{64/51} \propto m_*^{(395+117\zeta)/408} \propto m_*^{1.03}$ for $\zeta=0.21$.

\begin{figure}
	\centering 
			\includegraphics[width=1\textwidth]{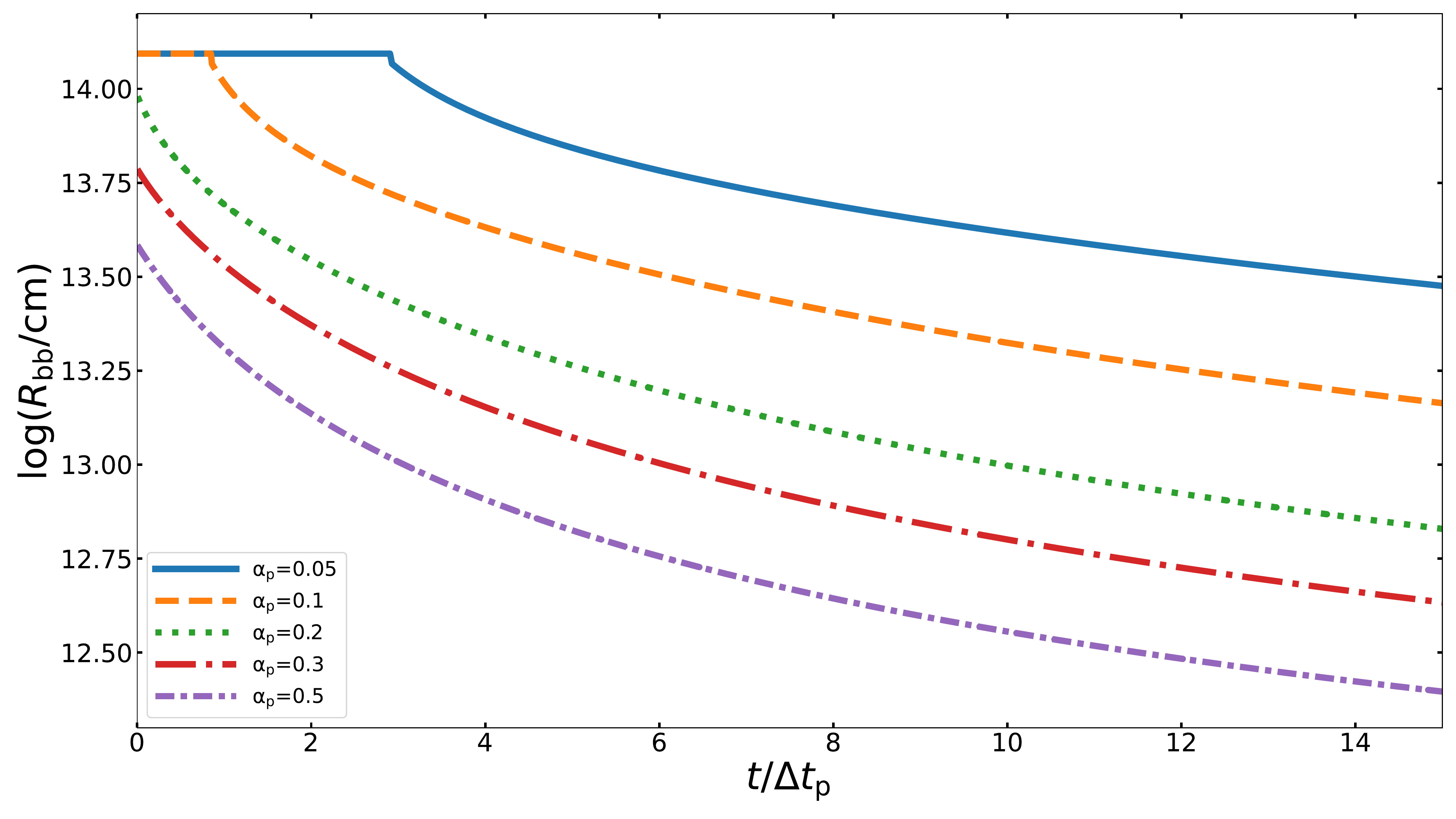} 
	\caption{Effective blackbody radius as a function of time for different viscosity parameter $\alpha_{\rm p}$. 
	The effective blackbody radius is calculated with $M_{\rm BH} = 10^6 M_\odot$, $\beta_* = 1$, 
	$r_{\rm ms} =2 r_{\rm S}$, $\beta_{\rm g} = 1$,  $r_{\rm p} = r_{\rm p*}$, $m_*=0.3$, $r_* = m_*^{1-\zeta}$ 
	with $\zeta=0.21$, $f_{\rm T} =1.56$, and $n=5/3$. The time starts at the peak accretion 
	rate. 
		\label{fig:Rbbtvis}}
\end{figure}
%
\begin{figure}
	\centering 
			\includegraphics[width=1\textwidth]{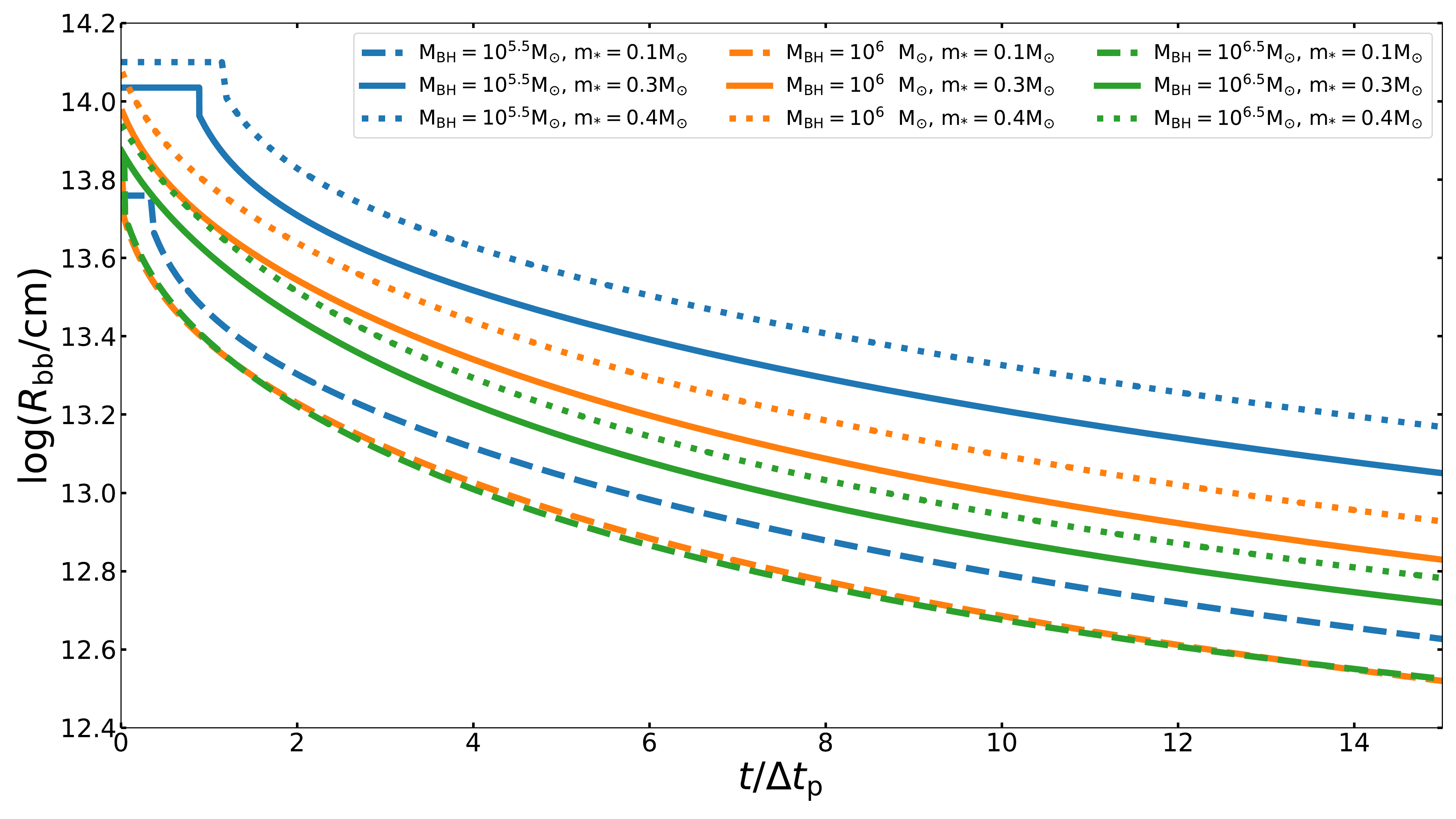} 
	\caption{Effective blackbody radius as a function of time for different BH and stellar masses. 
	The effective blackbody radius is calculated with $\alpha_{\rm p}=0.2$, $\beta_* = 1$, $r_{\rm ms} =2 
	r_{\rm S}$, $\beta_{\rm g} = 1$,  $r_{\rm p} = r_{\rm p*}$, $r_* = m_*^{1-\zeta}$ with $\zeta=0.21$, $f_{\rm 
	T} =1.47 (0.80+0.26 M_6^{0.5})$, and $n=5/3$. The time starts at the peak accretion rate. 
		\label{fig:Rbbt}}
\end{figure}
%

Figure~\ref{fig:Rbbtvis} gives the effective blackbody radius as a function of time for different viscosity parameter
$\alpha_{\rm p}$, and Figure~\ref{fig:Rbbt} shows the variations of the effective blackbody radius with time for
different BH masses $M_{\rm BH} = 10^{5.5} M_\odot$, $10^6 M_\odot$, and $10^{6.5} M_\odot$ and the mass 
of the star $m_*= 
0.1$, 0.3, and 0.4. The effective blackbody radius significantly depends on the  accretion rate $R_{\rm bb}
\propto (\dot{M}/\dot{M}_{\rm p})^{0.662}$ and decreases with time $R_{\rm bb}\propto t^{-75/68} \propto
t^{-1.10}$ for $n=5/3$, which is very different from the expectation of constant radius of the circular accretion 
disk or shock model for TDEs. The power law of index $0.662$, because of the slight dependence of temperature 
on accretion rate, is higher than the index $0.5$, which is expected with constant blackbody temperature 
$T_{\rm bb}$. Because accretion rate given in equation~(\ref{eq:accr}) depends on the power-law index 
$n$ and the structure and age of the star, we would suggest to observe the effective blackbody radius 
$R_{\rm bb}$ as a function of both accretion rate $\dot{M}$ (or luminosity) and time to measure the 
power-law index of $n$, which depends on the age and structure of a star.

\subsection{Bright TDEs with constant radiation radius of apocenter}\label{sec:critaccrate}

Equation~(\ref{eq:radradiuscent2}) shows that the radiation radius $r_{\rm 0}$ varies with accretion rate 
$\dot{M}$ and should increase with time before peak brightness.  Because the radiation radius cannot 
be larger than the apocenter radius $(1+e) a_{\rm d}$, we have a critical accretion rate $\dot{M}_{\rm 
cr}$. For $\dot{M} <\dot{M}_{\rm cr}$ the radiation radius $r_{\rm 0}$ varies with accretion rate and is given 
by equation~(\ref{eq:radradiuscent2}), while for $\dot{M} \geq \dot{M}_{\rm cr}$ the radiation radius $r_{\rm 
0}$ does not change with accretion rate and remains constant with $r_{\rm 0} = (1+e) a_{\rm d}$. 
Letting $r_{\rm 0} = (1+e) a_{\rm d}$, $a=a_{\rm d}$, and $r_{\rm p} = r_{\rm p*}$ and from 
equation~(\ref{eq:radradiuscent}), we obtain 
\begin{eqnarray}
{1+e \over 1-e} &\simeq & A_{\rm 0}^{-(11\gamma-13)/2(11\gamma-7)} B_{\rm 0}^{8/(11\gamma-7)}
     C_{\rm 0}^{-(11\gamma-13)/(11\gamma-7)} C_{\rm p}^{(11\gamma-25)/2(11\gamma-7)} 
                D_{\rm p}^{-26/(11\gamma-7)} \times \cr
  && \left({45\pi \over 8 }f \right)^{-8/(11\gamma-
            7)} \left({20 \over 3 \alpha_{\rm p}}f\right)^{18/(11\gamma-7)} \left({1+e\over 
            2}\right)^{-2/(11\gamma-7)} \times \cr
   &&\left\{\left({\beta_{\rm g}^{-1} k_{\rm B} \over \mu m_{\rm H}}\right)^{-11/(11\gamma-7)}  
              \sigma_{\rm SB}^{8/(11\gamma-7)} \kappa_{\rm es}^{-4/(11\gamma-7)} 
              r_{\rm S}^{2/(11\gamma-7)} 
              c^{-2/(11\gamma-7)} \kappa_{\rm 0}^{6/(11\gamma-7)}\right\}\cr
     &&\beta_*^{12/(11\gamma-7)} 
              \left({r_{\rm t} \over r_{\rm S}}\right)^{-12/(11\gamma-7)} \left({1- e \over 
              2}\right)^{6/(11\gamma-7)} \times \cr
 &&\left({\dot{M}_{\rm p}\over \dot{M}_{\rm Edd}}\right)^{10/(11\gamma-7)}   \left({\dot{M}_{\rm 
     cr} \over \dot{M}_{\rm p}}\right)^{10/(11\gamma-7)} ,
\label{eq:criticalaccrate}
\end{eqnarray}
where we have used $1-(r_{\rm 0}/2a) = (1-e)/2$. From equation~(\ref{eq:criticalaccrate}), we have
\begin{eqnarray}
{\dot{M}_{\rm cr}\over \dot{M}_{\rm p}}&\simeq&  \left({\dot{M}_{\rm p}\over \dot{M}_{\rm Edd}}
        \right)^{-1} \left({1+e \over 2}\right)^{(11\gamma-3)/5} \left[{2\delta (1+\Delta_*) \over 
        4}\right]^{-(11\gamma-1)/10} \times \cr
 &&A_{\rm 0}^{(11\gamma-13)/20}  B_{\rm 0}^{-4/5} C_{\rm 0}^{(11\gamma-13)/10} 
        C_{\rm p}^{-(11\gamma-25)/20} D_{\rm p}^{13/5} \times \cr
&& \left({45\pi \over 8 } f\right)^{4/5} \left({20 \over 3 
          \alpha_{\rm p}}f\right)^{-9/5} \left\{\left({\beta_{\rm g}^{-1} k_{\rm B} \over \mu m_{\rm H}}\right)^{11/10}  
              \sigma_{\rm SB}^{-4/5} \kappa_{\rm es}^{2/5} 
              r_{\rm S}^{-1/5} c^{1/5} \kappa_{\rm 0}^{-3/5}\right\} \times \cr
              && \beta_*^{-6/5}  \left({r_{\rm t} \over r_{\rm S}}\right)^{6/5} ,
\label{eq:criticalaccrate1}
\end{eqnarray}
where we have used $1-e^2 = 2\delta (1+\Delta_*)$.  For $\gamma=5/3$, equation~(\ref{eq:criticalaccrate1}) 
gives
\begin{eqnarray}
{\dot{M}_{\rm cr}\over \dot{M}_{\rm p}}&\simeq& \left({\dot{M}_{\rm p}\over \dot{M}_{\rm Edd}}
        \right)^{-1}  \left({1+e \over 2}\right)^{46/15} \left[{2\delta (1+\Delta_*) \over 
        4}\right]^{-26/15} \times \cr
                &&A_{\rm 0}^{4/15} B_{\rm 0}^{-4/5}  C_{\rm 0}^{8/15} C_{\rm p}^{1/3}  D_{\rm p}^{13/5} 
                \left({45\pi \over 8 } f\right)^{4/5} \left({20 \over 3 \alpha_{\rm p}}f\right)^{-9/5} \times \cr
&& \left\{\left({\beta_{\rm g}^{-1} k_{\rm B} \over \mu m_{\rm H}}\right)^{11/10}  
     \sigma_{\rm SB}^{-4/5} \kappa_{\rm es}^{2/5} 
      r_{\rm S}^{-1/5} c^{1/5} \kappa_{\rm 0}^{-3/5}\right\} \beta_*^{-6/5}  \left({r_{\rm t} \over
           r_{\rm S}}\right)^{6/5} \cr
&\simeq& 0.822  A_{\rm 0}^{4/15} B_{\rm 0}^{-4/5} C_{\rm 0}^{8/15} C_{\rm p}^{1/3} 
            D_{\rm p}^{13/5} f^{-1} \alpha_{\rm -1}^{9/5} \times \cr
 &&  \beta_{\rm g}^{-11/10} \beta_*^{8/15} \left({f_{\rm T} \over 1.56}\right)^{89/15} 
         \left(1+\Delta_*\right)^{-26/15}  \left({1+e \over 2}\right)^{46/15} \times \cr
&&   \left[{3(n-1) \over 2}\right]^{-1} \left({m_*\over 0.3}\right)^{-(25+243\zeta)/90} M_6^{97/90}
   \label{eq:criticalaccrate53}
\end{eqnarray}
with $\zeta=0.21$. 
	
Equation~(\ref{eq:criticalaccrate53}) shows that the peak accretion rate of optical/UV TDEs with typical stellar
mass $M_{\rm BH} \ga 10^6 M_\odot$ and viscosity parameter $\alpha_{\rm p} \ga 0.2$ is less than the critical 
accretion rate. For TDEs with $\dot{M}_{\rm cr} > \dot{M}_{\rm p}$, the radiation radius $r_0$ is given with 
equation~(\ref{eq:radradiuscent2}) and the effective blackbody radius $R_{\rm bb}$ is calculated with 
equation~(\ref{eq:envradius53}). Our elliptical accretion disk model suggests that both the radiation radius 
$r_0$ and effective blackbody radius $R_{\rm bb}$ should closely follow the change of accretion rate or the
luminosity with some possible delay of peak radius relative to the peak accretion rate because of the term 
$\left[1-(r_0/2a)\right]$. The blackbody temperature is given with equation~(\ref{eq:tempbb3}) and would 
change with accretion rate near the peak of the accretion rate or at the peak time $t\sim 0$, if the peak accretion
rate is about the critical accretion rate $\dot{M}_{\rm p} \sim \dot{M}_{\rm cr}$, as shown in 
Figures~\ref{fig:tbbrp} and \ref{fig:tbbt}.

Equation~(\ref{eq:criticalaccrate53}) suggests that for TDEs with BH mass $M_{\rm BH} \la 10^6 M_\odot$ 
or viscosity parameter $\alpha _{\rm p} \la 0.2$ the peak accretion rate may be larger than the critical accretion 
rate. For $\dot{M} \geq \dot{M}_{\rm cr}$, the total luminosity of TDEs decreases with accretion rate and the 
radiation radius remains constant with $r_{\rm 0} = (1+e) a_{\rm d}$. From Equation~(\ref{eq:tempbb1}), we 
have 
\begin{eqnarray}
T_{\rm bb} & \simeq &  \left(5 \pi f\right)^{1/4}  \left({1-e^2 \over 4}\right)^{1/2} 
      \left({1+e \over 2}\right)^{-3/4} D_{\rm p}^{1/4} \left\{\kappa_{\rm es}^{-1/4} \sigma_{\rm SB}^{-1/4} r_{\rm 
        S}^{-1/4} c^{3/4}\right\} \times\cr
&&    \beta_*^{3/4} \left({r_{\rm t} \over r_{\rm S}}\right)^{-3/4} \left({\dot{M}_{\rm p}\over 
       \dot{M}_{\rm Edd}}\right)^{1/4} \left({r_{\rm p} \over r_{\rm p*}}\right)^{-3/4} \left({\dot{M}\over 
       \dot{M}_{\rm p}}\right)^{1/4} \cr
& \simeq& 3.69 \times 10^4 \, ({\rm K})\,  f^{1/4} D_{\rm p}^{1/4}  
     \beta_*^{1/4} \left({f_{\rm T} \over 1.56}\right)^{-2} r_*^{-9/8} m_*^{11/12}   M_6^{-7/24}  \times \cr
&&  \left[{3(n-1) \over 2}\right]^{1/4} \left(1+\Delta_*\right)^{1/2} \left({1+e \over 2}\right)^{-3/4}   \left({r_{\rm p} 
      	\over r_{\rm p*}}\right)^{-3/4} \left({\dot{M}\over \dot{M}_{\rm p}}\right)^{1/4} 
\label{eq:tempbbcr}
\end{eqnarray}
for $r_{\rm mb} \leq r_{\rm p} \leq r_{\rm p*}$, where we have used $r_{\rm p} = (1-e) a$, $1-e^2 = 2 \delta 
(1+ \Delta_*)$, and $B_0 = 3/2$. The elliptical accretion disk with accretion rate $\dot{M} \ga \dot{M}_{\rm 
cr}$ has a distribution of effective temperature with pericenter radius $r_{\rm p}^{-3/4}$ and increases with 
accretion rate $\dot{M}^{1/4}$, which are the same as those of a standard thin accretion disk \citep{fra02} 
but have much lower peak value. Equation~(\ref{eq:tempbbcr}) 
shows that the elliptical accretion disk with accretion rate $\dot{M}_{\rm p} \ga \dot{M}_{\rm cr}$ has a 
distribution of temperature from $T_{\rm bb} \simeq 3.7\times 10^4 \, {\rm K}$ at $r_{\rm p} = 23.545 r_{\rm S}$ 
to the maximum temperature $T_{\rm bb} \simeq 8.2 \times 10^4 \, {\rm K}$. 
 
Our results suggest that for optical/UV TDEs during accretion rate $\dot{M}_{\rm p} \ga \dot{M}_{\rm cr}$ the 
effective blackbody radius would remain constant, as is shown in Figure~\ref{fig:Rbbtvis} for viscosity parameter 
$\alpha_{\rm p} =0.05$ and 0.1 and in Figure~\ref{fig:Rbbt} for BH mass $M_{\rm BH}= 10^{5.5} M_\odot$.  The
blackbody temperature changes with accretion rate as a power law $T_{\rm bb} \propto \dot{M}^{1/4}$ and
would increase with time before peak brightness and decrease afterward until $\dot{M} < \dot{M}_{\rm 
cr}$, as is shown in Figure~\ref{fig:tbbt} for viscosity parameter $\alpha_{\rm p}= 0.05$ and $0.1$
and in Figure~\ref{fig:tbbrp} for less massive SMBH mass $M_{\rm BH} = 10^{5.5} M_\odot$. When the accretion rate
$\dot{M}$ decreases to smaller than the critical rate $\dot{M}_{\rm cr}$ at late times, the SED becomes a 
blackbody spectrum of nearly single and constant temperature as given by equation~(\ref{eq:tempbb3}), 
and the blackbody radius decreases with time as suggested by equation~(\ref{eq:envradius53}).


\section{Comparisons with the observations of optical/UV TDEs}\label{sec:compobs}

In this section, we compare the expectations of the elliptical accretion disk model with the observations 
of optical/UV TDEs. The low radiation efficiency of an elliptical accretion disk and its implications for the observations 
of the peak luminosity, total accreted mass, and the measurements of BH masses of TDEs have been discussed in 
recent work \citep{zho20}.

\subsection{Blackbody SEDs of single temperature of all optical/UV TDEs}

One of the puzzling observations of optical/UV TDEs is that the SEDs can be well fitted with 
blackbody of nearly single temperature and the blackbody temperature ranges from $1\times 10^4 \, 
{\rm K}$ to $6\times 10^4 \, {\rm K}$ \citep{gez12,hol14,wev17,wev19,van20}. The effective blackbody 
temperature does not correlate with the estimated BH masses of optical/UV TDEs \citep{wev17,wev19}. 

Equation~(\ref{eq:tempbb3}) shows that the temperature $T_{\rm bb}$ depends only weakly on the pericenter
radius, $T_{\rm bb} \propto r_{\rm p}^{-6/17} \propto r_{\rm p}^{-0.35}$ with a power-law index much 
smaller than the index $0.75$ of the standard thin or slim accretion disk \citep{abr88,fra02}. For a typical
tidal disruption of optical/UV TDEs by SMBHs of mass $10^6 M_\odot$ and penetration factor $\beta_* \simeq 1$,
the effective temperature $T_{\rm bb}$ increases only by about 70\% (1.7 times), when pericenter radius $r_{\rm p}$ 
decreases from the outer boundary $r_{\rm p}\simeq 23.545 r_{\rm S}$ to $r_{\rm p} \simeq 3 r_{\rm S}$. 
The effective surface temperature of the standard thin accretion disk increases by about 370\% (4.7 times) 
for the same range of radius, neglecting the effect of the inner boundary condition. The small variation of 
effective blackbody temperature of the elliptical accretion disk would radiate with a blackbody spectrum of 
nearly single temperature.  

Equation~(\ref{eq:tempbb3}) suggests that the blackbody temperature of the elliptical accretion disk 
is nearly independent of the BH mass $T_{\rm bb}\propto M_6^{11/272} \propto M_6^{0.040}$, which is 
well consistent with the observations of optical/UV TDEs \citep{wev17,wev19}. The blackbody temperature 
weakly depends on the orbital penetration factor and the mass of a star $T_{\rm bb} \propto \beta_*^{6/17} 
r_*^{-63/272} m_*^{-3/68} \propto \beta_*^{0.35} m_*^{-(75-63\zeta)/272} \propto \beta_*^{0.35}
m_*^{-0.23}$ for $\zeta=0.21$ but varies with the viscosity parameter $T_{\rm bb} \propto \alpha_{\rm 
p}^{81/136} \propto \alpha_{\rm p}^{0.596}$. The recent observations of the disk-dominated
late-time UV luminosity of optical/UV TDEs suggest that the disk viscosity parameter is roughly between
$0.07$ and $0.6$ \citep{van19}. The estimates of the viscosity parameter are based on a circular disk 
model of radial size $2 r_{\rm p*}$  \citep{van19}, and the viscosity parameter of the 
elliptical disk model is for the viscous pericenter region of the elliptical disk of the radial size of about $r_{\rm 
p*}$ and azimuthal span $\sim \pi$. The inferred $\alpha$ values cannot be exactly applicable, but it is 
reasonable to expect that they are suitable to the elliptical disk model within orders of magnitude and that 
we have $0.01\la \alpha_{\rm p} \la 1$ with typical value $\alpha_{\rm p}\sim 0.2$. For the 
range of the viscosity parameters $0.05\la \alpha_{\rm p} \la 0.5$, the blackbody temperature $T_{\rm bb}$ 
ranges from $1 \times 10^4\, {\rm K}$ to $8 \times 10^4\, {\rm K}$, well consistent with the observations. 

Because the effective blackbody radius $R_{\rm bb}$ also depends on the viscosity parameter 
$\alpha_{\rm p}$, the elliptical accretion disk model predicts a strong correlation between the effective
blackbody temperature $T_{\rm bb}$ and blackbody radius $R_{\rm bb}$, which will be 
discussed in Section~\ref{subsec:corptpr}.

\subsection{Time (in)dependence of blackbody temperature}\label{sec:obs_cons_temp}

It is well known that the blackbody temperature of optical/UV TDEs changes little with time 
\citep{gez12,gez17,hol14,hol19,van19,van20,hin20}. Table~6 of \citet{van20} gave the measurements of 
the blackbody temperature and its variations with time (${\mathrm{d}{T}_{\rm bb}/\mathrm{d}{t}}$) of 17 
optical/UV TDEs. The measurements of ${\mathrm{d}{T}_{\rm bb}/\mathrm{d}{t}}$ have a very large scatter 
and ranges from $-0.85\times 10^2 \, {\rm K\; day^{-1}}$ to $1.95\times 10^2 \, {\rm K\; day^{-1}}$
with an average $\langle {\mathrm{d}{T}_{\rm bb}/\mathrm{d}{t}}\rangle_{\rm ob} \sim 0.47\times 10^2 \, 
{\rm K\; day^{-1}}$. 

Equation~(\ref{eq:tempbb2}) gives the variation of the blackbody temperature with the accretion rate
\begin{equation}
T_{\rm bb} \propto \left[{3(n-1) \over 2}\right]^{11(\gamma -2)/4(11\gamma-7)}
        \left(1-{r_{\rm 0}\over 2a}\right)^{(11\gamma-25)/8(11\gamma-7)} \left({\dot{M}
        \over \dot{M}_{\rm p}}\right)^{11(\gamma -2)/4(11\gamma-7)} ,
\end{equation}
where the radiation radius $r_{\rm 0}$ given with equation~(\ref{eq:radradiuscent1}) changes with accretion 
rate. The term $\left(1-{r_{\rm 0}\over 2a}\right)$ is important when $\dot{M} \sim \dot{M}_{\rm cr}$ and 
$r_{\rm 0} \sim (1+e) a_{\rm d}$. From equation~(\ref{eq:accr}), we have 
\begin{eqnarray}
T_{\rm bb} &\propto& \left[{3(n-1) \over 2}\right]^{11(\gamma -2)/4(11\gamma-7)} 
 \left(1-{r_{\rm 0}\over 2a}\right)^{(11\gamma-25)/8(11\gamma-7)} \left({t+\Delta{t}_{\rm 
       p} \over \Delta{t}_{\rm p}}\right)^{-11n(\gamma-2)/4(11\gamma-7)}  \cr
       &\propto& \left(1-{r_{\rm 0}\over 2a}\right)^{-5/68} \left({t+ \Delta{t}_{\rm p} \over \Delta{t}_{\rm 
       p}}\right)^{0.13} \qquad {\rm for\, } n=5/3  , \cr
       &\propto& \left(1-{r_{\rm 0}\over 2a}\right)^{-5/68} \left({t+ \Delta{t}_{\rm p} \over \Delta{t}_{\rm 
       p}}\right)^{0.18} \qquad {\rm for\, } n=9/4  .
\label{eq:dlnTbbdlnt}
\end{eqnarray}
To obtain the equation~(\ref{eq:dlnTbbdlnt}), we have adopted the typical polytropic index $\gamma=5/3$.   
Because $r_{\rm 0}$ decreases with time, the blackbody temperature decreases with time for $\dot{M} 
\la \dot{M}_{\rm cr}$ and then increases slowly with time at later. The expected change of the blackbody 
temperature at late time is 
\begin{eqnarray}
{\mathrm{d}{T}_{\rm bb} \over \mathrm{d}{t}} &=& -n{11(\gamma-2)\over 4(11\gamma-7)} \left({T_{\rm 
      bb} \over \Delta{t}_{\rm p}}\right) \left(1+{t_{\rm f}\over \Delta{t}_{\rm p}}\right)^{-1} \cr
	    &\simeq&0.652 \times 10^2\, ({\rm K\; day^{-1}})\, 
    \alpha_{-1}^{81/136} \beta_{\rm g}^{-99/272} \beta_*^{6/17} f_{\rm T}^{-423/136} 
     r_*^{-471/272} m_*^{65/68} \times \cr
     && M_6^{-125/272} \left(1+{t_{\rm f} \over \Delta{t}_{\rm p}}\right)^{-353/408}     
\label{eq:dTdtmdot}
\end{eqnarray}
for $\gamma=5/3$ and $n=5/3$, where $t_{\rm f}$ is the time at the end of the observational campaign. 
Equation~(\ref{eq:dTdtmdot}) shows that $\mathrm{d}{T}_{\rm bb}/\mathrm{d}{t}$ depends on the 
indices $\gamma$ and $n$, the masses of the SMBH and the star, the viscosity parameter $\alpha_{\rm p}$, 
and the duration of the observational campaign. Because of the differences of the parameters $M_{\rm BH}$, 
$M_*$, $\alpha_{\rm p}$,  and the ratio of the observational time $t_{\rm f}$ to $\Delta{t}_{\rm p}$
among TDEs, a large scatter of the measurements of the change rate of blackbody temperature is 
expected. Therefore, we would suggest to measure 
\begin{equation}
{\mathrm{d}(\ln{T}_{\rm bb}) \over \mathrm{d}[\ln(t+\Delta{t}_{\rm p})]} \simeq -n {11(\gamma-2)
    \over 4(11\gamma-7)} ,
\end{equation}
which depends only on the polytropic index $\gamma$ and the power-law index of fallback rate $n$.

To compare the expectations of the elliptical accretion disk and the observations of optical/UV TDEs 
in \citet{van20}, we need $t_{\rm f}/\Delta{t}_{\rm p}$. To obtain $t_{\rm f}/\Delta{t}_{\rm 
p}$, we use their fitting results of the $\Delta{t}_{\rm p}$ in Table~6 of \citet{van20} for $n=5/3$. From 
their figure~5, we have the average $\langle t_{\rm f}/\Delta{t}_{\rm p}\rangle \sim 1.03$. 
From equation~(\ref{eq:dTdtmdot}), we have the model expectation ${\mathrm{d}{T}_{\rm bb}/
\mathrm{d}{t}} \sim 0.35 \times 10^2 \, ({\rm K\; day^{-1}}) \, \alpha_{-1}^{81/136} f_{\rm 
T}^{-423/136} m_*^{-0.412} M_6^{-125/272}  \sim 0.32
\times 10^2 (\alpha/0.3)^{81/136} \, {\rm K\; day^{-1}}$ for $\zeta=0.21$, $\beta_{\rm g}\simeq 1$,
$\beta_*\simeq 1$, $M_6=1$, $f_{\rm T} = 1.5$, and $m_* \simeq 0.3$. To compare 
the model expectations with the average of observations, we adopt the typical mass of a star, $ m_* 
\simeq 0.3$ for typical initial mass function (IMF). Taking into account the large scatters of the
observations, we conclude that the model expectation of the decay rate ${\mathrm{d}{T}_{\rm bb}/ 
\mathrm{d}{t}} \sim 0.32 \times 10^2 (\alpha/0.3)^{81/136} \, {\rm K\; day^{-1}}$ is consistent with 
the observations $\langle {\mathrm{d}{T}_{\rm bb}/\mathrm{d}{t}}\rangle_{\rm ob} \sim 0.47\times 
10^2 \, {\rm K\; day^{-1}}$. 

We have adopted the adiabatic index $\gamma = 5/3$ as the fiducial value, because the gradient of the 
temperature in the $z$-direction is expected to be ${\delta{T} \over T} \sim 0$ at $r_{\rm p}$ owing to 
the strong compressing shocks near pericenter, and the emission in the regions $r\ll r_{\rm 0}$ is 
negligible. Because the radiation at $r \ll r_{\rm 0}$ is mainly in soft X-rays, no significant emission 
is expected for polytropic process with $\gamma=5/3$. If the radiation cooling in soft X-rays in the region
of the ellipse $r_{\rm p} \la r \ll r_{\rm 0}$ is significant, the polytropic index $\gamma$ would be larger
than the adiabatic index $\gamma =5/3$. Equation~(\ref{eq:dTdtmdot}) shows that a larger polytropic 
index $\gamma$ results in a smaller increase of the blackbody temperature with time. If the radiation 
cooling in soft X-rays is comparable to the optical/UV luminosity and $\gamma \simeq 2$, we would have 
a constant blackbody temperature with ${\mathrm{d}{T}_{\rm bb}/\mathrm{d}{t}} = 0$, while for $\gamma>2$
the blackbody temperature would decrease with time, ${\mathrm{d}{T}_{\rm bb}/\mathrm{d}{t}} < 0$. The 
elliptical accretion disk model predicts that optical/UV TDEs with significant X-ray radiation would have
constant or even decreasing blackbody temperature with time. The X-ray-bright ($L_{\rm X}\sim L_{\rm opt}$)
optical TDEs ASASSN-14li, with a rather constant temperature with $\mathrm{d}T_{\rm bb}/\mathrm{d}t \simeq 0$
\citep{hol16a}, and AT2019ehz, with a decaying temperature with $\mathrm{d}T_{\rm bb}/\mathrm{d}t \simeq
- 0.24\times 10^2 \, {\rm K\; day^{-1}}$ \citep{van20}, are consistent with the expectation. TDE AT2019dsg 
is the first TDE candidate associated with a neutrino event source and is detected in X-ray with a ratio of the X-ray 
to optical luminosities $L_{\rm x}/L_{\rm opt} \simeq 0.1$ \citep{ste20}. The source has a moderate relativistic jet. 
The change rate of the blackbody temperature is $\mathrm{d}T_{\rm bb}/
\mathrm{d}t \simeq 0.24\times 10^2 \, {\rm K\; day^{-1}}$ \citep{van20}, consistent with the elliptical accretion
disk model for $\gamma=5/3$.

\subsection{Large and evolving blackbody radius}\label{sec:obs_BB_rad}

The observations of optical/UV TDEs \citep[e.g.][]{hol14,hol19,lel19,gom20,hin20,sho20,van20} show that the 
blackbody radii generally follow the luminosity to increase before peak brightness and reach a maximum 
near or soon after the peak brightness. The maximum of the effective blackbody radius is in the range 
$10^{14.18}\, {\rm cm} \la R_{\rm bb} \la 10^{15.47}\, {\rm cm}$ \citep{wev19,van20}. After the peak, 
the effective blackbody radii generally decrease with the decay of luminosity.  

\begin{figure}
	\centering 
	\includegraphics[width=1\textwidth]{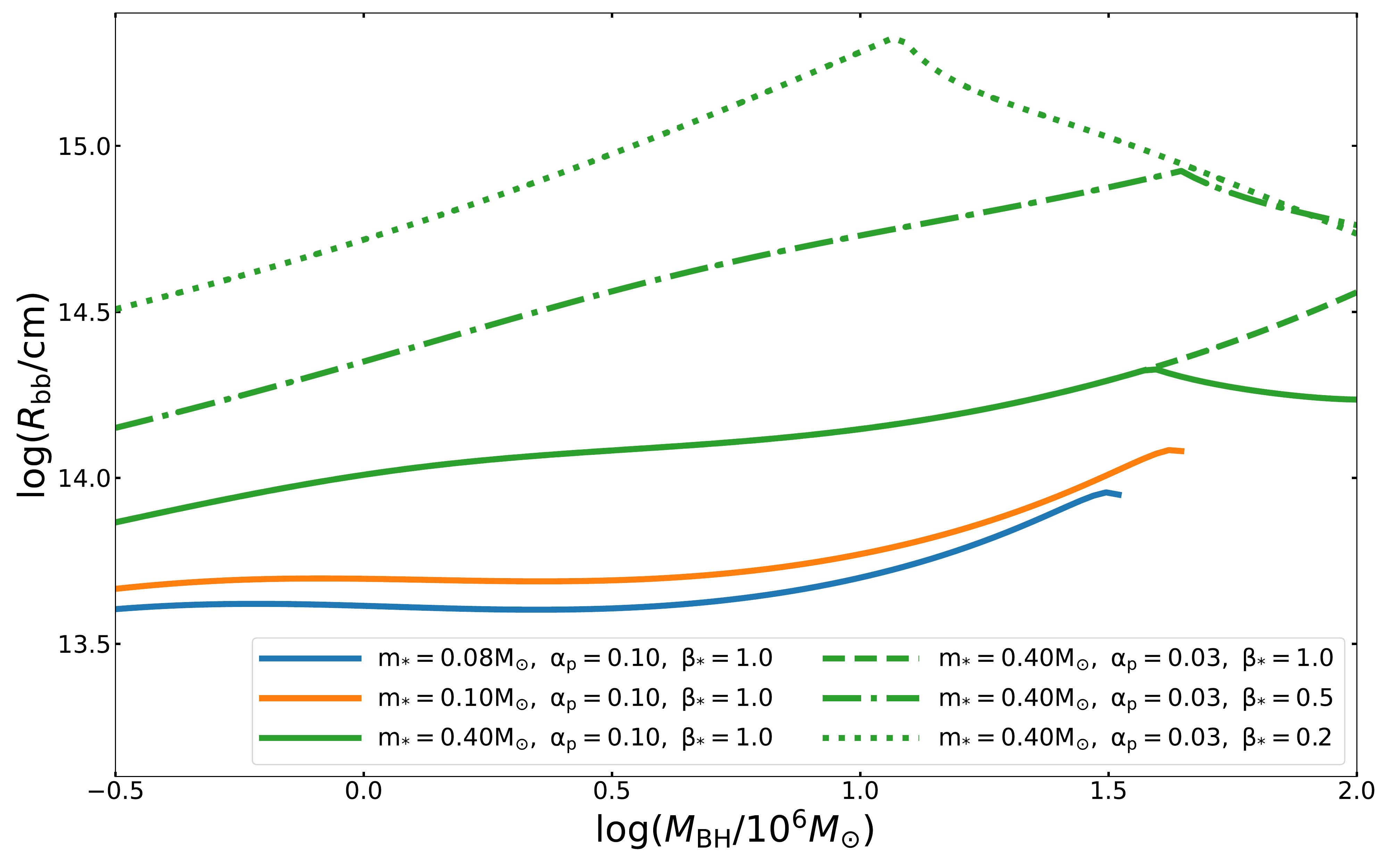} 
	\caption{Peak blackbody radius as a function of the BH mass. The peak blackbody radius is calculated 
	with $n=9/4$ for $\beta_* < 1$ and $n= 5/3$ for $\beta_* \geq 1$. In the calculations, the other parameters are 
	$r_{\rm ms} =2 r_{\rm S}$, and $f_{\rm T} = f_* (0.80+0.26 M_6^{0.5})$ with $f_*=1.212$ \citep{ryu20}. 
	}
		\label{fig:rbbmbh}
\end{figure}

The equation~(\ref{eq:envradius53}) shows that the effective blackbody radius changes with the accretion 
rate
\begin{eqnarray}
R_{\rm bb} &\simeq& 1.87 \times 10^{15}\, ({\rm cm})\, \alpha_{\rm -1}^{-81/68} f^{11/68} 
		\left({2 r_{\rm S} \over r_{\rm ms}}\right)^{1/2} \left(1+\Delta_*\right)^{1/2} 
		\beta_*^{-41/34} \times \cr
&& f_{\rm T}^{-121/68}   r_*^{-39/136} m_*^{64/51}  M_6^{-203/408} \left[{3(n-1) \over 2}\right]^{45/68}
      \left(1-{r_{\rm 0}\over 2a}\right)^{5/34}  \left({\dot{M}\over \dot{M}_{\rm p}}\right)^{45/68} \cr
&\simeq& 1.87 \times 10^{15}\, ({\rm cm})\, \alpha_{\rm -1}^{-81/68} f^{11/68} 
		\left({2 r_{\rm S} \over r_{\rm ms}}\right)^{1/2} \left(1+\Delta_*\right)^{1/2} 
		\beta_*^{-41/34} \times \cr
&& f_{\rm T}^{-121/68}   r_*^{-39/136} m_*^{64/51}  M_6^{-203/408} \left[{3(n-1) \over 2}\right]^{45/68}
      \left(1-{r_{\rm 0}\over 2a}\right)^{5/34}  \left({ t + \Delta{t}_{\rm p} \over \Delta{t}_{\rm 
		p}}\right)^{-45n/68} ,
		\label{eq:envradiuscobsn}
\end{eqnarray}
which gives 
\begin{eqnarray}
R_{\rm bb} &\simeq& 1.87 \times 10^{15}\, ({\rm cm})\, \alpha_{\rm -1}^{-81/68} f^{11/68} 
		\left({2 r_{\rm S} \over r_{\rm ms}}\right)^{1/2} \left(1+\Delta_*\right)^{1/2} 
		\beta_*^{-41/34} \times \cr
&& f_{\rm T}^{-121/68}   r_*^{-39/136} m_*^{64/51}  M_6^{-203/408} \left(1-{r_{\rm 
       0}\over 2a}\right)^{5/34}  \left({ t + \Delta{t}_{\rm p} \over \Delta{t}_{\rm 
		p}}\right)^{-1.10} 
\label{eq:envradiuscobs}
\end{eqnarray}
for $n=5/3$ and 
\begin{eqnarray}
R_{\rm bb} &\simeq& 2.83 \times 10^{15}\, ({\rm cm})\, \alpha_{\rm -1}^{-81/68} f^{11/68} 
		\left({2 r_{\rm S} \over r_{\rm ms}}\right)^{1/2} \left(1+\Delta_*\right)^{1/2} 
		\beta_*^{-41/34} \times \cr
&& f_{\rm T}^{-121/68}   r_*^{-39/136} m_*^{64/51}  M_6^{-203/408} \left(1-{r_{\rm 
       0}\over 2a}\right)^{5/34}  \left({ t + \Delta{t}_{\rm p} \over \Delta{t}_{\rm 
		p}}\right)^{-1.49} 
\label{eq:envradiuscobs49}
\end{eqnarray}
for $n=9/4$, where $r_{\rm 0}$ is given with equation~(\ref{eq:radradiuscent2}). The effective blackbody radius
decreases significantly with time, consistent with the observations. Both equations~(\ref{eq:envradiuscobs}) 
and (\ref{eq:envradiuscobs49}) show that the peak  blackbody radius depends on  both the mass of the star and the 
effective viscosity parameter, $R_{\rm bb} \propto m_*^{(395+117\zeta)/408} \alpha_{\rm p}^{-81/68} \propto 
m_*^{1.08} \alpha_{\rm p}^{-1.19}$ for $\zeta=0.21$. The peak blackbody radius  depends also on the orbital 
penetration factor of the star, $R_{\rm bb} \propto \beta_*^{-41/34}.$ For the BH mass $10^{5.5} M_\odot \la
M_{\rm BH} < 10^8 M_\odot$ and the star mass $0.08 \la m_* \la 1$, the elliptical accretion disk model with 
the ranges of the orbital penetration factor 
$0.2 \la \beta_* \la 3$ and effective viscosity parameter $0.01 \la \alpha_{\rm p} \la 1$ could give a peak 
blackbody  radius consistent with the observations $10^{14} \, {\rm cm} \la R_{\rm bb} \la 10^{15.5} \, {\rm cm}$.
Figure~\ref{fig:rbbmbh} gives the  peak blackbody radius $R_{\rm bb}$ as a function of the BH mass for different
stellar masses, the orbital penetration factor $\beta_*$ of the star, and the effective viscosity parameter $\alpha_{\rm 
p}$. Figure~\ref{fig:rbbmbh} shows that the peak blackbody radius increases with the mass of the star and inversely 
with the orbital penetration factor of the star. When the peak accretion rate $\dot{M}_{\rm p}$ is large 
and the radiation radius $r_0$ is determined by $r_0 \simeq (1+e) a_{\rm d}$, the peak blackbody radius 
increases with the BH mass and is independent of the effective viscosity parameter $\alpha_{\rm p}$. When the 
peak accretion rate $\dot{M}_{\rm p}$ decreases with the BH mass (see equation~(\ref{eq:pmass})) until the 
radiation radius $r_0$ at the peak accretion rate is $r_0 < (1+ e) a_{\rm d}$ and is given with 
equation~(\ref{eq:radradiuscent2}), the peak blackbody radius decreases with the BH mass. The 
critical BH mass depends on the effective viscosity parameter $\alpha_{\rm p}$.

%

\subsection{Anticorrelation of the blackbody temperature and blackbody radius}
\label{subsec:corptpr}

The recent observations with the sample of 39 optical/UV TDEs showed that the blackbody temperature at
the peak brightness strongly anticorrelates with the peak spherical blackbody radius 
\begin{equation}
L_{\rm bb} = 4\pi R_{\rm bb}^2 \sigma_{\rm SB} T_{\rm bb}^4 
\label{eq:corrtemrademp}
\end{equation}
with a scatter of about 0.3 dex and the best fit $L_{\rm bb} \simeq 10^{44.05} \, {\rm erg\; s^{-1}}$ 
\citep{van20}. From equation~(\ref{eq:corrtemrademp}), we have the empirical correlation of the blackbody 
temperature and the effective blackbody radius 
\begin{eqnarray}
T_{\rm bb} &=& \left({L_{\rm bb} \over 4\pi 10^6 r_{11}^2 \sigma_{\rm SB}}\right)^{1/4} \left({R_{\rm bb} \over
	10^3 r_{\rm 11}}\right)^{-1/2} \cr
&\simeq& 3.67\times 10^4\, {\rm K}\, \left({R_{\rm bb} \over 10^3 r_{\rm 11}}\right)^{-1/2} .
\label{eq:corrtemrademp1}
\end{eqnarray}

\begin{figure}
	\centering 
	\includegraphics[width=1\textwidth]{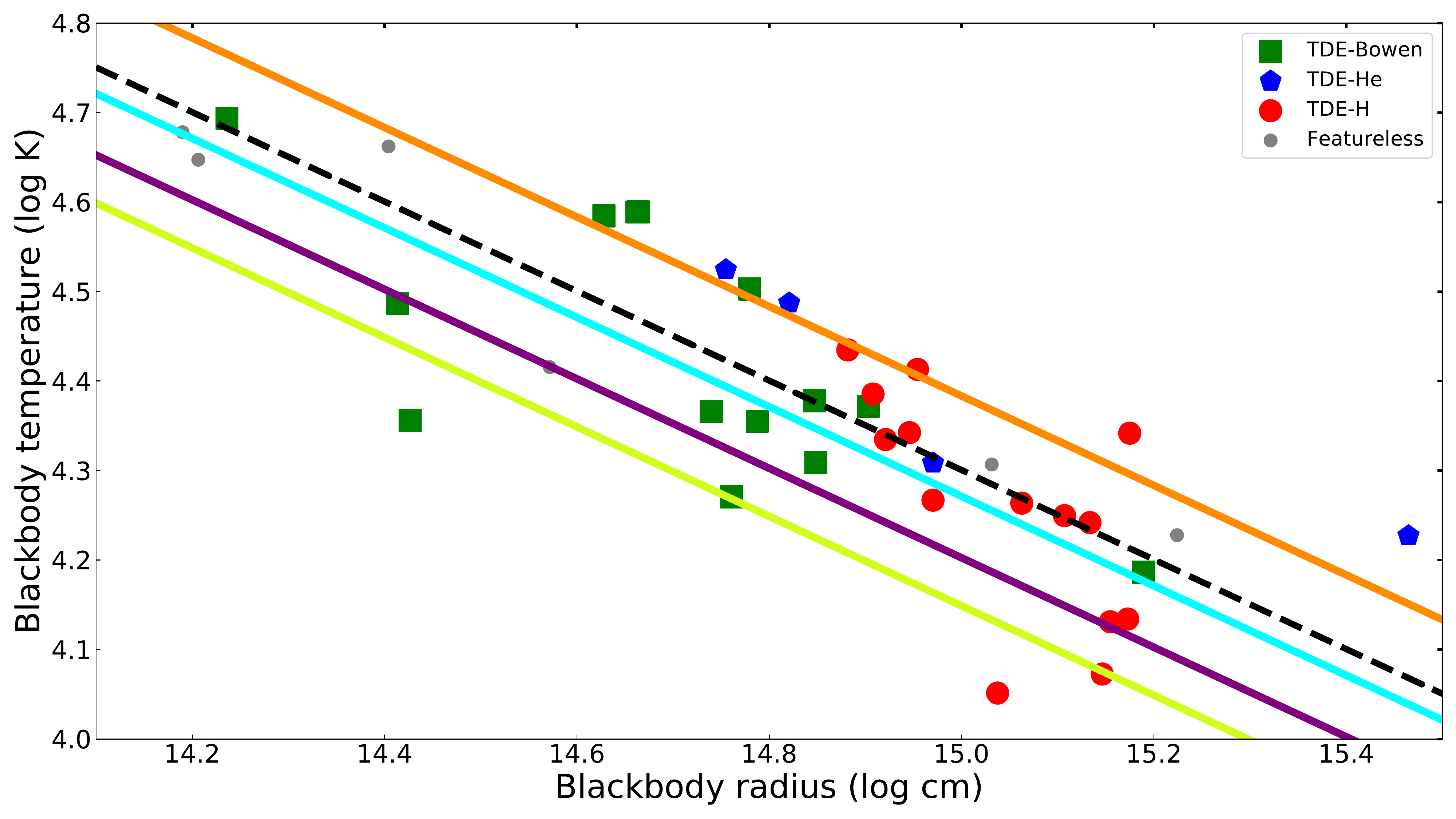} 
	\caption{Expected correlation of the blackbody temperature and the effective 
	blackbody radius vs the observations. The observational data and the best fit (black dashed) 
	are taken from Figure~8 of \citet{van20}. The theoretical correlations are, respectively, for SMBH mass 
	$M_{\rm BH} = 10^{5.5} M_\odot$ (orange solid), $10^6 M_\odot$ (cyan solid), $10^{6.5} M_\odot$
	(purple solid), and $10^7 M_\odot$ (lime yellow solid). In the calculations, the other parameters are 
	$n=5/3$, $r_{\rm ms} =2 r_{\rm S}$, $\beta_* = 1$, $m_*=0.4$, and $f_{\rm T} =f_* (0.80+0.26 
	M_6^{0.5})$ with $f_*=1.212$ \citep{ryu20}. The slope and correlation coefficients of the 
	observations appear naturally in the elliptical accretion disk model.
		\label{fig:tbbrbb}}
\end{figure}
%

Equations~(\ref{eq:tempbb3}) and (\ref{eq:envradius53}) show that both the blackbody temperature 
$T_{\rm bb}$ and the effective blackbody radius $R_{\rm bb}$ depend mainly on the viscosity parameter 
$\alpha_{\rm p}$. With equation~(\ref{eq:envradius1}), we eliminate the viscosity parameters $\alpha_{\rm 
p}$ from equation~(\ref{eq:tempbb2}) and obtain the correlation of the blackbody temperature and  
radius at the peak accretion rate  
\begin{eqnarray}
T_{\rm bb} &\simeq&3.27 \times 10^4 \, {\rm K}\,   \left({1+e \over 2}\right)^{-1/4} \left({2r_{\rm S} \over r_{\rm ms}}\right)^{1/4} 
      \left[1-\delta {(1+\Delta_*) \over (1+e)} \left({r_{\rm S} \over r_{\rm ms} - r_{\rm S}}\right)\right]^{1/4} 
      \left(1+\Delta_*\right)^{1/4} \times \cr
&&  \beta_*^{-1/4} \left({f_{\rm T}\over 1.56}\right)^{-1} r_*^{-3/8} m_*^{7/12} 
      M_6^{-5/24} \left[{3(n-1)\over 2}\right]^{1/4} \left({R_{\rm bb}\over 10^3 r_{\rm 11}}\right)^{-1/2} .
\label{eq:cortempradth}
\end{eqnarray}
Equation~(\ref{eq:cortempradth}) can also be obtained from equation~(\ref{eq:envradius0}) with a bit 
more algebraic calculations. The correlation is independent of both the polytropic index $\gamma$ and of
the physical mechanism driving the variations of the  blackbody temperature. Figure~\ref{fig:tbbrbb} overplots 
the expected correlation and intrinsic scatter given by equation~(\ref{eq:cortempradth}) on the observations 
of optical/UV TDEs \citep{van20}. The theoretical correlation in Figure~\ref{fig:tbbrbb} is obtained with 
$n=5/3$, $r_{\rm ms}=2 r_{\rm S}$, $\beta_*=1$, $M_{\rm BH} =10^{5.5} M_\odot$, $10^6 M_\odot$, 
$10^{6.5} M_\odot$, and $10^7 M_\odot$ with $f_{\rm T} = f_* (0.80 + 0.26 M_6^{0.5} )$, and $m_* = 0.4$. Here 
we use $f_* = 1.212$ for $m_* = 0.4$ \citep{ryu20}. Equation~(\ref{eq:cortempradth}) gives a strong anticorrelation 
of the blackbody temperature and radius with a small scattering because of the weak dependence of the BH 
mass ($\propto M_6^{-0.21}$) and the stellar mass ($\propto r_*^{-3/8} m_*^{7/12} \sim m_*^{(5+9\zeta)/24} 
\sim m_*^{0.287}$). For BHs of mass $10^{5.5} M_\odot \la M_{\rm BH} \la 10^{7} M_\odot$, stars of 
mass $0.08 \leq m_* \la 3 $, and orbital penetration factor $0.5\la \beta_* \la 3$, we have an intrinsic scatter 
$\sim 0.33$ dex. Figure~\ref{fig:tbbrbb} and equation~(\ref{eq:cortempradth}) show that the elliptical accretion 
disk can reproduce not only the anticorrelation of blackbody temperature and blackbody radius but also the  
intrinsic scatter of the empirical correlation and suggest that the intrinsic scatter of the correlation is 
mainly due to the differences of the masses of BHs and stars and possibly  of the orbital penetration factor 
$\beta_*$. The slope of the logarithmic correlation of the temperature and blackbody radius is the result of the 
assumption that no strong outflows emerge from the accretion disk and the radial advection cooling of 
the heat across the ellipse is negligible, resulting in the luminosity closely following the mass fallback rate.
The normalization of the correlation and its dependence on the masses of the BHs and stars and on the 
orbital penetration factor result from the assumptions that the accretion disk is elliptical with nearly uniform 
eccentricity over the disk and that the eccentricity is determined jointly by the location of the self-intersections
and the conservations of the angular momentum of the streams.

\subsection{TDE-Bowen TDEs have smaller blackbody radius and larger blackbody temperature}
\label{sec:Bowen}

Recent studies \citep{van20} have shown that TDEs with both broad Balmer emission lines and Bowen 
fluorescence emission lines (TDE-Bowen TDEs) may have larger blackbody temperatures and smaller 
blackbody radii at peak brightness than TDEs with Balmer line features only (TDE-H TDEs). The two spectroscopic 
classes of TDEs have similar blackbody luminosity. Because the Bowen fluorescence mechanism requires 
both a high flux of EUV photons and a high gas density, \citet{van20} interpret the observations to 
suggest that the TDE-Bowen class has higher gas density, larger blackbody temperature, and smaller 
blackbody radius than the TDE-H population.

In the elliptical accretion disk model, the broad emission lines are suggested to originate in the elliptical 
accretion disk \citep{liu17,cao18}. The elliptical accretion disk model is able to fit well the double-peaked 
broad H$\alpha$ profiles of the TDEs PTF09djl \citep{liu17} and AT 2018hyz/ASASSN-18zj \citep{hun20,sho20}, 
the single-peaked broad H$\alpha$ profiles of ASASSN-14li \citep{cao18},  and the flat-topped Balmer lines 
of AT2018zr/PS18kh \citep{hol18} because of the strong dependence of the emission-line profiles on the 
orientation and shape of the elliptical disk \citep{cao18}, and it can explain the flat Balmer 
decrement of a number of TDEs \citep{sho20}. The disk origin of broad emission lines of TDEs requires that the 
accretion disk of TDE-Bowen TDEs has higher mass density than the accretion disk of TDE-H TDEs does. From 
equation~(\ref{eq:rhorr53}),  we have the mass density of TDEs, $\rho \propto \alpha_{\rm -1}^{-2/17} 
r_*^{-57/34} m_*^{14/17} M_6^{-3/34} \left[3(n-1)/2\right]^{3/17} \left({\dot{M}/ \dot{M}_{\rm 
p}}\right)^{3/17}  \left({r_{\rm p}/ r}\right)$, which is nearly independent of the accretion rate, the BH mass, 
and the viscosity parameters. Because the gas density inversely correlates with the stellar mass, $\rho \propto 
m_*^{-(29-57\zeta)/34} \propto m_*^{-0.501}$, TDE-Bowen TDEs are expected to have smaller  masses of 
stars with respect to the TDE-H population. Equations~(\ref{eq:tempbb3}) and (\ref{eq:envradius53}) show
that a smaller mass of a star implies a higher blackbody temperature and smaller blackbody radius of 
TDE-Bowen TDEs, which are consistent with the observations \citep{van20}. To give quantitative 
comparison of the observations and the disk expectations, detailed radiative transfer calculations of the 
broad emission lines are needed, which is beyond the scope of this paper. Because 
the gas density is nearly independent of the mass of BHs, $\rho \propto M_6^{-3/34}$, and the large 
intrinsic scatter of the host galaxy correlation of the $M_{\rm BH}$-$M_{\rm tot}$ \citep{har04,kor13,mcc13}, 
no correlation between the spectroscopic classification of TDEs and the total mass of host galaxy $M_{\rm tot}$ 
is expected. The prediction is in line with the observations \citep{van20}.

\subsection{Steep decrease of TDE event rate with the effective blackbody radius}

The observations show that the TDE-Bowen class has low optical luminosity at the peak but has been detected in 
equal numbers to the H-only class \citep{van20}. The low luminosity implies a higher intrinsic rate. Because the 
TDE-Bowen class has smaller blackbody radii at peak relative to the H-only class, the observations suggested a 
steep decrease of the event rate of TDEs with the blackbody radius at peak brightness, $\mathrm{d}\dot{N}_{\rm
TDE}/\mathrm{d}R_{\rm bb} \propto R_{\rm bb}^{-3}$ \citep{van20}. \citet{van20} showed that the correlation
between the event rate of TDEs and blackbody radius at peak could be explained with a typical IMF of the 
stellar population, e.g., $\mathrm{d}N_*/\mathrm{d} M_* \propto M_*^{-2.3}$ \citep{kro01}, provided that 
the blackbody radius of TDEs would be proportional to the mass of the star and the stars of the TDE-Bowen class 
have small mass.  

As was discussed in Section~\ref{sec:Bowen},  a high mass density is required to produce Bowen emission
lines, and the stars of the TDE-Bowen class should have smaller masses, consistent with the requirement of the 
observations. From equation~(\ref{eq:envradius53}), the effective blackbody radius is $R_{\rm bb} \propto 
r_*^{-39/136} m_*^{64/51} \propto m_*^{(395+117\zeta)/408}$. Because $\zeta \simeq 0.21$ for $0.1\la 
m_* \leq 1$ and $\zeta\simeq 0.44$ for $1 < m_* \leq 150$ \citep{kip12}, we have $R_{\rm bb} \propto 
m_*^{1.03}$ for $0.1 \la m_* \leq 1$ and $R_{\rm bb} \propto m_*^{1.09}$ for $1 < m_* \leq 150$, exactly 
as required by the observations.

\section{Discussion and conclusions}\label{sec:dis}

Liu and colleagues \citep{liu17,cao18} suggested that the broad optical emission lines of TDEs originate in 
an elliptical accretion disk and showed that the broad double-peaked profiles of H$\alpha$ 
emission lines of TDE PTF09djl  imply a large and highly eccentric elliptical accretion disk of nearly
uniform eccentricity. A highly eccentric 
accretion disk would convert only a small fraction of matter into radiation \citep{liu17,cao18,zho20} -- 
see also \citet{svi17} and \citet{pir15} for a summary of a parallel, independent work by Piran and collaborators, 
who stressed that optical/UV TDEs are powered by the shocks owing to the self-intersections of streams near 
apocenter. The predicted luminosities at peak and total radiation energies and the inferred 
BH masses with the elliptical disk model with uniform eccentricity are well consistent, respectively, with the 
observations of TDEs and the host galaxy properties \citep{zho20}. In this paper, we  investigate the hydrodynamic 
structures and SED of the elliptical accretion disk of uniform eccentricity, based on the analytical 
treatments of fluid hydrodynamics, viscosity, radiative transfer, the heat generation and cooling, and 
the general relativistic effects, captured with the generalized Newtonian potential.

Our results show that the highly eccentric elliptical accretion disk has distinctive hydrodynamic properties 
and SED with respect to the circular accretion disk because of the significant variations of the hydrodynamics 
and radiative transfer around the eccentric ellipse. The elliptical accretion disk cannot reach vertical hydrostatic 
equilibrium, and the flows are laminar because of the variations of the vertical gravitational potential 
around the ellipse. The elliptical accretion disk is geometrically thin and optically thick. The surface density is 
nearly constant around the ellipse as in the circular accretion disk, but the gas density decreases linearly with 
radius $r$ along the ellipse.

Because of the large electron scattering opacity, the soft X-ray photons generated at about the pericenter
are well trapped inside the accretion disk and advected around the eccentric ellipse with little emission. When the 
trapped soft X-ray photons move with the fluids around the ellipse, they are absorbed owing to bound-free 
and free-free absorptions and reemitted in emission lines and low-frequency continuum because of recombination 
and bremsstrahlung radiation. Electron scattering does not reprocess the soft X-ray photons into low frequency, 
but it significantly increases the diffusive path of photons and the effective bound-free and free-free opacities. 
Because the effective Rosseland mean opacity significantly increases with radius, the vertical diffusion
timescale increases with radius and exceeds the dynamical timescale of fluids at the photon-trapping radius 
$r_0$. For $r > r_0$, the low-frequency continuum photons are trapped and advected with the fluids through the 
apocenter and back to $r_0$. The elliptical accretion disk emits mainly in the region of size of about the 
photon-trapping radius $r_0$  and the radiation from region $r> r_0$ is small. The photon-trapping radius $r_0$ is 
the radiation radius and self-regulates owing to the balance of generation and cooling of heat.

Because the photon-trapping radius is self-regulated and changes with the accretion rate, the temperature of 
both the disk center and surface of the radiation radius is nearly independent of the BH mass, accretion rate, 
and the mass of the star. The radiation temperature is determined mainly by the effective viscosity parameter and 
is typically $T_{\rm bb} \simeq 3.0 \times 10^4 \, {\rm K}\, \left({\alpha_{\rm p}/ 0.2}\right)^{81/136} \left[3(n 
-1)/2\right]^{-11/136} \beta_*^{6/17} f_{\rm T}^{-15/136} M_6^{11/272} m_*^{-(75-63\zeta)/272} \left({r_{\rm p} 
/ r_{\rm p*}}\right)^{-6/17} \left({\dot{M} / \dot{M}_{\rm p}}\right)^{-11/136}$ with $\zeta=0.21$ for $0.1\la m_* 
\la 1$. The radiation temperature $T_{\rm bb}$ varies weakly with pericenter radius $r_{\rm p}$ with a power-law 
index of $0.353$, much smaller than the canonical power-law index of $0.75$ in the standard thin accretion disk. The 
radiation blackbody temperature varies only by a small amount, and the SED of an elliptical accretion disk has a 
blackbody spectrum of nearly single temperature. The blackbody temperature $T_{\rm bb}$ is typically about 
$3 \times 10^4\, {\rm K}$ and ranges from $1\times 10^4\, {\rm K}$ to $8 \times 10^4\, {\rm K}$. 
Because the BH mass and accretion rate affect the TDE temperature and SEDs only slightly and the stellar mass
and the viscosity parameter are more influential, it would be difficult to infer BH mass or accretion from real 
observations of any particular event, but it would be easier  to constrain the stellar mass and the viscosity 
parameter.

The elliptical accretion disk has very low  radiation efficiency and gives rise to a typical total luminosity of 
optical/UV TDEs, $L_{\rm bol} \simeq 0.71\times 10^{44} \, {\rm ergs}\, \beta_*^{-1} \left(f_{\rm T}/
1.56\right)^{-4} m_*^{(5+9\zeta)/6} M_6^{-5/6} (1+\Delta_*) \left[3(n -1)/2\right] \left({2r_{\rm S} /r_{\rm 
ms}}\right) \left({\dot{M} / \dot{M}_{\rm p}}\right)$. Most TDEs with stellar mass $m_*<1$ have 
sub-Eddington peak luminosities and are expected to have luminosities closely following the mass fallback rate,
$L_{\rm bol} \simeq 0.49 \beta_*^{-1} \left(f_{\rm T}/1.56\right)^{-4} m_*^{(5+9\zeta)/6} 
M_6^{-11/6} (1+\Delta_*) \left[3(n -1)/2\right] \left({2r_{\rm S} / r_{\rm ms}}\right) \left({\dot{M} /
\dot{M}_{\rm p}}\right) L_{\rm Edd}$. Provided the total luminosity $L_{\rm bol}$ and the blackbody 
temperature $T_{\rm bb}$, we have the effective blackbody radius of the elliptical accretion disk, $R_{\rm 
bb} \simeq 10^{14.57} \, ({\rm cm})\, \left({\alpha_{\rm p} \over 0.2}\right)^{-81/68} \beta_*^{-41/34} 
\left(f_{\rm T}/1.56\right)^{-121/68} m_*^{(395+117\zeta)/408} M_6^{-203/408} \left[3(n -1)/2\right]^{45/68}
(1+\Delta_*)^{1/2} \left({2r_{\rm S} / r_{\rm ms}}\right)^{1/2} \left({\dot{M}/ \dot{M}_{\rm 
p}}\right)^{45/68}$. The effective blackbody radius changes with accretion rate and decrease with time, 
$R_{\rm bb}\propto \left({\dot{M}/ \dot{M}_{\rm p}}\right)^{45/68} \propto t^{-1.10}$ for $n=5/3$ 
and $R_{\rm bb} \propto t^{-1.49}$ for $n=9/4$.

The elliptical accretion disk has a blackbody radiation spectrum of nearly single and time-independent 
temperature and a large and varying effective blackbody radiation radius with accretion rate, which are
the unique characteristics of optical/UV TDEs and are often adopted to identify them among the nuclear 
transients in the literature. We quantitatively compare the expectations of the elliptical 
accretion disk and the observations of optical/UV TDEs. Our results show that the expected blackbody SEDs 
of nearly single temperature, blackbody temperature and variations with time, effective blackbody radius
and dependence on the accretion rate, anticorrelations of the blackbody temperature and 
blackbody radius at peak, and steep decrease of TDE event rate with blackbody radius at peak are all 
well consistent with the observations of optical/UV TDEs. 

Observations show that the TDE-Bowen class with both broad Balmer and Bowen fluorescence 
emission lines has larger blackbody temperatures and smaller blackbody radii at peak brightness than the
TDE-H population with Balmer line features only \citep{van20}. The Bowen fluorescence mechanism 
requires that the TDE-Bowen class should have high gas density. The observations of spectral TDE classes
can be explained, based on the accretion disk origin of the broad emission lines of optical/UV TDEs 
\citep{liu17,cao18}. Because the gas density of the elliptical accretion disk is nearly independent of the 
BH mass, the accretion rate, and the viscosity parameter and is mainly determined by the mass of the star, 
$\rho \propto \alpha_{\rm -1}^{-2/17} m_*^{-(29-57\zeta)/34} M_6^{-3/34} \left[3(n -1)/2\right]^{3/17}
\left({\dot{M} / \dot{M}_{\rm p}}\right)^{3/17} \propto m_*^{-0.501}$ for $\zeta=0.21$, the observations imply 
that the stars of the TDE-Bowen class have small masses. The small masses of the stars would lead to the TDE-Bowen 
class relative to the TDE-H population having a high blackbody temperature, $T_{\rm bb} \propto r_*^{-63/272} 
m_*^{-3/68} \propto m_*^{-(75-63\zeta)/272} \propto m_*^{-0.23}$ for $\zeta=0.21$, and a small effective 
blackbody radius, $R_{\rm bb} \propto r_*^{-39/136} m_*^{64/51} \propto m_*^{(395+117\zeta)/408} 
\propto m_*^{1.03}$ for $\zeta=0.21$. The expectations of the elliptical accretion disk are well consistent 
with the spectroscopic observations of TDEs. The expectation for the correlation of the effective blackbody 
radius and the masses of the stars is well consistent with the observation that the event rate of TDEs steeply 
decreases with the blackbody radius at peak brightness as $\mathrm{d}\dot{N}_{\rm TDE}/\mathrm{d}R_{\rm 
bb} \propto R_{\rm bb}^{-3}$ and the steep decrease of number density of a star with the stellar masses in 
the typical IMF, $\mathrm{d}N_*/\mathrm{d} M_* \propto M_*^{-2.3}$. The expectation for small masses of 
stars of the TDE-Bowen class is consistent with that the TDE-Bowen class relative to the TDE-H population has 
a higher intrinsic event rate.

Most optical/UV TDEs have strong broad optical emission lines but show little or no X-ray emission
\citep[see][for a recent review]{kom15}. In the accretion disk model for the broad optical emission lines, 
an extended soft X-ray source is required to power the broad emission lines of optical/UV TDEs, and
the line emissivity of the accretion disk can be approximated with a broken  power law in radius
\citep{liu17,cao18}. In this paper, we showed that the soft X-ray photons are generated in the pericenter 
region and trapped in disk fluids owing to the extremely large electron scattering opacity. When the soft 
X-ray photons are advected and move around the eccentric ellipse, they would be absorbed mainly as 
a result of photon ionization and free-free absorption and reprocessed into emission lines and low-frequency 
continuum via recombinations and bremsstrahlung radiation. The soft X-ray source is extended up to the 
entire accretion disk and powers the broad emission lines underneath the atmosphere of the accretion disk. 
Because of the high gas density and temperature, the collisional excitation may make some contributions 
to the line emissivity. The broken radius of the broken power law of disk line emissivity may be associated 
at some extent with the photon-trapping radius of the low-frequency continuum. To determine the broken
radius of the disk line emissivity and the correlation with the photon-trapping radius of the continuum, 
numerical simulations with full radiative transfer are needed.

No significant X-ray radiation is expected to be emitted from the disk surface of an elliptical accretion disk. The 
observations show that the soft X-ray radiation is detected in a couple of optical/UV TDEs. The origin of the soft 
X-ray emissions of optical/UV TDEs and the relationship between the optical/UV and soft X-rays TDEs will be 
discussed in the next work. 

Following \citet{liu17} and \citet{cao18}, we assumed for simplicity that the eccentric accretion disk has 
a uniform eccentricity. In reality, the eccentricity may change with radius \citep{bon16,svi17}, in particular 
when the orbital pericenter radius of the fluid in the disk is about the innermost stable circular orbit (ISCO), 
$r_{\rm p} \sim r_{\rm ISCO} = 3 r_{\rm S}$. It is shown that the orbits of the fluid elements at the inner edge 
may even be closer to parabolic \citep{svi17,cha18}. However, our results do not change significantly and 
the conclusions are still valid, if the orbital eccentricity at $r_{\rm p} \sim r_{\rm ISCO}$ is $e \ga 0.6$.

\subsection{Summary}

In this paper, we investigate the dynamic structures and the disk SEDs of the elliptical accretion disks in the 
context of the TDEs. Our results show that such accretion flows have unique characteristics. The elliptical 
accretion disk is geometrically thin and optically thick and cannot reach vertical hydrodynamic static equilibrium. 
The flow is laminar. The surface density is nearly constant around the ellipse, but the gas density and temperature 
significantly vary. The heat and soft X-ray photons are generated at pericenter and nearby and are advected 
around the ellipse without escaping, because of the large electron scattering opacity and photon trapping. The 
soft X-ray photons are absorbed owing to the bound-free and free-free absorption and reprocessed into line emission
and low-frequency continuum via recombinations and free-free emission. 
Because of the rapid increase of the bound-free and free-free opacities with radius, the low-frequency continuum 
photons become trapped in the fluid at the photon-trapping radius and are advected through the apocenter
and back to the photon-trapping radius. The emission of the low-frequency continuum originates mainly at the 
photon-trapping radius. Because the photon-trapping radius is self-regulated and changes with accretion rate, the 
radiation temperature is nearly independent of both BH mass and accretion rate and depends weakly on 
the mass of the star and the viscosity parameter. The SED of the elliptical accretion disk resembles that of a 
single-temperature blackbody. Our results imply that it would be difficult to infer the BH mass from the real 
observations of any particular event, but it would be easier to constrain the stellar mass and the 
viscosity parameter. The predictions of our elliptical accretion disk model are well consistent with the 
observations of optical/UV TDEs.

\section*{Acknowledgements}

We would like to thank Tsvi Piran and Stefanie Komossa for insightful comments and discussions and Jiayi 
Tang for some technical help. We are very grateful to the anonymous referee for helpful comments. This 
work is supported by the National Natural Science Foundation of China (NSFC No.~11473003, NSFC No.~11721303),
the National Key R\&D Program of China (grant No.~2020YFC2201400), and the Strategic Priority Research 
Program of the Chinese Academy of Sciences (Grant No. XDB23010200 and No. XDB23040000). M.A. is supported 
in part by the inter-excellence project No.~LTI17018, aimed to strengthen international collaboration
of Czech scientific institutions. M.W. acknowledges the support of the Black Hole Initiative at Harvard University, 
which is funded by grants from the John Templeton Foundation and the Gordon and Betty Moore Foundation to 
Harvard University.



\begin{thebibliography}{}

\bibitem[Abramowicz et al.(1988)]{abr88} Abramowicz, M.A., Czerny, B., Lasota, J.P., \& Szuszkiewicz, E. 
    1988, \apj, 332, 646

\bibitem[Abramowicz et al.(1978)]{abr78} Abramowicz, M.A., Jaroszynski, M., \& Sikora, M. 1978, \aap, 63, 221

\bibitem[Abramowicz et al.(2000)]{abr00} Abramowicz, M.A., Lasota, J.P., \& Igumenshchev, I.V. 2000, \mnras,
    314, 775

\bibitem[Arcavi  et al.(2014)]{arc14} Arcavi, I., Gal-Yam, A., Sullivan, M., et al. 2014, \apj, 793, 38

\bibitem[Asplund et al.(2009)]{asp09} Asplund, M., Grevesse, N., Sauval, A.J., \& Scott, P. 2009, \araa,
           47, 481

\bibitem[Blagorodnova et al.(2019)]{bla19} Blagorodnova, N., Cenko, S.B., Kulkarni, S.R., et al. 2019, 
      \apj, 873, 92

\bibitem[Bonnerot et al.(2016)]{bon16} Bonnerot, C., Rossi, E. M., Lodato, G., \& Price, D. J. 2016, \mnras, 
455, 2253

\bibitem[Cao et al.(2018)]{cao18} Cao, R., Liu, F.K., Zhou, Z.Q., Komossa, S., \& Ho, L.C. 2018, \mnras, 
    480, 2929

\bibitem[Chan et al.(2018)]{cha18} Chan, C.-H.,  Krolik, J.H., \&  Piran, T., 2018, \apj, 856, 12

\bibitem[Coughlin \& Nixon(2019)]{cou19} Coughlin, E.R., \& Nixon, C.J. 2019, \apjl, 883, L17

\bibitem[Dai et al.(2015)]{dai15} Dai, L.,  McKinney, J.C.., \& Miller, M.C. 2015, \apjl, 812, L39

\bibitem[Dai et al.(2018)]{dai18} Dai, L., McKinney, J.C., Roth, N., Ramirez-Ruiz, E., 
      \& Miller, M.C. 2018, \apjl, 859, L20.

\bibitem[de Sitter(1916)]{des16} de Sitter, W. 1916, \mnras, 77, 155

\bibitem[Eddington(1918)]{edd18} Eddington, A.S. 1918, \apj, 48, 205

\bibitem[Evans \& Kochanek(1989)]{eva89} Evans, C.R., \& Kochanek, C.S. 1989, \apjl, 346, L13

\bibitem[Frank et al.(2002)]{fra02} Frank, J., King, A., \& Raine, D. 2002, Accretion Power in Astrophysics, 
    (Cambridge University Press: Cambridge)

\bibitem[Gaskell \& Rojas Lobos(2014)]{gas14} Gaskell, C.M., \& Rojas Lobos, P.A. 2014, \mnras, 438, L36

\bibitem[Gezari et al.(2017a)]{gez17} Gezari, S., Blagorodnova, N., Roth, N., et~al.\ 2017a, \apj, 842, 29

\bibitem[Gezari et al.(2017b)]{gez17b} Gezari, S., Cenko, S.B. \& Arcavi, I. 2017b, \apjl, 851, L47

\bibitem[Gezari et al.(2012)]{gez12} Gezari, S., Chornock, R., Rest, A., et al.\ 2012, \nat, 485, 217

\bibitem[Gomez et al.(2020)]{gom20} Gomez, S.,  Nicholl, M., Short, P., et al., 2020,  \mnras, 497, 1925

\bibitem[Guillochon et al.(2014)]{gui14} Guillochon, J., Manukian, H., \& Ramirez-Ruiz, E. 2014, \apj, 783, 23

\bibitem[Guillochon  \& Ramirez-Ruiz(2013)]{gui13} Guillochon, J. \& Ramirez-Ruiz, E. 2013, \apj, 767, 25

\bibitem[H\"aring \& Rix(2004)]{har04} H\"aring, N., \& Rix, H.-W. 2004, \apjl, 604, L89

\bibitem[Hayasaki \& Loeb(2016)]{hay16} Hayasaki, K. \& Loeb, A.\ 2016, NatSR, 6, 35629

\bibitem[Hayasaki et al.(2013)]{hay13} Hayasaki, K., Stone, N., \& Loeb, A.\ 2013, \mnras, 434, 909

\bibitem[Hills(1975)]{hil75} Hills, J.G. 1975, \nat, 254, 295

\bibitem[Hinkle et al.(2021)]{hin20a} Hinkle, J.T., Holoien, T.W.-S., Auchettl, K., et al. 2021, \mnras, 500, 1673

\bibitem[Hinkle et al.(2020)]{hin20} Hinkle, J.T., Holoien, T.W.-S., Shappee, B.J., et al. 2020, \apjl, 894, L10

\bibitem[Holoien et al.(2018)]{hol18} Holoien, T.W.-S., Brown, J.S.,  Auchettl, K., et al. 2018, \mnras, 480, 5689

\bibitem[Holoien et al.(2019)]{hol19} Holoien, T.W.-S., Huber, M.E., Shappee, B.J., et al. 2019, \apj,  880, 120 

\bibitem[Holoien et al.(2016a)]{hol16a} Holoien, T.W.-S., Kochanek, C.S., Prieto, J.L., et~al.\ 2016a, \mnras, 455, 2918

\bibitem[Holoien et al.(2016b)]{hol16b} Holoien, T.W.-S., Kochanek, C.S., Prieto, J.L., et~al.\ 2016b, \mnras,  463, 3813

\bibitem[Holoien et al.(2014)]{hol14} Holoien, T.W.-S., Prieto, J.L., Bersier, D., et~al.\ 2014, \mnras, 445, 3263

\bibitem[Hung et al.(2020)]{hun20} Hung, T., Foley, R.J., Ramirez-Ruiz, E., et al., 2020, \apj, 903, 31 

\bibitem[Ivanov \& Chernyakova(2006)]{iva06} Ivanov, P.B., \& Chernyakova, M. A. 2006, \aap, 448, 843

\bibitem[Jiang et al.(2016)]{jia16} Jiang, Y.-F., Guillochon, J., \& Loeb, A. 2016, \apj, 830, 125

\bibitem[Kajava et al.(2020)]{kaj20} Kajava, J.J. E., Giustini, M., Saxton, R.D., \& Miniutti, G. 2020, \aap, 639, A100

\bibitem[Kippenhahn \& Weigert(2012)]{kip12} Kippenhahn, R., \& Weigert, A. 2014, Stellar Structure and
Evolution (Berlin, New York: Springer-Verlag)

\bibitem[Kochanek(1994)]{koc94} Kochanek, C.S. 1994, \apj, 422, 508

\bibitem[Komossa(2015)]{kom15} Komossa, S. 2015, Journal of High Energy Astrophysics, 7, 148 

\bibitem[Komossa et al.(2008)]{kom08} Komossa, S., Zhou, H., Wang, T., et~al.\ 2008, \apjl, 678, L13

\bibitem[Kormendy \& Ho(2013)]{kor13} Kormendy, J., \& Ho, L.C. 2013, \araa, 51, 511

\bibitem[Krolik et al.(2020)]{kro20} Krolik, J., Piran, T., \& Ryu, T. 2020, \apj, 904, 68 

\bibitem[Krolik et al.(2016)]{kro16} Krolik, J., Piran, T., Svirski, G., \& Cheng, R.M. 2016, \apj, 827, 127

\bibitem[Kroupa(2001)]{kro01} Kroupa, P. 2001, \mnras, 322,  231

\bibitem[Leloudas et al.(2019)]{lel19} Leloudas, G., Dai, L.X., Arcavi, I., et al. 2019, \apj,  887, 218 

\bibitem[Liu et al.(2017)]{liu17} Liu, F.K., Zhou, Z.Q., Cao, R., Ho, L.C., \& Komossa, S.\  2017, \mnras,
472, L99

\bibitem[Liu et al.(2019)]{liu20} Liu, X.-L., Dou, L.-M., Shen, R.-F., \& Chen, J.H. 2019, \apj, submitted; 
arXiv:1912.06081

\bibitem[Lodato et al.(2009)]{lod09} Lodato, G., King, A.R., \& Pringle, J. E. 2009, \mnras, 392, 332

\bibitem[Lodato \& Rossi (2011)]{lod11} Lodato, G., \& Rossi, E.M. 2011, \mnras, 410, 359 

\bibitem[Lyubarskij et al.(1994)]{lyu94} Lyubarskij, Y.E., Postnov, K.A., \& Prokhorov, M.E., 1994, \mnras, 266, 583

\bibitem[McConnell \& Ma(2013)]{mcc13} McConnell, N.J., \& Ma, C.-P., 2013, \apj, 764, 184

\bibitem[Metzger \& Stone(2016)]{met16} Metzger, B.D., \& Stone, N.C. 2016, \mnras, 461, 948

\bibitem[Mockler et al.(2019)]{moc19} Mockler, B., Guillochon, J., \& Ramirez-Ruiz, E.\ 2019, \apj, 872, 151

\bibitem[Nicholl et al.(2019)]{nic19} Nicholl, M., Blanchard, P.K., Berger, E., et al. 2019, \mnras, 488, 1878 

\bibitem[Ogilvie(2001)]{ogi01} Ogilvie, G.I. 2001, \mnras, 325, 231

\bibitem[Ogilvie \& Barker(2014)]{ogi14} Ogilvie, G.I., \& Barker, A.J. 2014, \mnras, 445, 2621

\bibitem[Phinney(1989)]{phi89} Phinney, E. S. 1989, in IAU Symp. 136, The Center of the Galaxy, 
    ed. M. Morris, (Dordrecht: Kluwer), 543

\bibitem[Piran(2015)]{pir15a} Piran, T. 2015, in The Jerusalem Bagel Model - Elliptical Accretion, lecture at 
the Jerusalem TDE workshop (Jerusalem: Israel Institute for Advanced Studies, Hebrew Univ. Jerusalem), 
(http://astro-icore.phys.huji.ac.il/node/64)

\bibitem[Piran et al.(2015)]{pir15} Piran, T., Svirski, G., Krolik, J., Cheng, R. M., \& Shiokawa, H. 2015, \apj, 806, 164

\bibitem[Ramirez-Ruiz \& Rosswog(2009)]{ram09} Ramirez-Ruiz, E., \& Rosswog, S., 2009, \apjl, 697, L77

\bibitem[Rees(1988)]{ree88} Rees, M.J.\ 1988,  \nat, 333, 523

\bibitem[Roth et al.(2016)]{rot16} Roth, N., Kasen, D., Guillochon, J., \& Ramirez-Ruiz, E. 2016, \apj, 827, 3

\bibitem[Ryu et al.(2020a)]{ryu20} Ryu, T., Krolik, J., Piran, T., \& Noble, S.C. 2020a,  \apj, 904, 98 

\bibitem[Ryu et al.(2020b)]{ryu20b} Ryu, T., Krolik, J., Piran, T., \& Noble, S.C. 2020b,  \apj, 904, 99 

\bibitem[Sadowski et al.(2016)]{sad16} Sadowski, A., Tejeda, E., Gafton, E., et al. 2016, \mnras, 458, 4250

\bibitem[Saxton et al.(2018)]{sax18} Saxton, C.J., Perets, H.B., \& Baskin, A. 2018, \mnras, 474, 3307

\bibitem[Shakura \& Sunyaev(1973)]{sha73} Shakura, N.I., \& Sunyaev, R.A. 1973, \aap, 24, 337

\bibitem[Shiokawa et al.(2015)]{shi15} Shiokawa, H., Krolik, J.H., Cheng, R.M., Piran, T., \& Noble, S.C.
  2015, \apj, 804, 85

\bibitem[Short et al.(2020)]{sho20} Short, P., Nicholl, M., Lawrence, A., et al. 2020,  \mnras, 498, 4119

\bibitem[Stein et al.(2020)]{ste20} Stein, R., van Velzen, V., Kowalski, M., et al. 2020, arXiv:2005.05340

\bibitem[Steinberg et al.(2019)]{ste19} Steinberg, E., Coughlin, E.R., Stone, N.C., \& Metzger, B.D. 2019,
    \mnras, 485, L146

\bibitem[Stone et al.(2013)]{sto13} Stone, N., Sari, R., \& Loeb, A. 2013, \mnras, 435, 1809

\bibitem[Strubbe \& Quataert(2009)]{str09} Strubbe, L.E., \& Quataert, E. 2009, \mnras, 400, 2070

\bibitem[Svirski et al.(2017)]{svi17} Svirski, G., Piran, T., \& Krolik, J. 2017, \mnras, 467, 1426

\bibitem[Syer \& Clarke(1992)]{sye92} Syer, D., \& Clarke, C.J. 1992, \mnras, 255, 92

\bibitem[Tejeda \& Rosswog(2013)]{tej13} Tejeda, E., \& Rosswog, S. 2013, \mnras, 433, 1930

\bibitem[van Velzen et al.(2021)]{van20} van Velzen, S., Gezari, S., Hammerstein, E., et al. 2021, \apj, 908, 4

\bibitem[van Velzen et al.(2019)]{van19} van Velzen, S., Stone, N.C., Metzger, B.D., et al. 2019, \apj, 878, 82

\bibitem[Wang et al.(2012)]{wan12} Wang, T.-G., Zhou, H.-Y., Komossa, S., et al. 2012, \apj, 749, 115

\bibitem[Wevers et al.(2019)]{wev19} Wevers, T., Stone, N.C., van Velzen, S., et al. 2019, \mnras, 487, 4136 

\bibitem[Wevers et al.(2017)]{wev17} Wevers, T., van Velzen, S., Jonker, P.G., et al. 2017, \mnras, 471, 1694

\bibitem[Zanazzi \& Ogilvie(2020)]{zan20} Zanazzi, J.J., \& Ogilvie, G.I. 2020, \mnras, 499, 5562

\bibitem[Zhou et al.(2021)]{zho20} Zhou, Z.Q., Liu, F.K., Komossa, S., et al. 2021, \apj, 907, 77

\bibliographystyle{aasjournal}


\end{thebibliography}
\end{document}